\documentclass[sigplan,9pt]{acmart}

\usepackage{booktabs} 
\usepackage{acm_information}
\usepackage{graphicx}
\usepackage{subfig}
\usepackage{blindtext}
\usepackage{adjustbox}
\usepackage{multirow}
\usepackage{color}
\usepackage{booktabs}
\usepackage{tabularx}
\usepackage{colortbl}

\usepackage{tikz}
\usepackage{listings}
\usepackage{etoolbox}
\usepackage{subfig}
\usepackage{url}
\usepackage{setspace}
\usepackage[british,english]{babel}
\usepackage{color, colortbl}
\usepackage{enumitem}
\usepackage{balance}

\definecolor{Gray}{gray}{0.9}

\newcommand{\circled}[2][]{\tikz[baseline=(char.base)]
	{\node[shape = circle, draw, inner sep = 1pt]
		(char) {\phantom{\ifblank{#1}{#2}{#1}}};%
		\node at (char.center) {\makebox[0pt][c]{#2}};}}
\robustify{\circled}

\newcommand{\ProfileSize}{100MB\xspace}
\newcommand{\PCA}{\texttt{PCA}\xspace}
\newcommand{\cs}{\texttt{cs}\xspace}
\newcommand{\KNN}{\texttt{KNN}\xspace}
\newcommand{\PCs}{\texttt{PCs}\xspace}
\newcommand{\PC}{\texttt{PC}\xspace}
\newcommand{\RDD}{\texttt{RDD}\xspace}
\newcommand{\RDDs}{\texttt{RDDs}\xspace}
\newcommand{\Quasar}{\textsc{Quasar}\xspace}
\newcommand{\DColocation}{\textsc{Pairwise}\xspace}
\newcommand{\Oracle}{\textsc{Oracle}\xspace}
\newcommand{\oracle}{\textsc{Oracle}\xspace}

\begin{document}
\title{Improving Spark Application Throughput Via Memory Aware Task Co-location: A Mixture of Experts Approach}


%

\author{Vicent Sanz Marco, Ben Taylor, Barry Porter, Zheng Wang}
\affiliation{%
  \institution{MetaLab, School of Computing and Communications, Lancaster University, U. K.}
} \email{{v.sanzmarco, b.d.taylor, b.f.porter, z.wang}@lancaster.ac.uk}

\renewcommand{\shortauthors}{V. S. Marco et al.}
\renewcommand{\shorttitle}{Improving Spark Application Throughput Via Memory Aware Co-location}

%
%
%
%

\begin{abstract}
Data analytic applications built upon big data processing frameworks such as Apache Spark are an important class of applications. Many of
these applications are not latency-sensitive and thus can run as batch jobs in data centers. By running multiple  applications on a
computing host, task co-location can significantly improve the server utilization and system throughput. However, effective task
co-location is a non-trivial task, as it requires an understanding of the computing resource requirement of the co-running applications,
in order to determine what tasks, and how many of them, can be co-located. State-of-the-art co-location schemes either require the user
to supply the resource demands which are often far beyond what is needed; or use a one-size-fits-all function to estimate the
requirement, which, unfortunately, is unlikely to capture the diverse behaviors of applications.

In this paper, we present a mixture-of-experts approach to model the memory behavior of Spark applications. We achieve this by learning,
off-line, a range of specialized memory models on a range of typical applications; we then determine at runtime which of the memory
models, or experts, best describes the memory behavior of the target application. We show that by accurately estimating the resource
level that is needed, a co-location scheme can effectively determine how many applications can be co-located on the same host to improve
the system throughput, by taking into consideration the memory and CPU requirements of co-running application tasks. Our technique is
applied to a set of representative data analytic applications built upon the Apache Spark framework. We evaluated our approach for system
throughput and average normalized turnaround time on a multi-core cluster.  Our approach achieves over 83.9\% of the performance
delivered using an ideal memory predictor. We obtain, on average, 8.69x improvement on system throughput and a 49\% reduction on
turnaround time over executing application tasks in isolation, which translates to a 1.28x and 1.68x improvement over a state-of-the-art
co-location scheme for system throughput and turnaround time respectively.
\end{abstract}


\copyrightyear{2017} \acmYear{2017} \setcopyright{acmlicensed} \acmConference{Middleware '17}{December 11-15, 2017}{Las Vegas, NV,
USA}\acmPrice{15.00}\acmDOI{10.1145/3135974.3135984} \acmISBN{978-1-4503-4720-4/17/12}

\maketitle

\section{Introduction}
\label{sect_introduction}

Big data applications built upon frameworks such as Hive~\cite{Thusoo2009}, Hadoop~\cite{Shvachko2010} and Spark~\cite{Zaharia2010} are
commonplace. Unlike interactive jobs, many of the data analytic applications are not latency-sensitive. Therefore, they often run as batch
jobs in a data center. However, how to effectively schedule such applications to improve the server utilization and
the system throughput remains a challenge.

Specifically, if an application task is given the entirety of main memory on each host to which it is deployed, it is effectively
preventing the host machine from being used for any other application until the current one has finished, even if the task does not use all
of the memory. Because many data analytic tasks do not use 100\% of the CPU during execution~\cite{7310708,Jiang2014} there is a
significant portion of unused processing capacity. An alternate approach is to share the computing host between multiple application tasks
(where each task does not use all of the memory), this could save time and energy by
co-locating processes more effectively on fewer machines.

Effective task co-locations require knowledge of the application's resource demand. For in-memory data processing frameworks like Apache
Spark, RAM consumption is a major concern~\cite{Li2015}. It is particularly important to understand the memory behavior of the application.
If we co-locate too many applications or give too much data to a single task, such that their total memory consumption exceeds the physical
memory of the host, we could cause memory paging onto the hard disk, or an ``out-of-memory" error, slowing down the overall system. To
achieve this we need a technique to predict the precise memory requirement of any given Spark application.

Existing task co-location schemes require either: the user to provide information of the resource
requirement~\cite{Hindman:2011:MPF:1972457.1972488}, or employ an analytical~\cite{Grandl:2014:MPC:2619239.2626334} or statistical
model~\cite{Luk:2009:QEP:1669112.1669121,gdwzlcpc13,Delimitrou:2014:QRQ:2541940.2541941} to estimate the resource requirement based on
historical jobs or runtime profiling.  These approaches, however, have significant drawbacks. Firstly, it is difficult for a user to give a
precise estimation of the application's requirement; and thus, the supplied information is often over-conservative, asking far more
resources than the application needs. Secondly, a one-size-fits-all function is unlikely to precisely capture behaviors of diverse
applications, and no matter how parameterized the model is, it is highly unlikely that a model developed today will always be suited for tomorrow.

In this paper, we present a generic framework to model the memory behavior of Spark applications. As a departure from prior work that uses
a fixed utility function to model the resource requirement, we use multiple linear and non-linear functions to model the memory
requirements of various applications. We then build a machine learning classifier to select which function should be used for a given
application and dataset at runtime. As the program implementation, workload and underlying hardware changes, different models will be
dynamically selected at runtime. Such an approach is known as \emph{mixture-of-experts}~\cite{ Jacobs:1991:AML:1351011.1351018}. The
central idea is that instead of using a single monolithic model, we use multiple models (\emph{experts}) where each expert is specialized
for modeling a subset of applications. Using this approach, each memory model is used only for the applications for which its predictions
are effective. One of the advantages of our approach is that new functions can easily be added and are selected only when appropriate. This
means that the system can evolve over time to target a wider range of applications, by simply inserting new functions. The result is a new
way of using machine learning for system optimization, with a generalized framework for a diverse set of applications.

We evaluate our approach on a 40-node multi-core cluster using 44 Spark applications that cover a wide range of application domains. We
show that the accurate memory-footprint prediction given by our approach enables the runtime scheduler to make better use of spare
computing resources to improve the overall system throughput via task co-location. We use two distinct metrics to quantify our results:
\emph{system throughput} and \emph{average normalized turnaround time}, and compare our approach against a state-of-art resource
scheduler~\cite{Delimitrou:2014:QRQ:2541940.2541941}. Experimental results show that our approach is highly accurate in predicting the
application's memory requirement, with an average error of 5\%. By better utilizing the memory resources of a host, our system achieves
8.69x improvement of system throughput and a 49\% reduction in application turnaround time. This translates to a 1.28x and 1.68x
improvement over the state-of-art respectively on throughput and turnaround time.

This paper makes the following contributions:

\begin{itemize}[leftmargin=5mm]
\item We present a novel machine learning based approach to automatically learn how to model the memory behavior of
Spark applications (Section~\ref{sect_model});

\item Our work is the first to employ mixture-of-experts for resource demand modeling. Our generic framework allows new models to be easily added to target a wider range of applications and performance metrics;

\item We show how to combine this resource modeling framework with runtime task co-location policies to improve
system throughput for Spark applications (Section~\ref{sec:runtimedeploy});

\item Our system is immediately deployable on real systems and does not require any modification to the application
source code.

\end{itemize}

\section{Background and Overview}

\subsection{Apache Spark}
Apache Spark is a general-purpose cluster computing framework~\cite{Zaharia2010}. with APIs in Java, Scala and Python and libraries for
streaming, graph processing and machine learning~\cite{Zaharia2010}. It is one of the most active open source projects for big data
processing, with over 2,000 contributors in 2016.

Each Spark application runs as an independent set of \emph{executor} processes, each with dedicated memory space for executing parallel
jobs within the application. The executors are coordinated by the \emph{driver} program running on a \emph{coordinating node}.  Input data
of Spark applications is stored in a shared filesystem and organized as \emph{resilient distributed datasets} (\RDDs) -- a collection of
objects that can be operated on in parallel. Each Spark executor allocates its own heap memory space for caching \RDDs. This work exploits
the data parallel property of \RDDs to characterize the application's memory behavior without wasting computing cycles.

\subsection{Problem Scope}
Our goal is to develop a framework to accurately predict the resource requirement of Spark applications for arbitrary inputs. In this work,
we focus on the memory requirement as RAM resources are a major concern for in-memory data processing frameworks like Apache
Spark~\cite{Li2015}. To demonstrate the usefulness of our approach, we apply it to perform task co-location for batched, data-analytic
Spark applications.  We do not consider latency-sensitive applications, such as search, as their stringent response time targets often
require isolated execution~\cite{Lo:2015:HIR:2749469.2749475}.

Our approach estimates the memory footprint of a Spark executor for a given input dataset. It then uses this information to determine if
there are enough spare resources (i.e. memory and CPU) to co-locate tasks; if there is, it calculates how many tasks could be co-located
and how much work should be given to each task. We exploit the fact that many big data applications do not spend all of their time at 100\%
CPU~\cite{Jiang2014} (in our case, the averaged CPU load is under 40\% -- see Section~\ref{sec:cpu_usage}). This observation suggests that there are opportunities to co-locate Spark
tasks without significantly increasing the CPU contention and slow down the performance of co-running applications (see also
Section~\ref{sec:interferences}). Our approach is applied to a simple task co-location policy in this work, yet the resulted scheme
outperforms the state-of-the-art task scheduling scheme. We want to stress that our framework can be used by other scheduling policies to
provide an estimation of the application's resource demand to support decision making.

Our current implementation is restricted to applications whose memory footprint is a function of their input size, this is a typical
behavior for many data analytical applications. In this work, we do not explicitly model disk and network I/O contention, because prior
research suggests that they have little impact on the performance on the type of the applications we
target~\cite{Ousterhout:2015:MSP:2789770.2789791}. It is to note that this observation may not hold for I/O intensive applications such as
database workloads~\cite{DBLP:journals/pvldb/RupprechtCP17}. We are also aware that not all applications' memory consumption is correlated
to the input size and would require adding new functions to make predictions based on other parameters, such as the model size of a machine
learning algorithm. Nonetheless, our framework is general and allows new models to be easily added
to target different applications, or other performance and resource metrics in the future.


\subsection{Overview of Our Approach}

\begin{figure}[t!]
	\centering
	\includegraphics[width=0.45\textwidth]{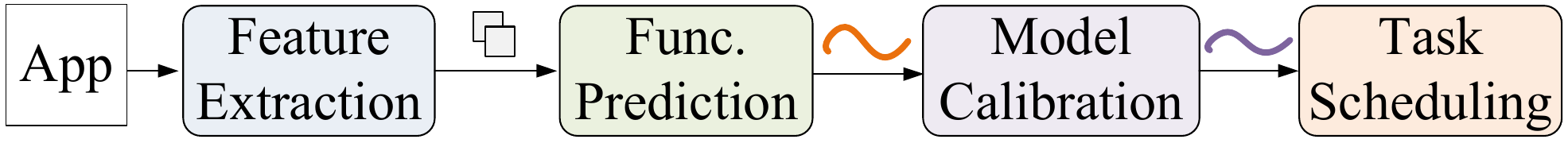}
	\caption{Overview of our approach.
     For an incoming application,
    our approach first extracts the features of the program.
    Based on the feature values, it predicts which of the off-line learned memory
    functions best describes the memory behavior of the application. It then
    instantiates the function parameters by profiling the application on some small
    sets of the input data items. A runtime scheduler then utilizes the
    memory function to perform task co-location.}
	\label{fig_4part_approach}
\end{figure}

Our approach, depicted in Figure~\ref{fig_4part_approach}, is \emph{completely automated}, and no modification to the
application source code is required.


Our mixture-of-experts framework for memory footprint prediction consists of a
range of distinct models built off-line. An expert selector decides which model should be invoked, based on the runtime
information of the application. To use our resource modeling framework to perform task co-location, a task scheduler
follows a number of steps described as follows.

For each ``\emph{new}" application that is ready to run, we predict which of the \emph{off-line} learned experts,
\emph{termed `memory function' in this paper}, best describes its memory behavior, i.e. how the memory footprint
changes as the input size varies. The selection of the memory function is based on runtime information of the program, such
as the number of L1 data and instruction cache misses. This information is collected by running the
application on a small portion (around \ProfileSize) of the input data items\footnote{We choose this modest input size
as an input of this size typically takes a short time to process, while at the same time, it is sufficiently large (i.e.
this often results in a working set that is larger than the size of the L3 data cache in most of the high-end CPUs) to capture the cache
behavior of the application.}.

We then calibrate the selected function to tailor its parameters to the target program and  input. We do so by first profiling the application with two small different-sized parts of the application input to instantiate two function
parameters; we then use the measured memory footprints to instantiate the parameter values. The calibrated memory
function is then used to determine how many unprocessed data items should be allocated to an executor under a given
memory budget. During the profiling run, we also record the average CPU usage of the application.  After determining
which memory function to use and obtaining the CPU usage of the application, the runtime scheduler can spawn new
executors to run on servers that have spare memory, and if the aggregate CPU load of co-running tasks will go over
100\% (i.e. to avoid CPU contention).

Since runtime information collection and model calibration are all performed on some unprocessed data
items and contribute to the final output, no computing cycle is wasted on profiling.
Furthermore, we will re-run an executor process in isolation if it
fails because of an ``out-of-memory" error, but this was not observed in our experiments.



The key to our approach is choosing the right memory function and then using lightweight profiling to instantiate the
function parameters. An alternative is to use extensive profiling runs at runtime to find a model to fit the application's memory
behavior. However, doing so will incur significant overhead and  could outweigh the benefit (see Section~\ref{exp:online}).

In the next section, we will describe how supervised machine learning~\cite{Alpaydin:2010:IML:1734076} can be used to construct the memory functions
(experts) and the expert selector to choose which function to use  for any ``\emph{unseen}" applications.

\section{Predictive Modeling \label{sect_model}}

Our approach involves using multiple memory functions (experts) to capture the memory requirement of an application for a specific runtime input.
The set of memory functions are constructed offline on a set of example programs, and then an expert selector dynamically chooses the best expert to use
at runtime.

Our expert selector for determining the memory function is a K-nearest neighbour (\KNN)\footnote{We have also explored several alternative
classification techniques, including decision trees and neural networks. This is discussed in Section~\ref{sec:model_analysis}.}
classifier~\cite{Keller1985}. The input to the classifier is a set of runtime features. Its output is a label to the memory function that
describes the memory behavior of the target application and the specific dataset.


\subsection{Learning Memory Functions \label{sec:lfunc}}

\begin{figure}[t!]
	\centering
	\includegraphics[width=0.42\textwidth]{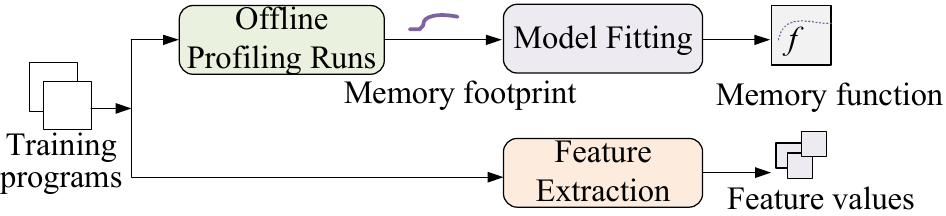}
	\caption{The training process.}
	\label{fig:training}
\end{figure}

\begin{table}[!t]
	\centering
	\caption{Memory functions used in this work}
	    \small
		\begin{tabular}{l l}
        \toprule
		\textbf{Modeling Technique} & \textbf{Formula}\\
		\midrule
        (Piecewise) Linear Regression & $y = m * x^b$ \\
		Exponential Regression & $y= m * (1 - e^{(-b*x)})$ \\
		Napierian Logarithmic Regression& $y= m + ln(x)*b $ \\
        \bottomrule
		\end{tabular}
	\label{memory_function}
\end{table}

\begin{figure}[!t]
	\centering
	\subfloat[Sort]{\includegraphics[width=0.22\textwidth]{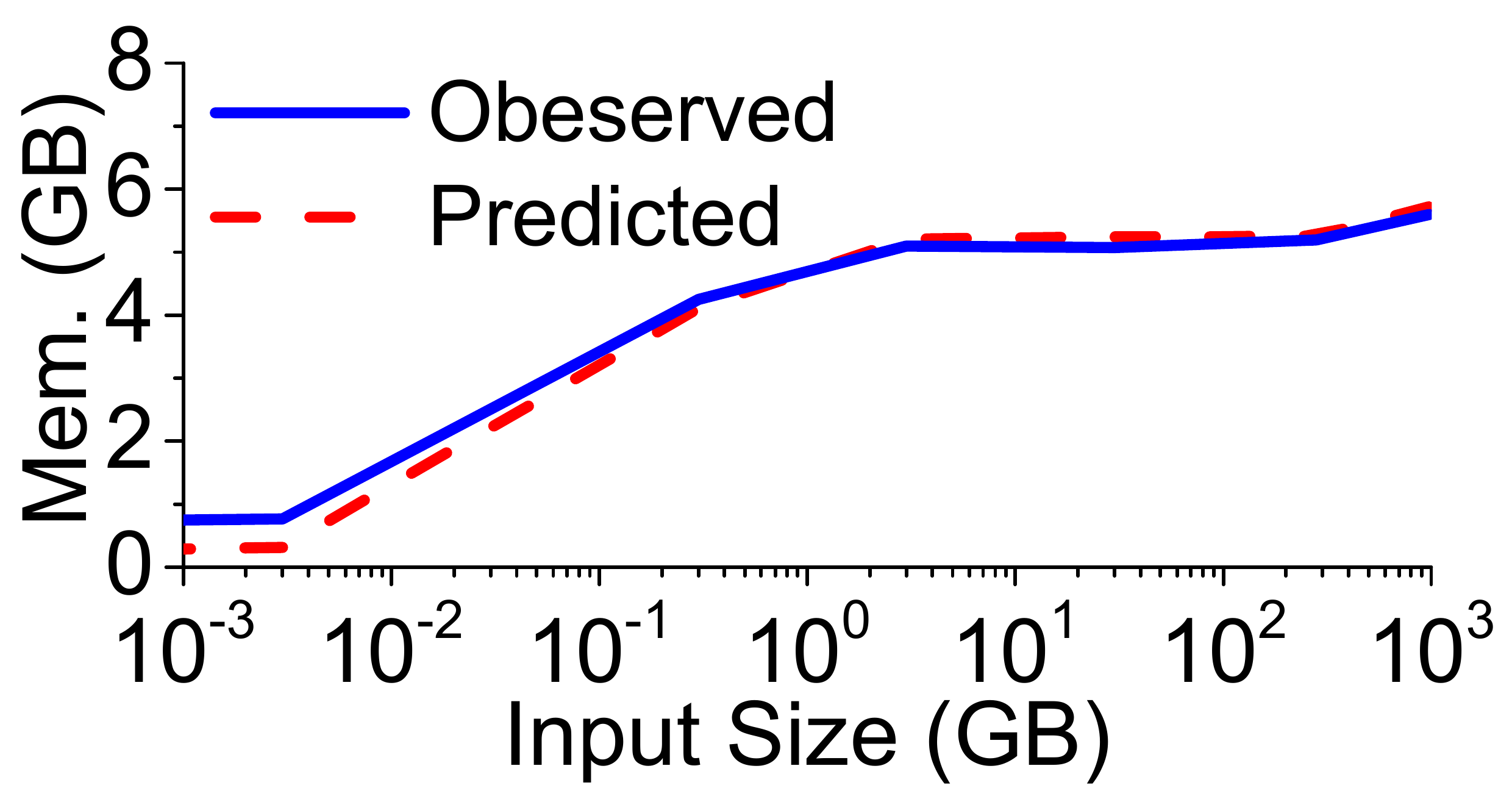}
	}
    \subfloat[PageRank]{\includegraphics[width=0.22\textwidth]{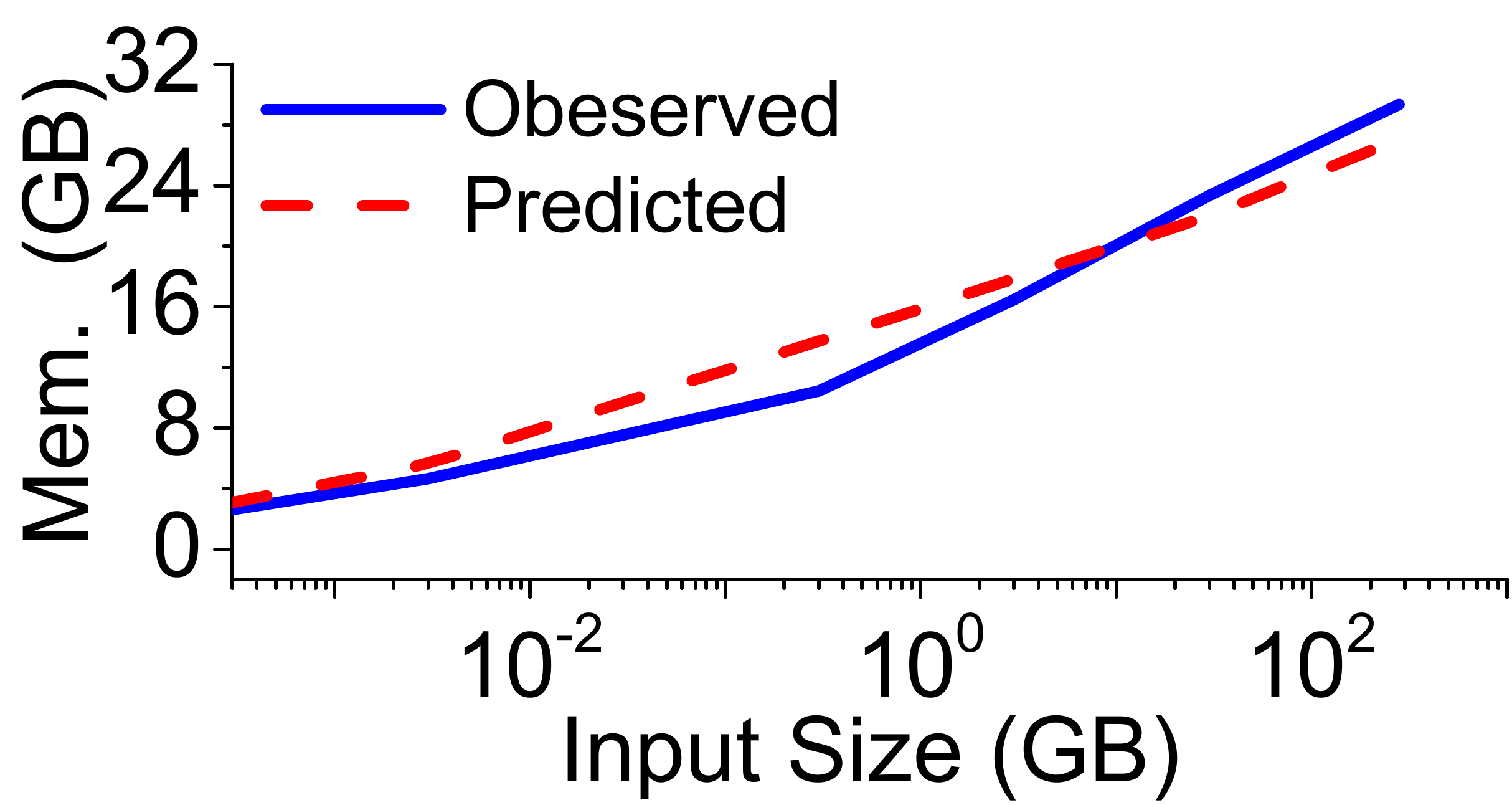}}
    \caption{ The observed and predicted memory footprints for \texttt{Sort} and \texttt{PageRank} from HiBench.
     The memory footprint of the two applications can be accurately described using one of the memory functions listed
     in Table~\ref{memory_function}.
    }
    \label{fig:memory_usage}
\end{figure}

Our memory functions and expert selector are trained \emph{off-line} using a set of training programs. The learned expert selector can then
be used to predict which memory function to use for any \emph{new, unseen} application. Figure~\ref{fig:training} depicts the process of
collecting training data to learn the memory functions to build the expert selector.  For each training program, we find a mathematical
function to model the application's memory footprint and collect its feature values.

During the training process, we run selected training programs in isolation on a computing host. We profiled each training application with
different sized inputs. For each program input, we record the memory footprint of the Spark executor process. Next, we try different
mathematical modeling techniques to discover which model best describes the relationship between input size and memory allocation, that is,
as the input size increases, how does the memory allocation change. In this training phase, we record the memory function used to describe
each training program. Our intuition is that the memory behavior for programs with similar characteristics will be similar. This hypothesis
is confirmed in Section~\ref{sec:model_analysis}.

We use a set of linear
and non-linear regression techniques to model the application's memory
behavior. Table~\ref{memory_function} gives the full list of modeling
techniques we used in this work. Each of our models has two parameters, $m$
and $b$, to be instantiated during runtime model calibration. Here $x$ and
$y$ are the input size (i.e. the number of \RDD objects in our case) and the
predicted memory footprint respectively. It is worth mentioning that all the
memory functions are automatically learned from training data, treating the
applications as black boxes; new applications would similarly be learned
automatically, potentially causing the addition of new memory functions.

\paragraph{Example.}
Figure~\ref{fig:memory_usage} shows the observed memory footprint and the
prediction given by our memory function for \texttt{Sort} and
\texttt{PageRank}. For these two  applications, the memory functions used in this work can
accurately model their memory behaviors.
Specifically, the memory footprint, $y$, of  \texttt{Sort} and
\texttt{PageRank} for a given input size,
$x$, can be precisely described using an exponential function , $y= m * (1 - e^{(-b*x)})$, where $m=5.768$, $b=4.479$ and
a Napierian logarithmic function, $y= m + ln(x)*b $, where $m=16.333$, $b=1.79$ respectively.

%


After building the memory functions, we need to have a mechanism to decide which of the functions to use. One of the
key aspects in building a successful expert predictor is finding the right features to characterize the input
application task. This process of feature selection is
described in the next section. This is followed by sections describing training data
generation and then how to use the expert selector at runtime.

\begin{table*}[!t]
	\caption{Raw features, sorted by their importance}
	\setlength{\tabcolsep}{2pt}
\small
	\label{table:candidate_features}
	\centering
	\begin{tabular}{llllllll }
		\toprule
		\textbf{Abbr.} & \textbf{Desc.} & \textbf{Abbr.} & \textbf{Desc.} & \textbf{Abbr.} & \textbf{Desc.} & \textbf{Abbr.} & \textbf{Desc.}\\
		\midrule
		\rowcolor{Gray}    \texttt{L1\_TCM} & L1 total cache miss rate	&
                            \texttt{L1\_DCM} & L1 data cache miss rate  &
                           \texttt{vcache} & \% of memory used as cache &
                            \texttt{L1\_STM} & L1 cache store miss rate \\

                           \texttt{bo} & \# blocks sent (/s) &
                           \texttt{L2\_TCM} & L2 data cache miss rate &
                           \texttt{L3\_TCM} & L2 total cache miss rate &
						   \cs &  \# context switches / s \\

        \rowcolor{Gray}     \texttt{FLOPs} & \# floating point operations /s &
                            \texttt{in} & \# interrupts / s &
                            \texttt{L2\_DCM} & L3 cache total miss rate &
						    \texttt{L2\_LDM} & L2 cache load miss rate \\

                            \texttt{L1\_ICM} & L1 instr. cache miss rate &
                            \texttt{swpd} & \% of virtual memory used &
                            \texttt{L2\_STM} & L2 cache store miss rate &
                            \texttt{IPC} & instruction per cycle \\

       \rowcolor{Gray}      \texttt{L1\_LDM} & L1 cache load miss rate &
							\texttt{L2\_ICM} & L2 instr. cache miss rate &
                            \texttt{ID} & \% of idle time&
                            \texttt{WA} & \% of time on IO waiting \\
                            \texttt{US} & \% spent on user time &                             \texttt{SY}&\% spent on kernel time \\
		\bottomrule
	\end{tabular}
\end{table*}

\subsection{Runtime Features}
\label{sect_run_feat}

\paragraph{Raw Features.} Expert selection is based on runtime characteristics of the
application task. These characteristics, called \emph{features}, are collected using system-wide profiling tools:
\texttt{vmstat}, Linux \texttt{perf} and performance counter tool \texttt{PAPI}. Collected feature values are
encoded to a vector of real values.  We considered 22 raw features in this work, which are given in
Table~\ref{table:candidate_features}. Some of these features are selected based on our intuition, while others are
chosen based on prior work~\cite{Yang:2015:CPM:2749469.2750401}. All these features can
be automatically and externally observed, without needing access to the source code.


\paragraph{Feature Scaling.} Supervised learning typically works better if the feature values lie in a certain
range. Therefore, we scaled the value for each of our features between the range of 0 and 1. We record the maximum and minimum value
of each feature found at the training phase, and use these values to scale features extracted from a new application
during runtime
deployment.

\begin{figure}[!t]
	\centering
	\subfloat[Principal components]{\includegraphics[width=0.20\textwidth]{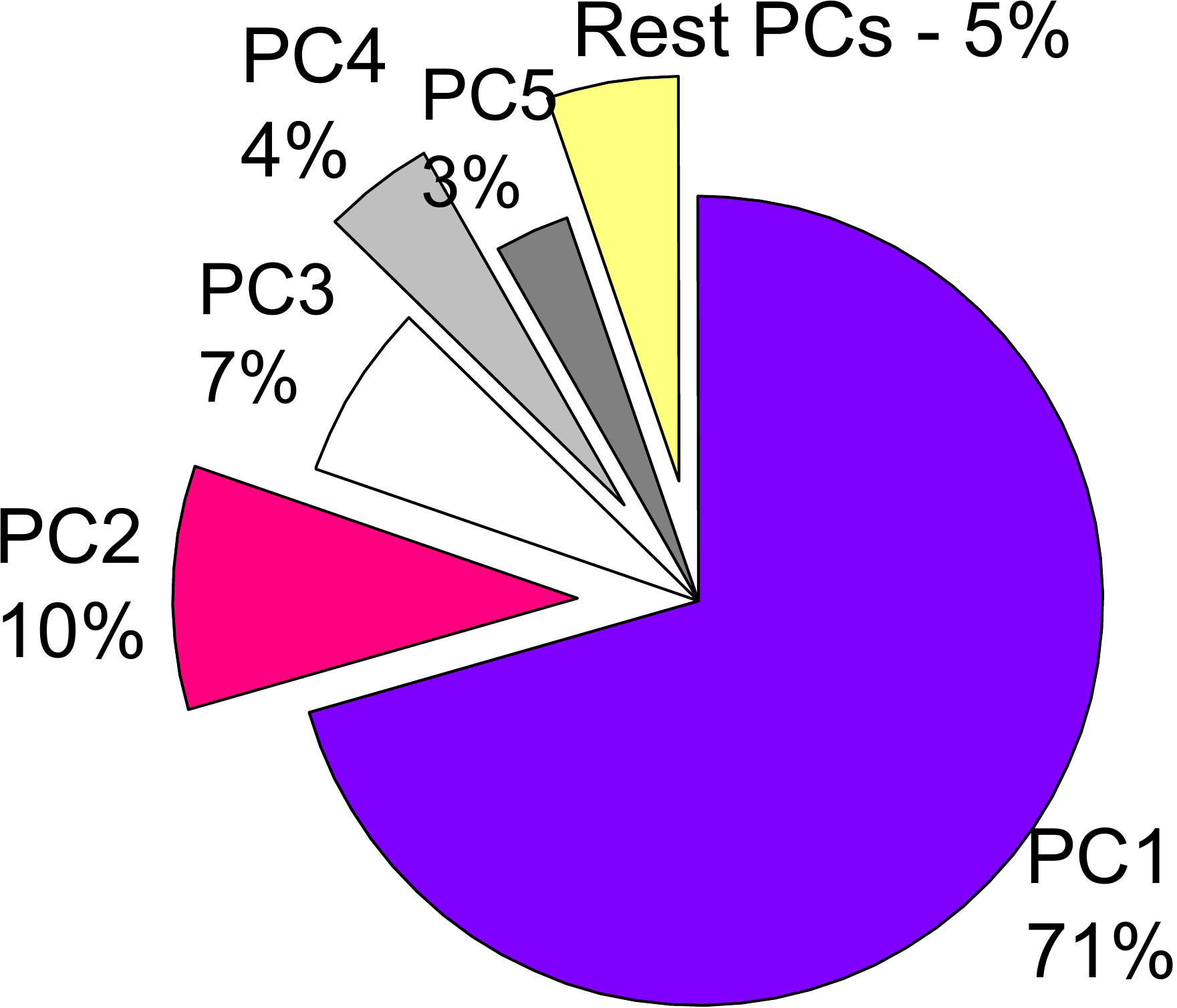}
	}\hfill
    \subfloat[Most important raw features]{\includegraphics[width=0.20\textwidth]{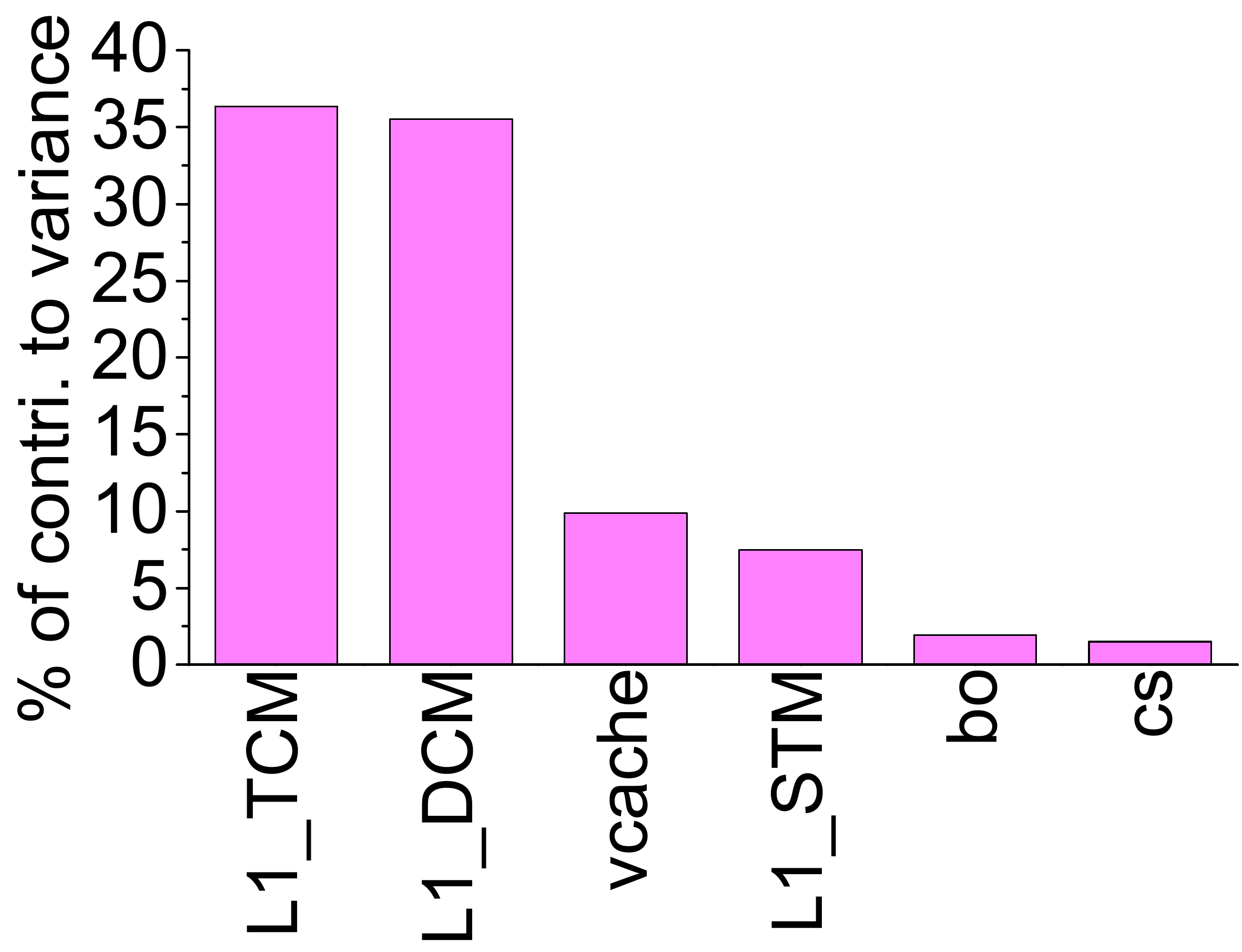}}
    \caption{The percentage of principal components (\PCs) to the overall feature variance (a), and
    contributions of the 5 most important raw features in the \PCA space (b).
    }
    \label{fig:feature_demo}
\end{figure}

\paragraph{Feature Reduction.} Given the relatively small number of training applications, we need
to find a compact set of features in order to build an effective predictor. Feature reduction is automatically
performed through applying Principal Component Analysis (\PCA) on the scaled raw features. This technique removes the
redundant features by linearly aggregating features that are highly correlated. After application of \PCA, we use
the top 5 principal components (\PCs) which account for 95\% of the variance of the original feature space.
We record the \PCA transformation matrix and use it to transform the raw features of the target application to \PCs
during runtime deployment.
Figure~\ref{fig:feature_demo}a illustrates how much feature variance that each component accounts for.
This figure shows that prediction can accurately draw upon a subset of aggregated feature values.

\paragraph{Feature Analysis.} To understand the usefulness of each raw feature, we apply the Varimax rotation~\cite{manly2004multivariate}
to the \PCA space. This technique quantifies the contribution of each feature to each \PC. Figure~\ref{fig:feature_demo}b shows the top 5
dominant features based on their contributions to the \PCs. Cache features, \texttt{L1\_TCM}, \texttt{L1\_DCM} and \texttt{L1\_STM}, are
found to be important for describing memory behaviors. This is not supervising as cache hit/miss rates are shown to be useful in
characterizing the application behavior in prior works~\cite{ Cavazos:2007:RSG:1251974.1252540,Singh:2009:RTP:1577129.1577137}. Other
features of virtual memory usage (\texttt{vcache}), I/O (\texttt{bo}) and thread contention (\texttt{cs}) are also considered to be useful,
but are less important compared to  cache features. Using this technique, we sort the raw features listed in
Table~\ref{table:candidate_features} according to the importance. The advantage of our feature selection process is that it  automatically
determines what features are useful when targeting a new computing environment where the importance of features may change.
Later in Section~\ref{sec:model_analysis}, we quantify the similarity of programs mapped to the same memory function, which
provides additional evidences to justify the choice of features.

Recently, deep neural networks, such as Long short-term memory (LSTM), are shown to be powerful in extracting features from program source
code~\cite{pact17}. Since our current implementation only relies on runtime information, techniques like \cite{pact17} could be used to
provide additional features obtained from the source code. It is to note that to train an effective deep neural network will require a
significantly larger number of training examples than we used in this work.

\subsection{Collect Training Data}



We use \emph{cross-validation} to construct memory functions and the \KNN classifer to select which function to use. This standard
evaluation technique works by picking some target programs for testing and using the remaining ones for training. In this work, we use
benchmarks from the HiBench~\cite{Huang2011} and BigDataBench~\cite{Gao2013} suites to build the memory models. Later we show that our
approach works well on benchmarks from the Spark-Perf~\cite{SparkPerf2016} and the Spark-Bench~\cite{Li2015} suites, although we did not
directly train our models on them. The process of collecting training data is described in Figure~\ref{fig:training}. To collect training
data, we first extract the feature values of each training program by running a single executor process in isolation, using inputs with an
average size of \ProfileSize. Next, we run each training program with different sized inputs (ranging from $\sim$300MB to $\sim$1TB) and
record the observed memory footprints. We then find a memory function to closely fit the curve. For each training program, we record its
principal component values and the memory function. Since training is only performed once, it is a \emph{one-off} cost.

Like any supervised learning based approaches, the effectiveness of our approach relies having a sufficient volume of high-quality training data. We find the
set of training benchmarks gives good performance in this work, because the benchmarks already cover a wide range of typical Spark applications. However, we remark that when
moving to a new application domain, additional benchmarks may need in order to have an adequate sampling over the program space. In this
case, one will need to add more training programs or using an automatic benchmark generator~\cite{cummins2017synthesizing} to automatically synthesize these programs.

\subsection{Modeling Other Metrics and Program Phases} We believe our approach can be extended to model other metrics, e.g. CPU contention. This involves
finding appropriate raw-features, modelling techniques for the experts and the expert selector, and employing a multi-objective scheduling
policy like~\cite{Cooper}. Besides these, the rest of our approach for automatic feature selection and model generation remains the same.
Furthermore, while not explored in this work, our approach can model changing program phases by e.g. treating a long-running phase as an
individual application.

\section{Runtime Deployment \label{sec:runtimedeploy}}
Once we have learned the memory functions as described above, we can
use a KNN algorithm to choose an appropriate function to estimate the memory footprint for any \emph{unseen}
applications with a given input, and to use the prediction to co-locate Spark executor tasks at runtime.

\begin{figure}[t!]
  \centering
  \includegraphics[width=.35\textwidth]{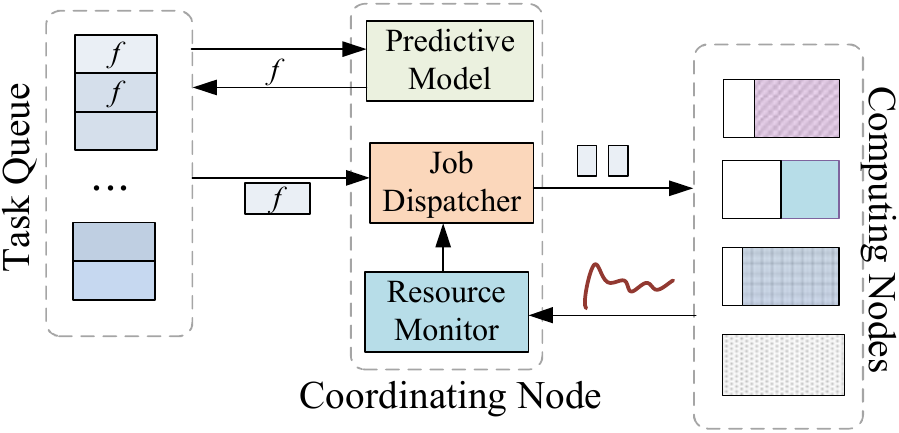}\\
  \caption{Our system predicts the memory function for each application and monitors the memory resources of computing nodes.
  The runtime scheduler creates new executors to run on computing nodes that have spare memory, and uses the memory function to determine how many data items should be given to
  the executor under a memory budget. }\label{fig:runtime_component}
\end{figure}

Our runtime system built upon YARN~\cite{Vavilapalli:2013:AHY:2523616.2523633}, a task and resource manager for Spark.
The co-location scheme will be triggered when more than one Spark application is waiting to be scheduled.
Figure~\ref{fig:runtime_component} illustrates the architecture of our system.
For each application task, we predict its memory function for its input dataset, and then use the memory function to co-locate Spark executor processes whenever possible.

\subsection{Memory Requirement Prediction}
To determine the memory function for an application task, a runtime system follows two steps, described as follows.

\paragraph{Memory Function Prediction.} We run the incoming application on a small set of the input \RDD objects (with an aggregated size
of around \ProfileSize) to collect and normalize feature values, and to perform the \PCA transformation. We then calculate the Euclidean
distance between the transformed input program feature vector and the feature vector of each training program to find out the \emph{nearest
neighbor}, i.e. the training program that is closest to the input program in the feature space (see also Section~\ref{sec:model_analysis}).
We use the memory function of the nearest neighbor as the prediction. One advantage of using \KNN is that the Euclidean distance used for
closeness evaluation can be used to measure the prediction confidence, which essentially provides a degree of soundness guarantee. For
example, if an application is too far from any training program, we can fall back to the default scheme to run the application, and
simultaneously re-train a new memory function for future. Our current implementation performs feature extraction by running the application
on the lightly-loaded coordinating node (where the driver program runs). The results generated in the feature extraction phase will
contribute to the final output of the application.

\paragraph{CPU Load.} We also record the average CPU usage during the profiling run, and use this information later to determine whether
co-location will cause CPU contention among co-running tasks.

\paragraph{Model Calibration.}
After we have determined the memory function, we need to instantiate the function coefficients (i.e. $m$ and $b$ in Table~\ref
{memory_function}). We calculate these by running the application on two sets of unprocessed input data items, where the first and the
second sets contain 5\% and 10\% of the input items, respectively. To determine the function parameters, we measured the memory footprints
during profiling runs, and use them together with the corresponding input sizes (i.e. the number of data objects) to solve the memory
function equation. At this stage, we are only concerned with the application's memory footprint but not runtime. Therefore, profiling runs
can be performed by either grouping different application tasks to run on a single host or running the target application with other
latency-insensitive tasks. Again, the results generated during this phase will contribute to the final output of the application. Moreover,
since the input and output data of the Spark application typically stored in a shared filesystem, we do not
need to explicitly move the data in or out from the profiling host.

\subsection{Resource Monitor}
Each computing node runs a daemon that periodically reports to the resource monitor its memory usage and CPU load. Our
current implementation reports the average memory usage and system load within a 5-minute window. The information is
retrieved from the Linux ``\texttt{/proc}" system. Since this is performed at a coarse-grained level (i.e. minutes), the
overhead of monitoring and communication is negligible.
With this monitoring scheme in place, a task scheduler can respond to execution phase changes and load variations, avoiding
over-subscribing the computing resources.


\subsection{Job Dispatcher}
By default, we use the dynamic allocation scheme of Spark to determine how many
free server nodes to use to run an application. However, the Spark dynamic scheme is not perfect, so we utilize
spare memory to spawn \emph{additional} executors to run on servers that have spare resources. Also, instead of waiting for the
servers to become completely free, our approach starts executing waiting applications as soon as possible, reducing the
turnaround time.

Once we have the memory function of the highest-priority application, the job dispatcher will spawn a new executor for
the application to run on severs that have spare memory and if the aggregate CPU load of all co-running tasks will not
go over 100\%. The dispatcher uses  the memory function to determine how much memory is needed for the remaining input
(to allow us to co-locate more applications if possible), and how many data items can be cached by the executor under a
given memory budget. To estimate the aggregate CPU load, we add up the CPU load of the computing host (which is reported by the
resource monitor) and the average CPU usage of the application to be scheduled (which is obtained during the profiling run
for feature collection). Furthermore, the number of data items to give to the co-located executor is dynamically
adjusted over time, adapting to the changes of execution stages and memory resources. A naive alternative is to
statically set the executor heap size to the size of free memory. But doing so can over-subscribe the memory resources
than necessary and precludes
co-locating more than two applications (see Section~\ref{sec:overallp}).

To minimize the potential thread contention, we dynamically adjust the number of threads (tasks) created by each
executor to evenly distribute processor cores across currently-running executors on a single host. Furthermore, to
enforce a certain degree of fairness, it is important to make sure that the new co-running task does not use the
resources that are deemed to be essential for the currently running application. While fairness is not a focus of this
work, our prediction framework helps the scheduler in this endeavor.

\section{Experimental Setup}
\subsection{Platform and Benchmarks}

\paragraph{Hardware.}
We use a  multi-core cluster with 40 nodes, each has an 8-core Xeon E5-2650 CPU @ 2.6GHz (16 threads with hyper-threading),  64GB of DDR4
RAM, and 16GB of swap. Nodes have SSD storage and are connected through 10Gbps Ethernet, precluding disk and network contention.

\paragraph{Software.} Each computing node runs CentOS 7.2 with Linux kernel 3.12. We rely on the local OS to schedule processes and do not
bind tasks to  specific cores. We use Apache Spark 2.1.0 with Hadoop Yarn 2.6 as the cluster manager and HDFS as the Spark file management
system. We use the Oracle Java runtime, Java SE 8u. We run Spark in the cluster mode. We also use the dynamic resource allocation scheme,
so that memory will be given back to Spark when an application task completes. We run the Spark driver on a dedicated coordinating node and
try to run multiple Spark executors on a single host to improve the system throughput. Finally, we use the Spark default configuration for
memory management, but we dynamically adjust the the number cores and heap size per executor to match the available hardware resources.

\paragraph{Workloads.} We used 44 Java-based Spark applications from four widely used suites: HiBench~\cite{Huang2011},
BigDataBench~\cite{Gao2013}, Spark-Perf~\cite{SparkPerf2016} and Spark-Bench~\cite{Li2015}. We used the native Spark implementations from
these suites. These benchmarks implement the core algorithms used in real-life applications e.g. machine learning, image and natural
language processing, and web analysis.

\subsection{Evaluation Methodology \label{sec:eval_method}}

\begin{table}[!t]
\centering
\caption{Application task mixes used in the experiments}
\setlength{\tabcolsep}{4pt}
    \small
    \begin{tabular}{crcrcrcr}
    \toprule
    \textbf{Label} & \textbf{\#App.} & \textbf{Label} & \textbf{\#App.} & \textbf{Label} & \textbf{\#App.} & \textbf{Label} & \textbf{\#App.}\\
    \midrule
     \rowcolor{Gray} L1 & 2 & L2 & 6 & L3 & 7 & L4 & 9 \\
     L5 & 11 & L6 & 13 & L7 & 19 & L8 & 23 \\
      \rowcolor{Gray} L9 & 26 & L10 & 30 & & & &\\
    \bottomrule
    \end{tabular}
	\label{table_list_bench}
\end{table}

\paragraph{Runtime Scenarios.} We evaluated our scheme using ten runtime scenarios with a mix of
2 to 30 randomly selected applications, detailed in Table~\ref{table_list_bench}.
For each scenario, we try $\sim$100 different application mixes and make sure all benchmarks are included in each
scenario.  The input size ranges from small ($\sim$300MB) and medium ($\sim$30GB) to large ($\sim$1TB). Inputs were
generated using the input generation tool provided by each benchmark suite. In the experiments, all tasks are scheduled
on a first come first serve basis, but we stress that
our technique can be applied to \emph{any scheduling policy}.

\paragraph{Predictive Model Evaluation.} Our memory functions (experts) and expert selector are trained using 16 benchmarks from  HiBench
and BigDataBench. We then apply the trained models to all 44 benchmarks from the four benchmark suites. When there are benchmarks from
HiBench and BigDataBench present in the task group, we use \emph{leave-one-out-cross-validation} to exclude the target applications from
the training program set and use the remaining benchmarks from HiBench and BigDataBench to build our model. To provide a fair comparison,
when testing an application from one benchmark suite that has an equivalent implementation in the other suite, we also exclude the
benchmark from other suite from the training set. For example, when testing \texttt{Sort} from HiBench, we exclude
\texttt{Sort} from BigDataBench from training.

\paragraph{Performance Report.} For each test case, we report \emph{the geometric mean} performance across all configurations. We
replay the schedule decisions for each test case multiple times, until the difference between the upper and lower confidence bounds under a
95\% confidence interval setting is smaller than 5\%. Furthermore, the time spent on feature extraction, model calibration, and prediction
is included in our results.

\subsection{Evaluation Metrics}
We use two standard evaluation metrics for multi-programmed workloads:
\emph{system throughput} and \emph{turnaround time}. We use the definitions
given in~\cite{Eyerman:2010:PJS:1736020.1736033}, defined as follows.

\paragraph{1. System throughput (STP)} is a \emph{higher is better metric}. It
describes the aggregated progress of all jobs under co-location execution over running each job one by one
using isolated execution. This is calculated as:
\begin{equation}
  STP = \sum_{i=1}^{n}\frac{C_{i}^{is}}{C_{i}^{cl}}
\end{equation}
where \emph{n} is the number of application tasks to be scheduled, and $C_{i}^{is}$ and
$C_{i}^{cl}$ are the execution time for task $i$ under the isolated execution mode (\texttt{is}) where
the task uses all available memory; and the co-locating mode (\texttt{cl}) where there may be multiple tasks running on the same host.


\paragraph{2. Average normalized turnaround time (ANTT)} is a
\emph{smaller is better} metric. It quantifies the time
between a task being created and its completion, indicating the
average user-perceived delay.
 This metric is defined as:
\begin{equation}
  ANTT = \frac{1}{n} \sum_{i=1}^{n}\frac{C_{i}^{cl}}{C_{i}^{is}}
\end{equation}

\subsection{Comparative Approaches}

\paragraph{\Quasar.} This is a state-of-the-art co-location scheme~\cite{Delimitrou:2014:QRQ:2541940.2541941}.
\Quasar uses classification techniques to determine the characteristics of the application to perform resource allocation, and
task assignment and co-location. Similar to our dynamic scheme, \Quasar monitors workload performance to adjust
resource allocation and assignment when needed.
Unlike our approach, \Quasar uses a single model for resource estimation.
To provide a fair comparison, we have implemented the \Quasar
classification scheme using the same set of training programs that we used to build our models.

\paragraph{\DColocation.} This pairwise co-location scheme looks for servers with spare memory to
co-locate an additional task on the host. It sets the maximum heap size of the co-locating task to the size of free memory,
 and relies on the Spark default scheduler to determine how many \RDD data items to be allocated to the co-running task.
This represents the default resource allocation policy used by many co-location schemes~\cite{Cooper}.

\paragraph{\Oracle.}
We also compare our approach to the performance of an ideal predictor (\oracle) that gives the perfect memory prediction for an application.
This comparison indicates how close our approach is to the theoretically perfect solution.
The prediction given by the \Oracle scheme is obtained through profiling the application on a given set of input \RDD data items,
but the profiling overhead is not included in the results since we assume the \Oracle predictor has the ability to make prophetic prediction.
Using the \Oracle predictor, the runtime scheduler can then search for the optimal number of data items to be given to a co-running task.



\begin{figure}[!t]
	\centering
	\subfloat[STP]{\includegraphics[width=0.46\textwidth]{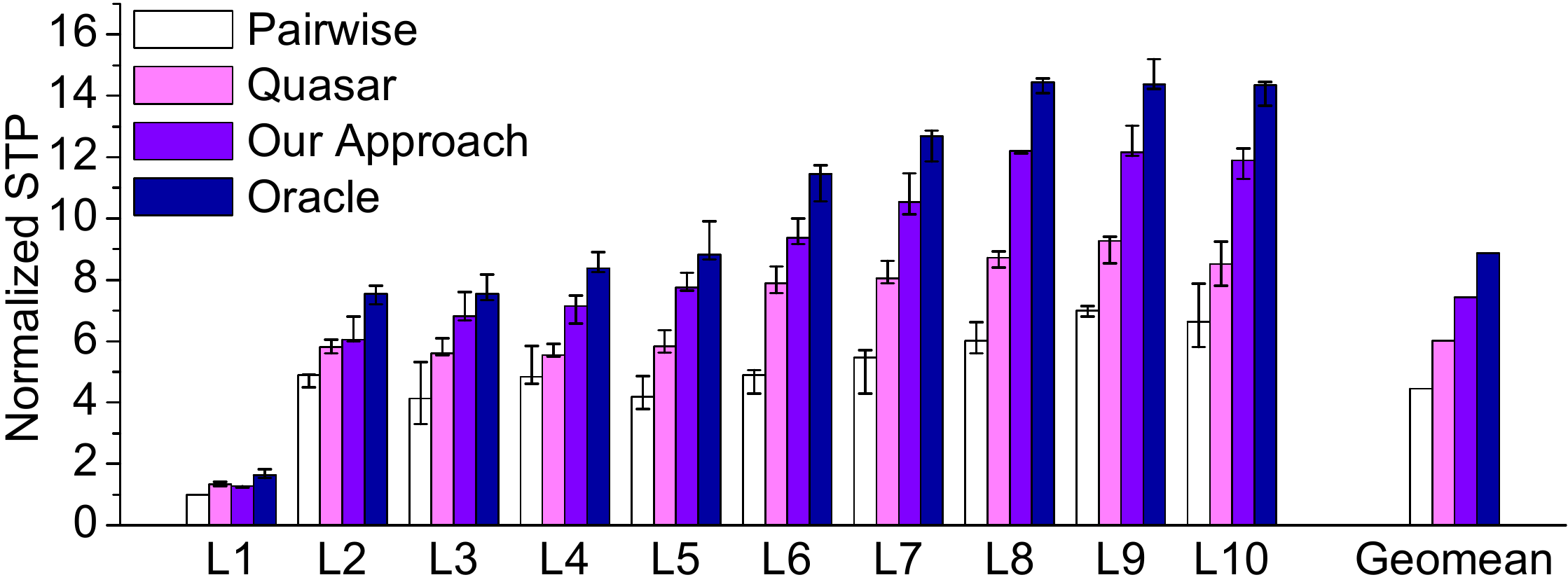}}
    \\
    \subfloat[ANTT]{\includegraphics[width=0.46\textwidth]{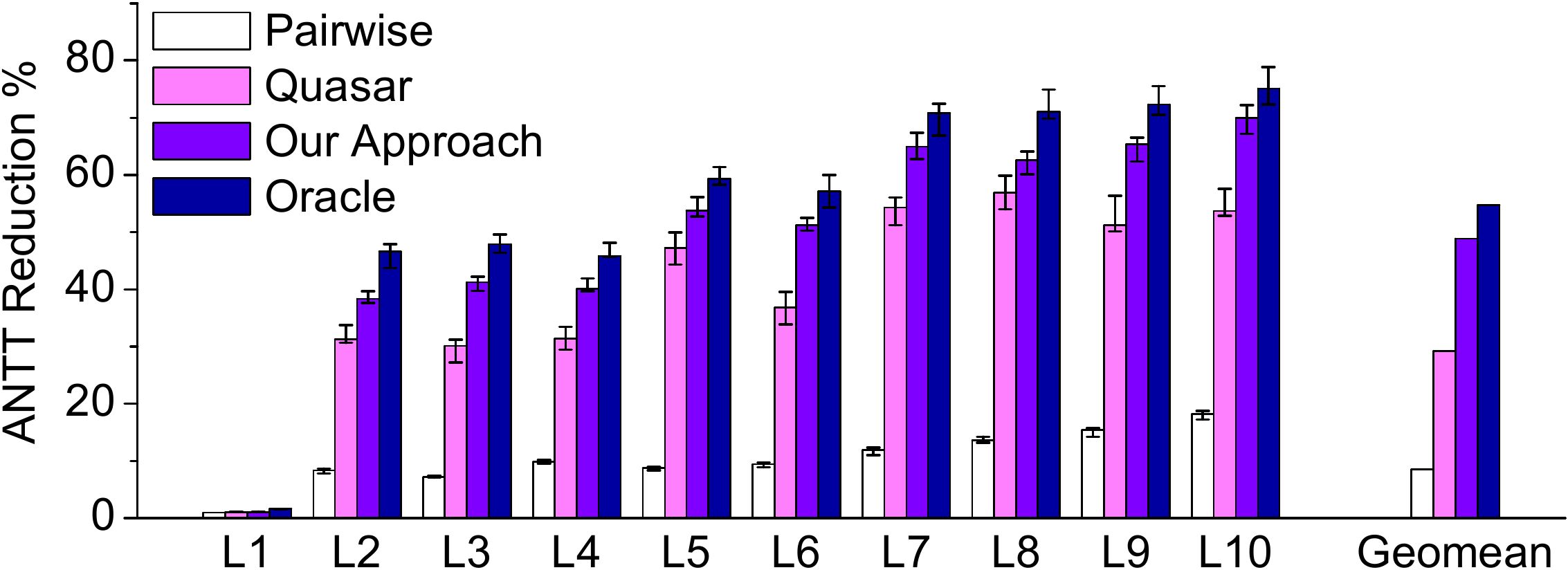}}
    \caption{Our approach outperforms \DColocation and \Quasar on STP (a) and ANTT (b).
     The baseline is running the applications one by one using isolated execution.
    The min-max bars show the range of performance achieved across task mixes for each runtime scenario. }
    \label{fig:overall}
\end{figure}

\begin{figure*}[t!]
	\centering
	\begin{tabular}{ccc}
    \subfloat[\DColocation]{	\includegraphics[width=0.30\textwidth]{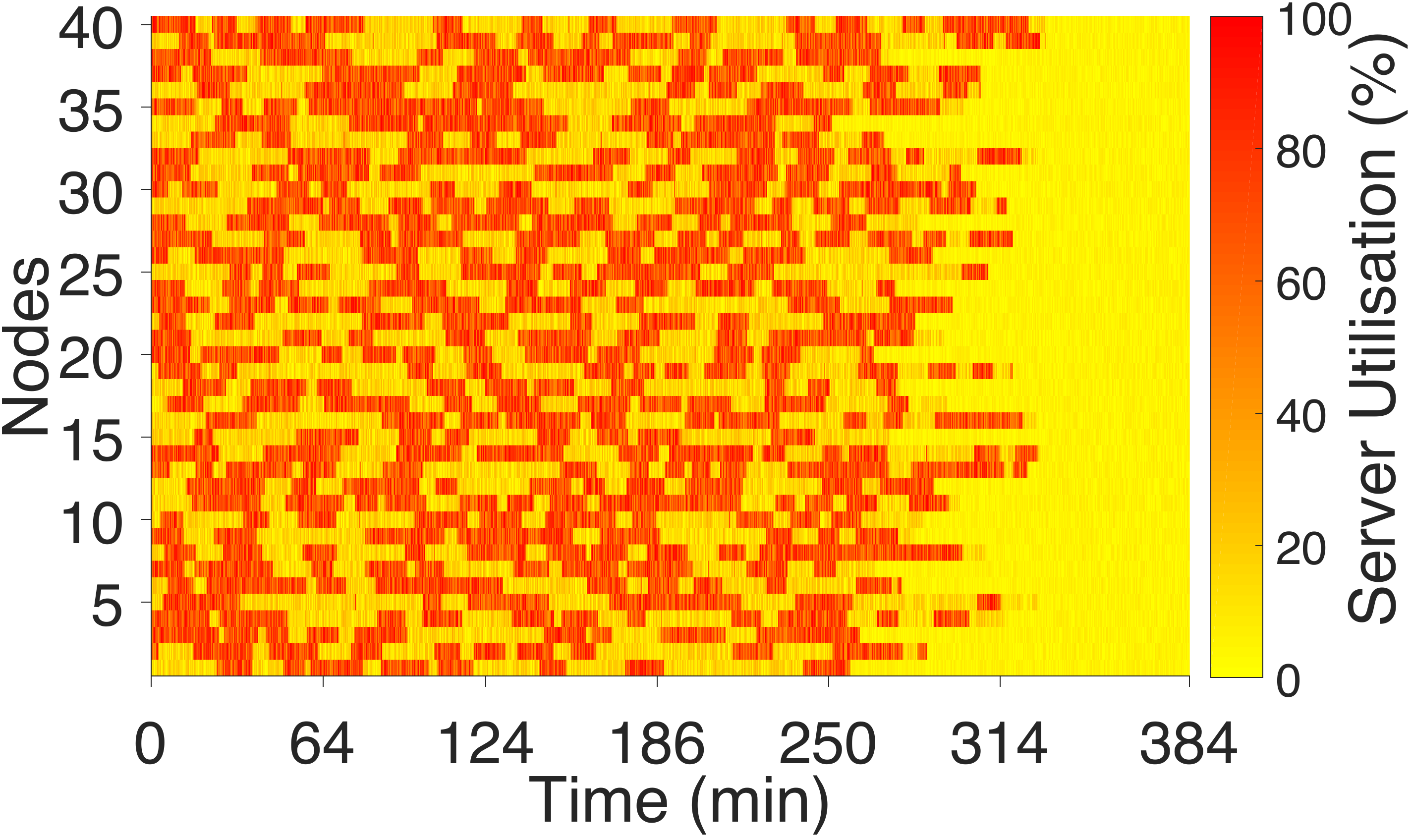}}
    &\subfloat[\Quasar]{	\includegraphics[width=0.30\textwidth]{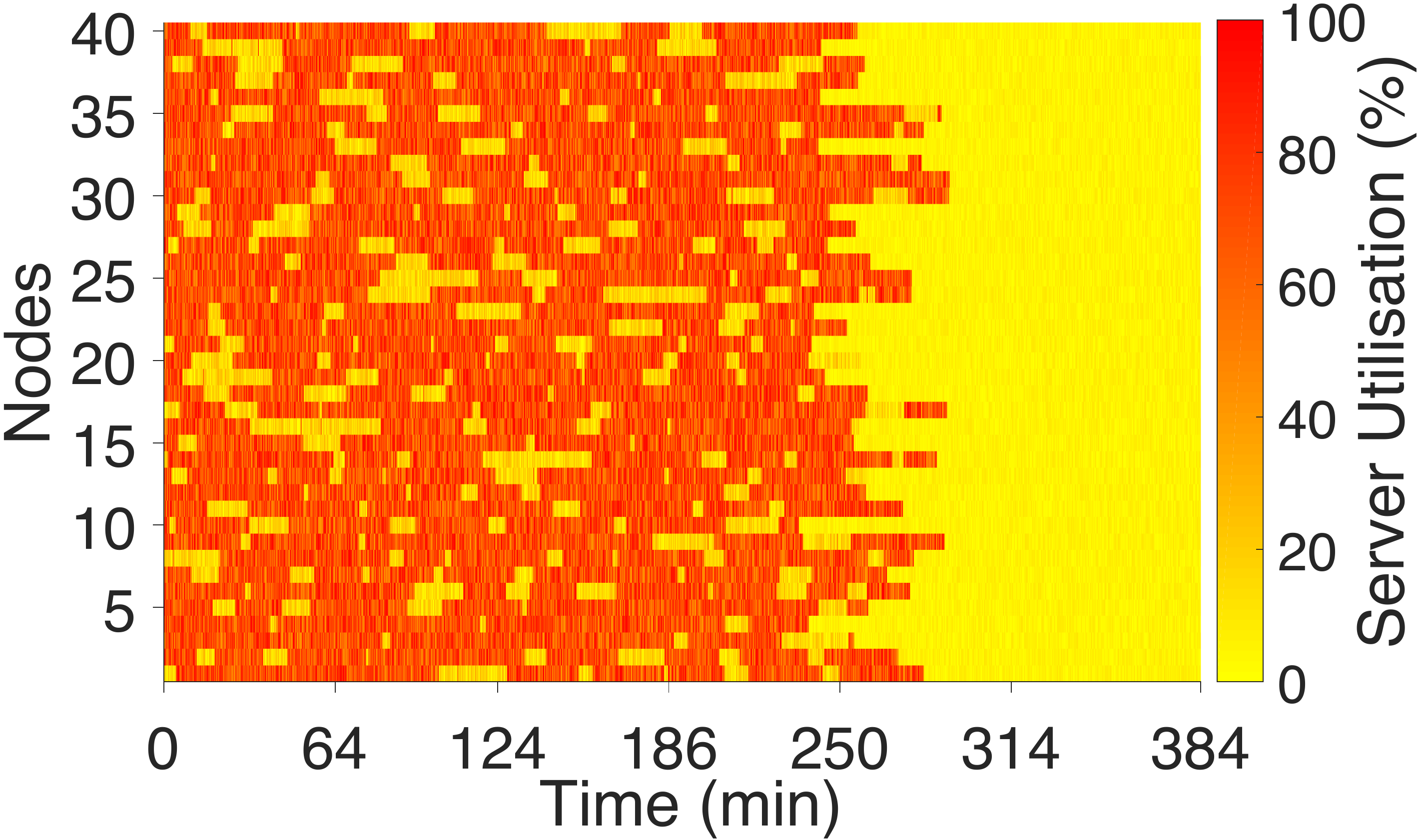}}
    &\subfloat[Our approach]{	\includegraphics[width=0.30\textwidth]{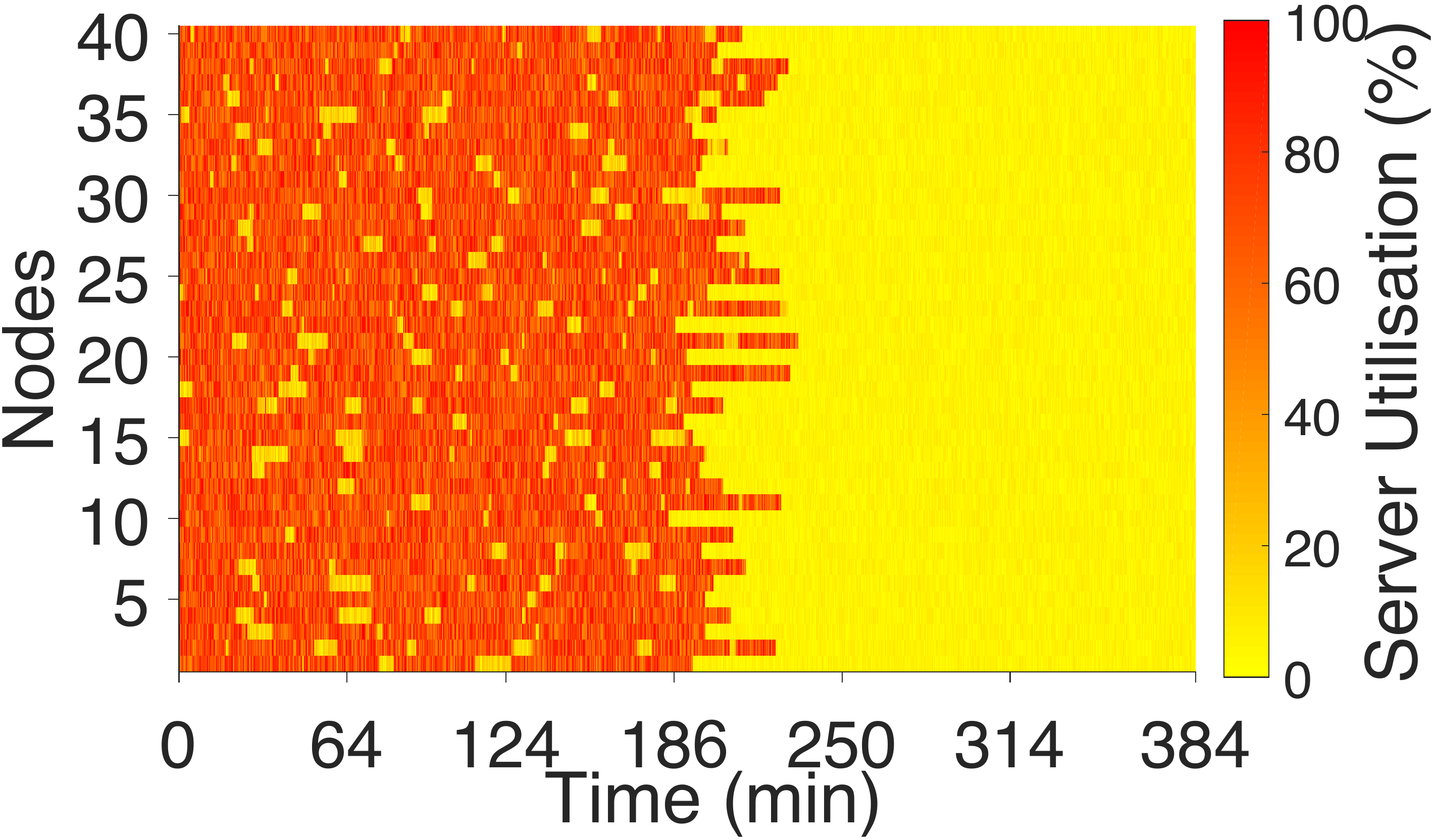}}
    \end{tabular}
    \caption{CPU utilization across servers when scheduling 30 Spark applications (L10). The right-most non-zero point indicates the time when all applications finish.
    Our approach leads to the highest server utilization and quickest turnaround time.
    }
	\label{fig:utilization}
\end{figure*}
\section{Experimental Results}
\label{sect_experiments}
%


\begin{table*}[!t]
\centering
\caption{Application mix for the experiment shown in Figures~\ref{fig:utilization} and \ref{fig:hmp}.}
\setlength{\tabcolsep}{4pt}
\scriptsize
    \begin{tabular}{crrcrrcrrcrrcrr}
    \toprule
    \textbf{Order} & \textbf{App.} & \textbf{In.Size} &
    \textbf{Order} & \textbf{App.} & \textbf{In.Size} &
		\textbf{Order} & \textbf{App.} & \textbf{In.Size} &
		\textbf{Order} & \textbf{App.} & \textbf{In.Size} &
		\textbf{Order} &\textbf{App.} & \textbf{In.Size}\\

    \midrule
    \rowcolor{Gray}1 & BDB.Wordcount & 30GB & 7 & HB.Scan & 30GB & 13 & SP.DecisionTree & 30GB & 19 & BDB.Kmeans & 30GB & 25 & SP.B.MatrixMult & 1TB \\
    2 & SP.Kmeans & 1TB & 8 & HB.TeraSort & 1TB & 14 & SP.Spearman & 1TB & 20 & HB.Sort & 1TB & 26 & BDB.Sort & 30GB\\
    \rowcolor{Gray}3 & SP.glm-classification & 1TB & 9 & SB.Hive & 1TB & 15 & SB.MatrixFact & 1TB & 21 & SP.CoreRDD & 300MB & 27 &SB.RDDRelation & 1TB\\
    4 & SP.glm-regression & 1TB & 10 & SP.NaiveBayes & 1TB & 16 & BDB.Grep & 1TB & 22 & SP.Gmm & 1TB & 28 & SP.Pearson & 1TB\\
    \rowcolor{Gray}5 & SP.Pca & 30GB & 11 & BDB.PageRank & 1TB & 17 & SB.LogRegre & 1TB & 23 & HB.Join & 1TB & 29 & SP.Chi-sq & 30GB\\
    6 & SB.SVD++ & 1TB & 12 & HB.PageRank & 30GB & 18 & BDB.NaivesBayes & 30GB & 24 & SP.Sum.Statis & 30GB & 30 & HB.Kmeans & 1TB\\
    \bottomrule
    \end{tabular}
	\label{table_list_task}
\end{table*}

In this section, unless stated otherwise, we report each approach's performance on STP and ANTT, by normalizing the results to a
\emph{baseline} that schedules the applications one by one with each application exclusively using all the memory of each allocated
computing node. The normalized STP and ANTT are referred to as \emph{normalized STP} and \emph{ANTT reduction} (shown in percentage) respectively.

\subsection{Highlights}
The highlights of our evaluation are as follows:
\begin{itemize}[leftmargin=5mm]
\item With the help of our mixture-of-experts approach, a simple task co-location scheme achieves, on average, a 8.69x
improvement on STP and a 49\% reduction on ANTT over isolated execution. This translates to a 1.28x and 1.68x
improvement on STP and ANTT respectively, when compared to \Quasar. See Section~\ref{sec:overallp};
\item Our approach is highly accurate in predicting the memory footprint of Spark applications, with an error of less than 5\% for most cases. See Section~\ref{sec:model_analysis};
\item Our scheme is low-overhead. The time spent on feature extraction and
model calibration is less than 10\% of the total application execution time, and the profiling runs contribute to the final results. See Section~\ref {sec:profil_overhead};
\item We thoroughly evaluate our scheme by comparing it against several alternative task co-location schemes and modeling techniques, and performing a detailed analysis on the working mechanism of the approach.
\end{itemize}

\subsection{Overall Performance \label{sec:overallp}}

\paragraph{STP.}
Figure~\ref{fig:overall} (a) confirms that task co-location improves system throughput. As the number of
tasks to be scheduled increases, we see an overall increase in the STP. \DColocation performs reasonably well for small
task groups, but it misses significant opportunities for large task groups.
For L9 and L10, \DColocation only delivers half of the \oracle performance.
This is because \DColocation does not scale up beyond pairwise co-location.
\Quasar performs significantly better than \DColocation by using a classifier model to coordinate resources among co-locating tasks,
but it is not as good as our approach. By employing multiple functions to model diverse applications,
our approach constantly outperforms \DColocation and \Quasar across all task groups.
For large task groups (L8 - L10),  our approach delivers over 1.72x and 1.48x improvement on the STP over \DColocation and \Quasar respectively.
Overall,  \Quasar gives on average 6.6x improvement on STP, which translates to 65.7\%  of the \oracle performance.
Our approach achieves 8.69x improvement on STP, which translates to a 1.28x improvement over \Quasar or 83.9\% of the \oracle performance.

\paragraph{ANTT.}
Figure~\ref{fig:overall} (b) shows the ANTT reduction over the baseline. By
maximizing the system throughput, task co-location in general leads to
favorable ANTT results, particularly for large task groups.
 \Quasar and our approach
outperforms \DColocation on ANTT by a factor of over 4x from L2 onward.
Our approach delivers better turnaround time over \Quasar, by avoiding
memory contention among co-locating Spark tasks. On average, our approach
reduces the turnaround time by 49\% across different task groups. This
translates to 93.4\% of the \oracle performance.
When compared with the 54\% \oracle performance given by \Quasar, our approach achieves 1.68x better turnaround time.



\paragraph{Summary.}
We achieve 83.9\% and 93.4\% of the \oracle performance for STP and ANTT respectively, outperforming \DColocation, a
widely used co-location policy, and \Quasar, a state-of-the-art co-location policy. The advantage of our approach is largely attributed to
its use of multiple models instead of just one to precisely capture an applications' memory behavior. Without this
accurate information, the alternative scheme often over- or under-provisions
resources, leading to worse performance.

%
%

\begin{figure}[t!]
	\centering
    \subfloat[STP (higher is better)]{	\includegraphics[width=0.25\textwidth]{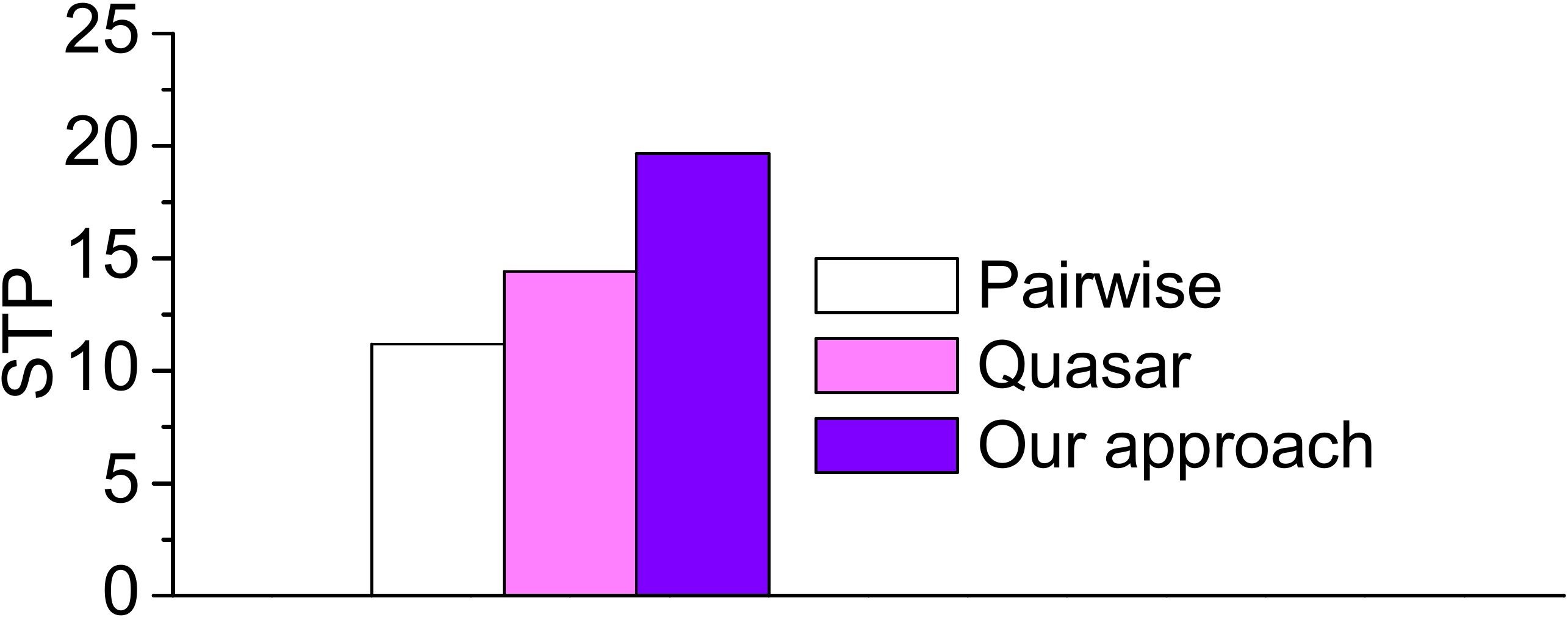}}
    \subfloat[Turnaround Time (lower is better)]{	\includegraphics[width=0.25\textwidth]{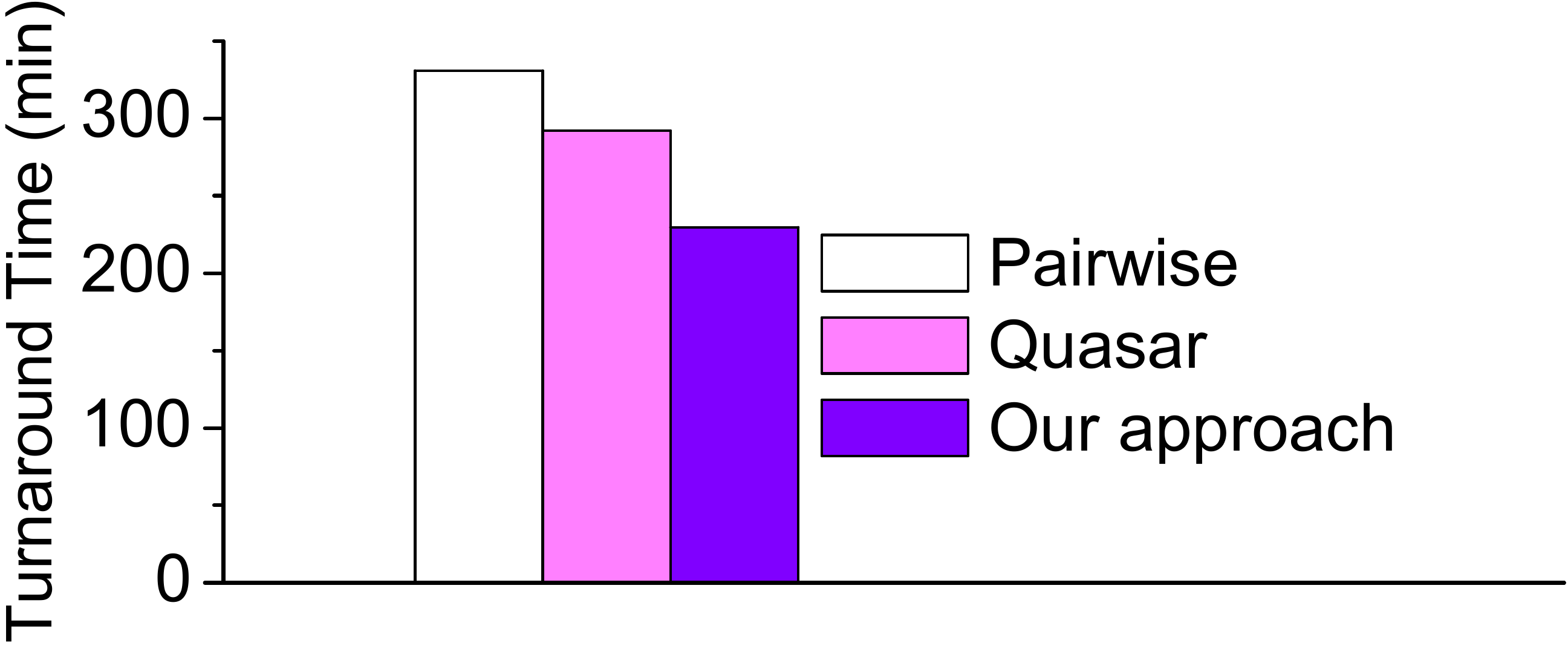}}
    \caption{Resultant STP (a) and turnaround time (b) for the scheduling scenario in Figure~\ref{fig:utilization}.
    Our approach gives better STP and faster turnaround time when compared with alternative co-location schemes.
    }
	\label{fig:hmp}
\end{figure}

\subsection{Server Utilization}
Figure~\ref{fig:utilization} shows the CPU utilization across 40 computing nodes for \DColocation, \Quasar and our approach when scheduling
30~Spark applications (L10) using different input sizes. Table~\ref{table_list_task} gives the application mix for this experiment, while
Figure~\ref{fig:hmp} presents the turnaround time (i.e. the wall clock time to finish the set of jobs) given by each approach.
Additionally, Table~\ref{table_list_task} shows the applications and the input size performed that where used to create
Figure~\ref{fig:utilization}. By carefully co-locating tasks using memory footprint predictions, our approach gives the best server
utilization, which in turn leads to the highest STP (1.81x and 1.39x higher STP over \DColocation and \Quasar respectively) quickest
turnaround time (1.46x and 1.28x faster turnaround time over \DColocation and \Quasar respectively).

\begin{figure}[t!]
	\centering
    \subfloat[STP]{	\includegraphics[width=0.45\textwidth]{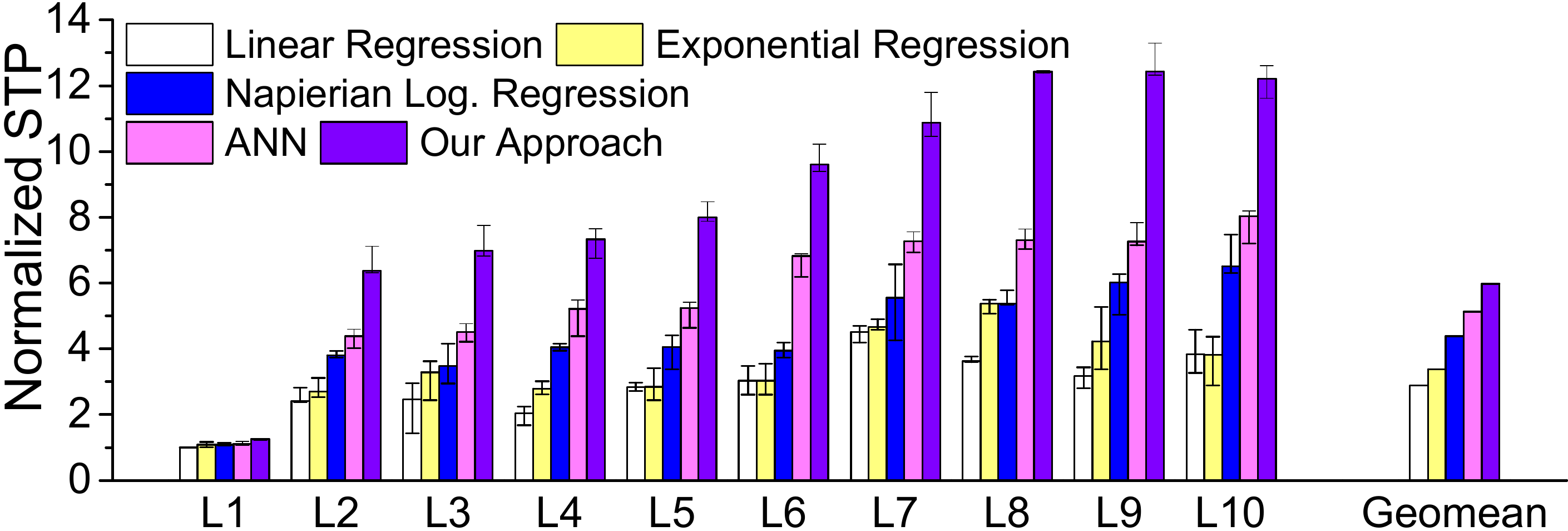}}\\
    \subfloat[ANTT]{\includegraphics[width=0.45\textwidth]{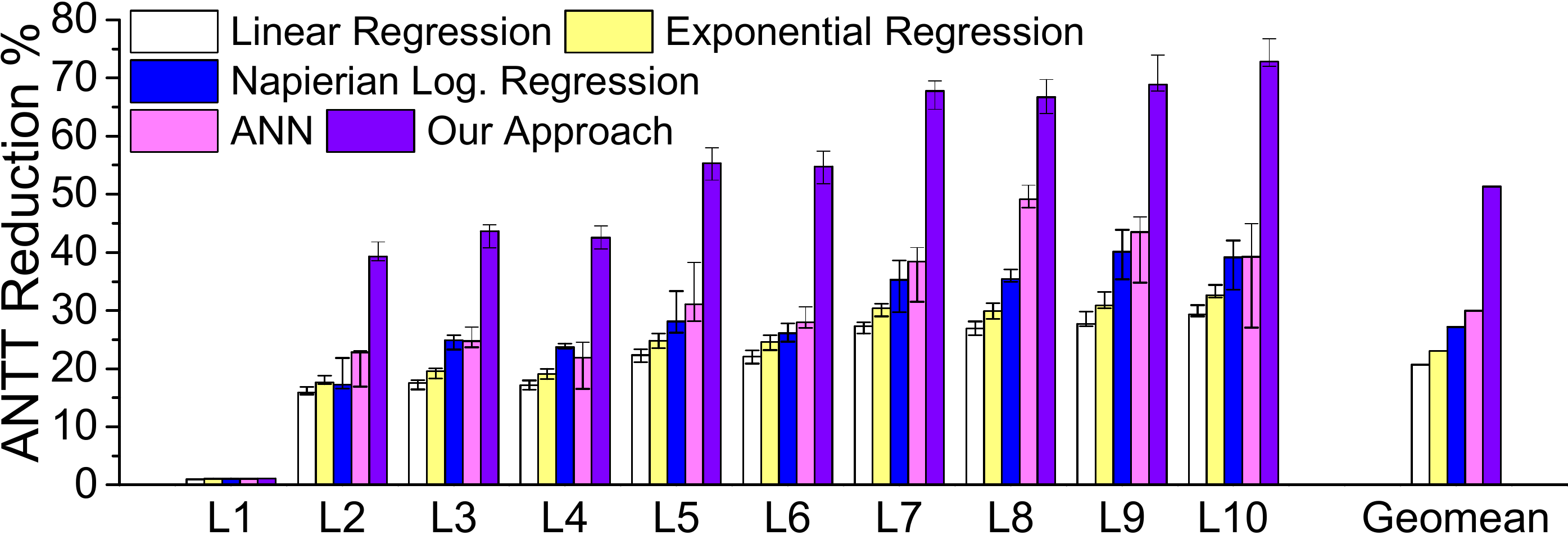}}
    \caption{Compare to unified model based approaches that use a single modeling technique to describe the application's memory behavior.    }
	\label{fig:um}
\end{figure}

\subsection{Compare to Unified Models}

Figure~\ref{fig:um} compares our scheme to approaches that use one modeling technique to predict the application's memory footprint.  In
addition to the three memory functions listed in Table~\ref{memory_function}, we also compare our scheme to a 3-layer artificial neural
network (ANN) trained using a backpropagation algorithm. We use the same training data to build the ANN model to predict the memory
footprint. The input to the ANN model is the same set of features used by our approach.  Among the single model approaches, the ANN gives
the best performance due to its ability to model linear and non-linear behaviors. Our approach outperforms ANN and all other approaches on
STP and ANTT. The results suggest the need for using multiple modeling techniques to capture the diverse application behaviors. This work
develops a generic framework to support this.

\subsection{Compare to Online Search \label{exp:online}}

\begin{figure}[t]
	\centering
    \subfloat[STP]{	\includegraphics[width=0.45\textwidth]{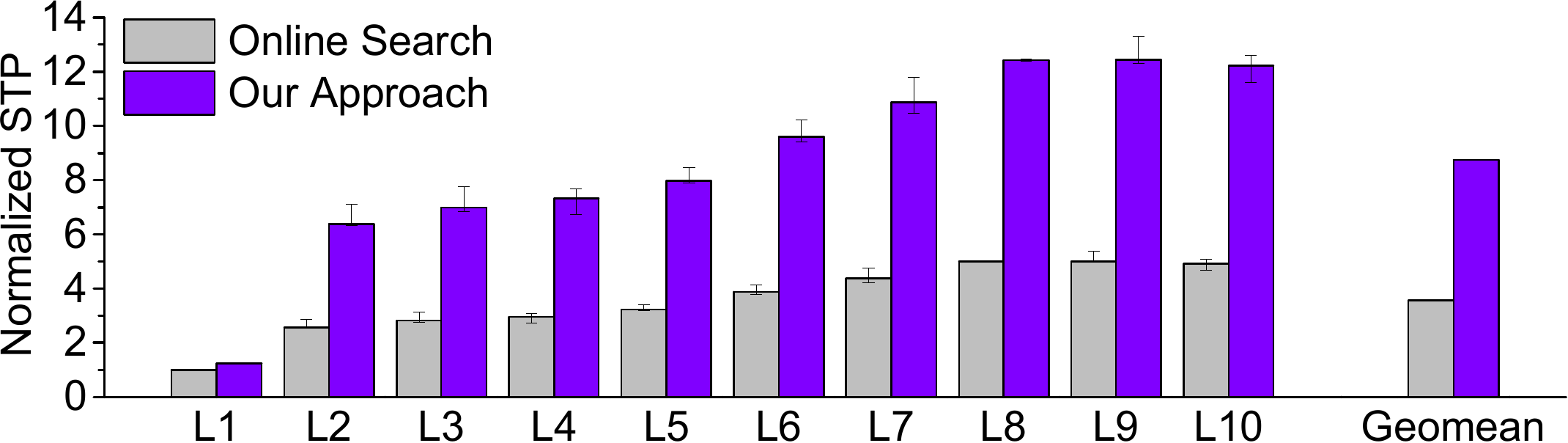}}\\
    \subfloat[ANTT]{\includegraphics[width=0.45\textwidth]{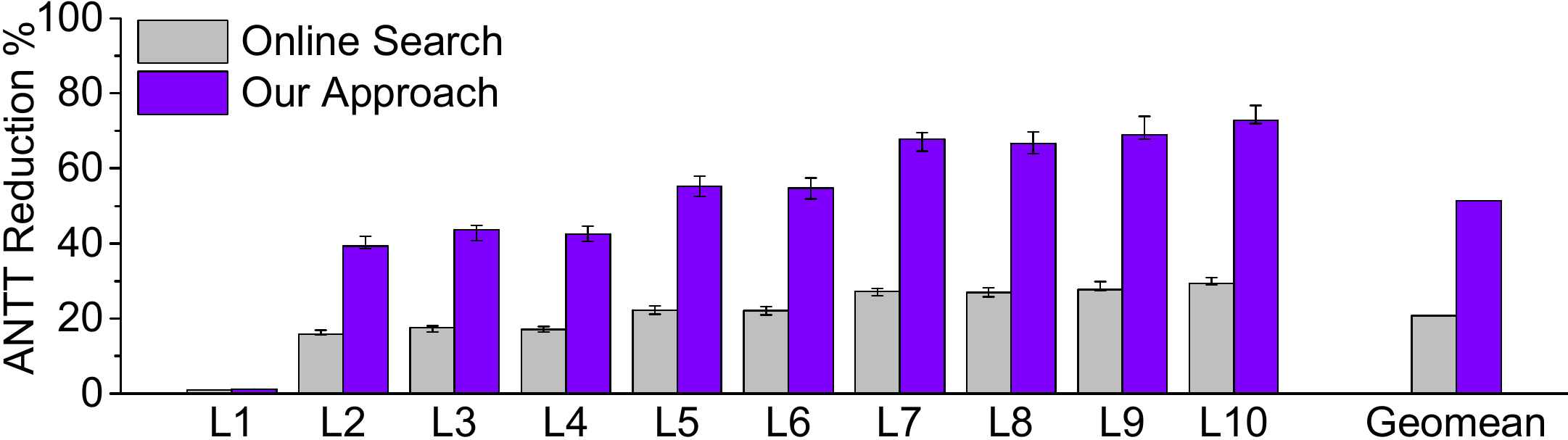}}
    \caption{Compare to using online search to allocate input for a given memory budget.  Our approach significantly
    outperforms the online search scheme, because it avoids the runtime overhead associated with finding the optimal
    number of data items to be given to the co-running task.   }
	\label{fig:os}
\end{figure}

Figure~\ref{fig:os} compares our approach to a method that uses descent gradient search to dynamically adjust the right input size for a
given memory budget. The online search based approach gives rather disappointing results due to the large overhead involved in finding the
right input size. Furthermore, this approach also suffers from a scalability issue, i.e. the searching overhead grows as the number of
computing nodes increases. Our approach avoids the overhead by directly predicting the memory footprint, leading to 2.4x and 2.6x better
performance on STP and ANTT respectively.

\subsection{Profiling Overhead \label{sec:profil_overhead}}
\begin{figure}[t!]
  \centering
  \includegraphics[width=0.45\textwidth]{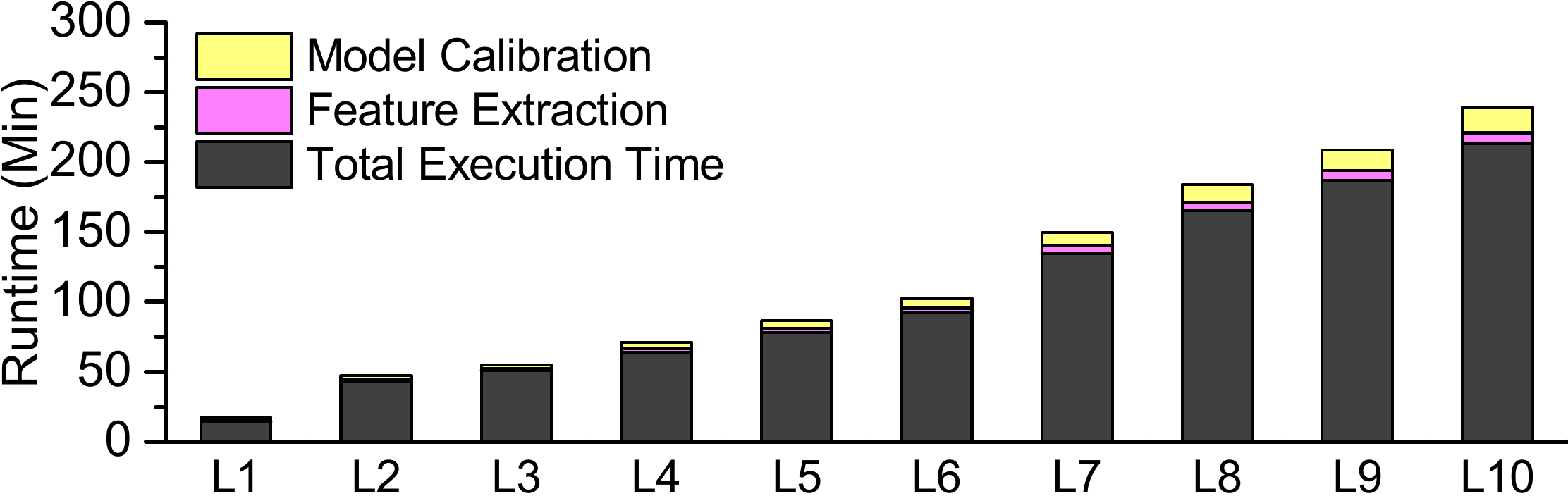}
  \caption{Average profiling time to total task execution time.
  It is to note that during feature extraction and model calibration, the application
   always executes a portion of the unprocessed data, no computing cycles are wasted.
  \label{fig:proftime}}
\end{figure}

\begin{figure}[t!]
  \centering
  \includegraphics[width=0.45\textwidth]{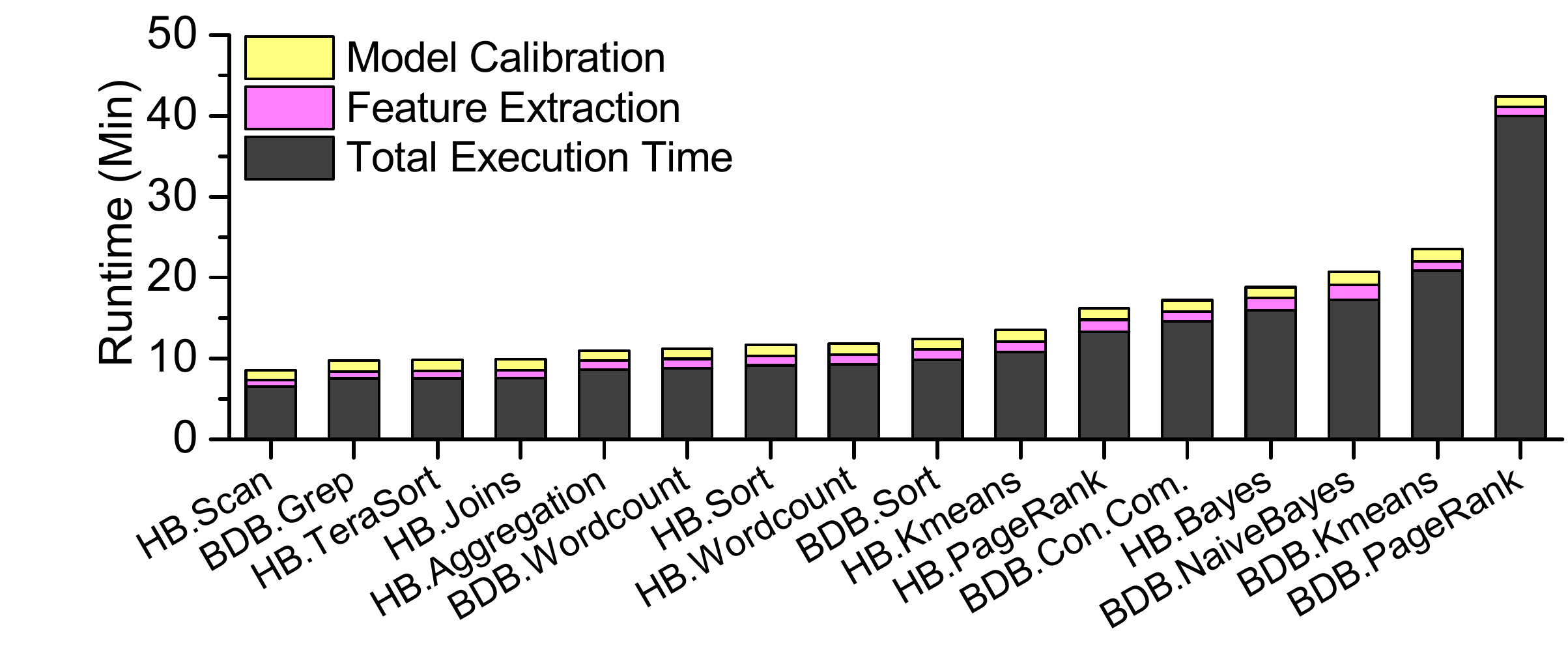}
  \caption{Average profiling time to total runtime per program for HiBench and BigDataBench.\label{fig:proftimeBench}}

\end{figure}

The stack chart in Figure~\ref{fig:proftime} shows the average time spent on feature extraction and model calibration
with respect to the total execution time per evaluation scenario.
Figure~\ref{fig:proftimeBench} gives a breakdown on per benchmark basis using an input size of around 280GB.
As profiling is performed on a single host (thus having little communication overhead) using small inputs, the cost is moderate.
Overall, the time spent on feature extraction and
model calibration accounts for 5\% and 8\% respectively to the total execution time. We stress that
profiling runs also contribute to the final output of the task, so no computing cycles are wasted; and
profiling is performed while the application is waiting to be scheduled.

\subsection{CPU Load in Isolation Mode \label{sec:cpu_usage}}
\begin{figure}[t!]
	\centering
	\includegraphics[width=0.45\textwidth]{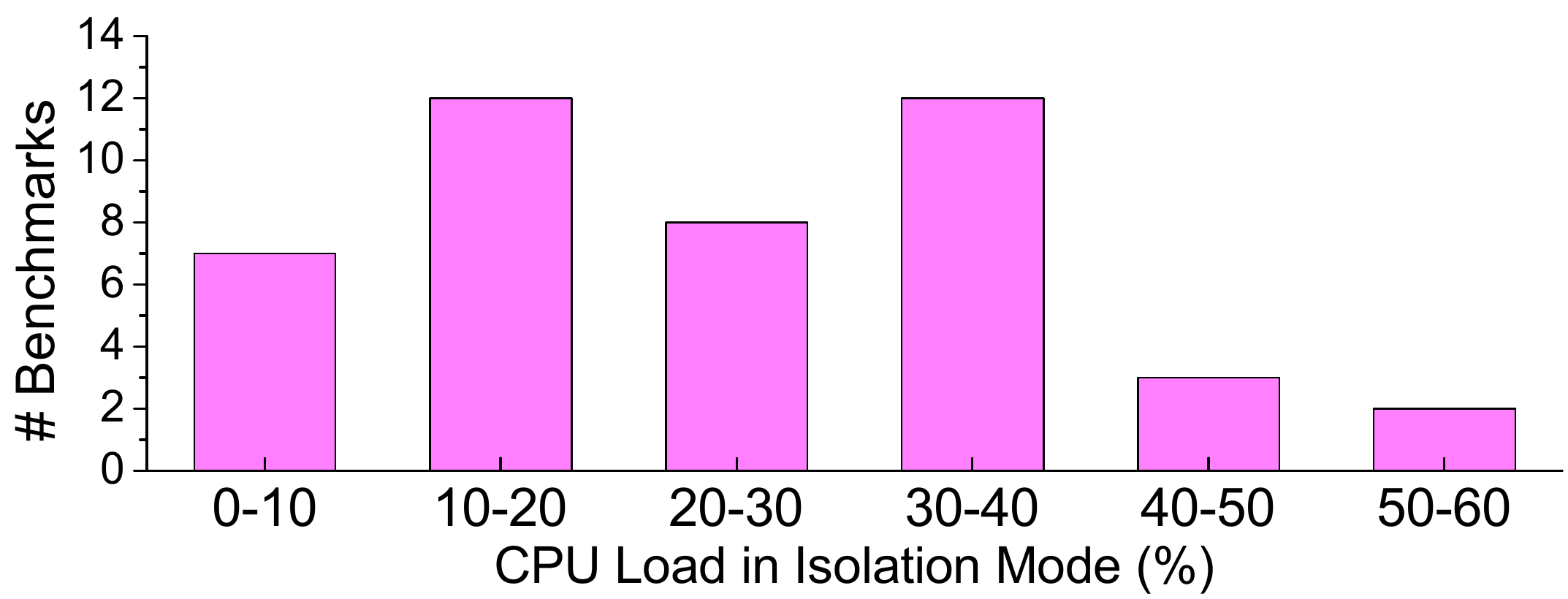}
	\caption{CPU load distributions across benchmarks when the application is executed in the isolation mode. }
	\label{fig_cpuImpact}
\end{figure}

Figure~\ref{fig_cpuImpact} shows the average CPU load when a benchmark is running in isolation using all the system's memory exclusively.
The CPU load for most of the 44 benchmarks is under 40\%. As a result, the CPU is often not fully utilized when just running on
application. This is in line with the finding reported by other researchers~\cite{7310708}. Our approach exploits this characteristic to
improve the system throughput through task co-location.

\subsection{Co-location Interferences  \label{sec:interferences}}
\begin{figure}[t!]
  \centering
  \includegraphics[width=0.45\textwidth]{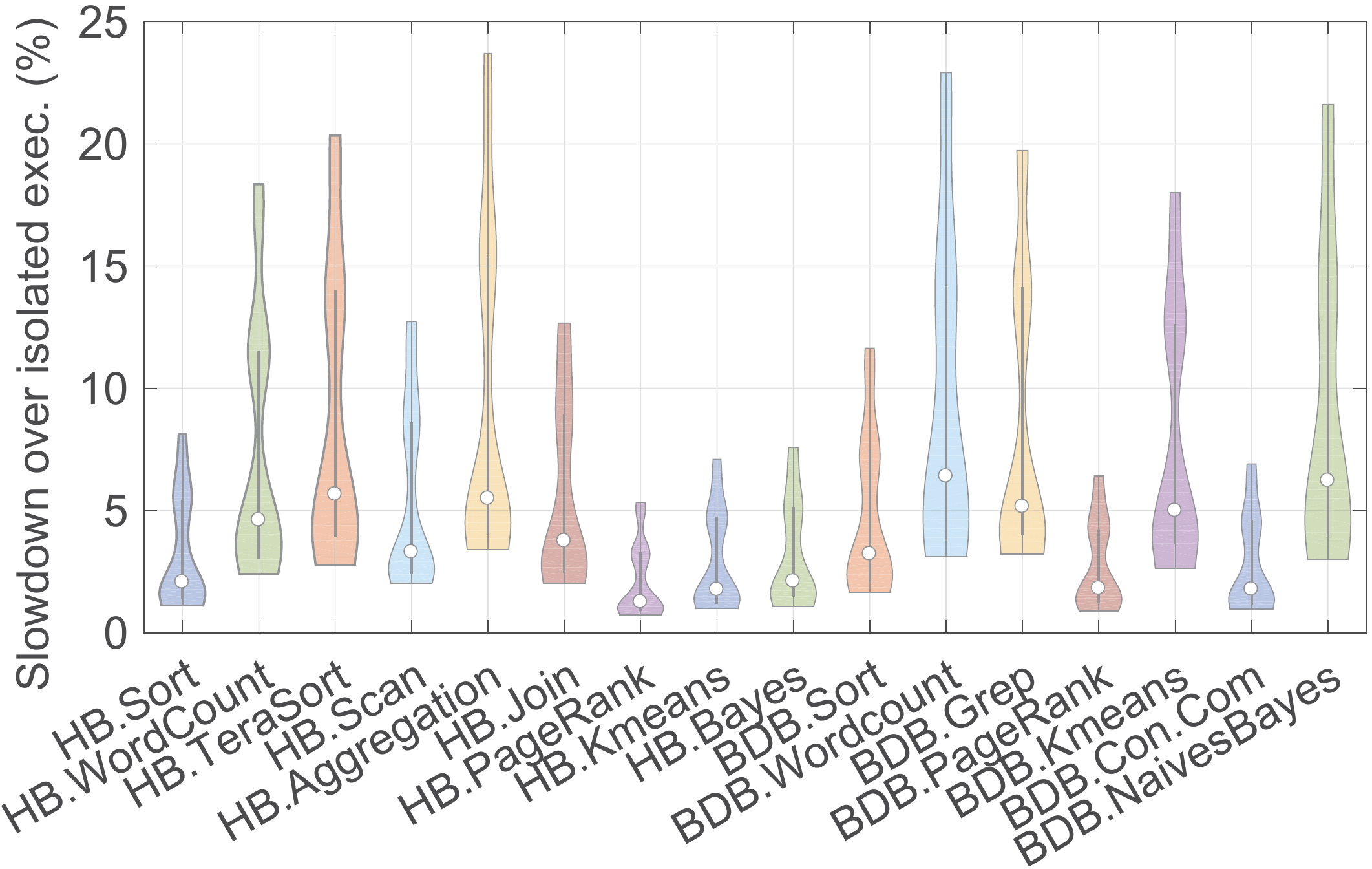}\\
  \caption{Violin plot showing the distribution of slowdown when using our scheme to co-locate the target benchmark
  with another application on a single host. The baseline is running the target application in isolation. Here we run
  each of the 16 target benchmarks from HiBench and BigDataBench along with each of the remaining 43 benchmarks.
  }\label{fig:clinterf}
\end{figure}

\paragraph{Interferences among Spark Benchmarks.} The violin plot in Figure~\ref {fig:clinterf} shows the
distribution of slowdown when running each of the 16 benchmarks from HiBench and BigDataBench along with each of the
remaining 43 benchmarks using our scheme. The shape of the violin corresponds to the slowdown distribution. The thick
black line shows where 50\% of the data lies. The white dot is the position of the median. In the experiment, we first
launch the target application and then use the spare memory to co-locate another competing workload. The input size of
the target program is $\sim$280GB.
As can be seen from the figure, the slowdown across applications is less than 25\%
and is less than 10\% on average. For applications with little computation demand, such as \texttt{HB.Sort}, the
slowdown is minor (less than 5\%). For benchmarks with higher computation demand, such as \texttt{HB.Aggregation}, we
observe greater slowdown due to competing of computing resources among co-locating tasks. Overall, our co-location
scheme has little impact on the
application's performance.

\begin{figure}[t!]
  \centering
  \includegraphics[width=0.45\textwidth]{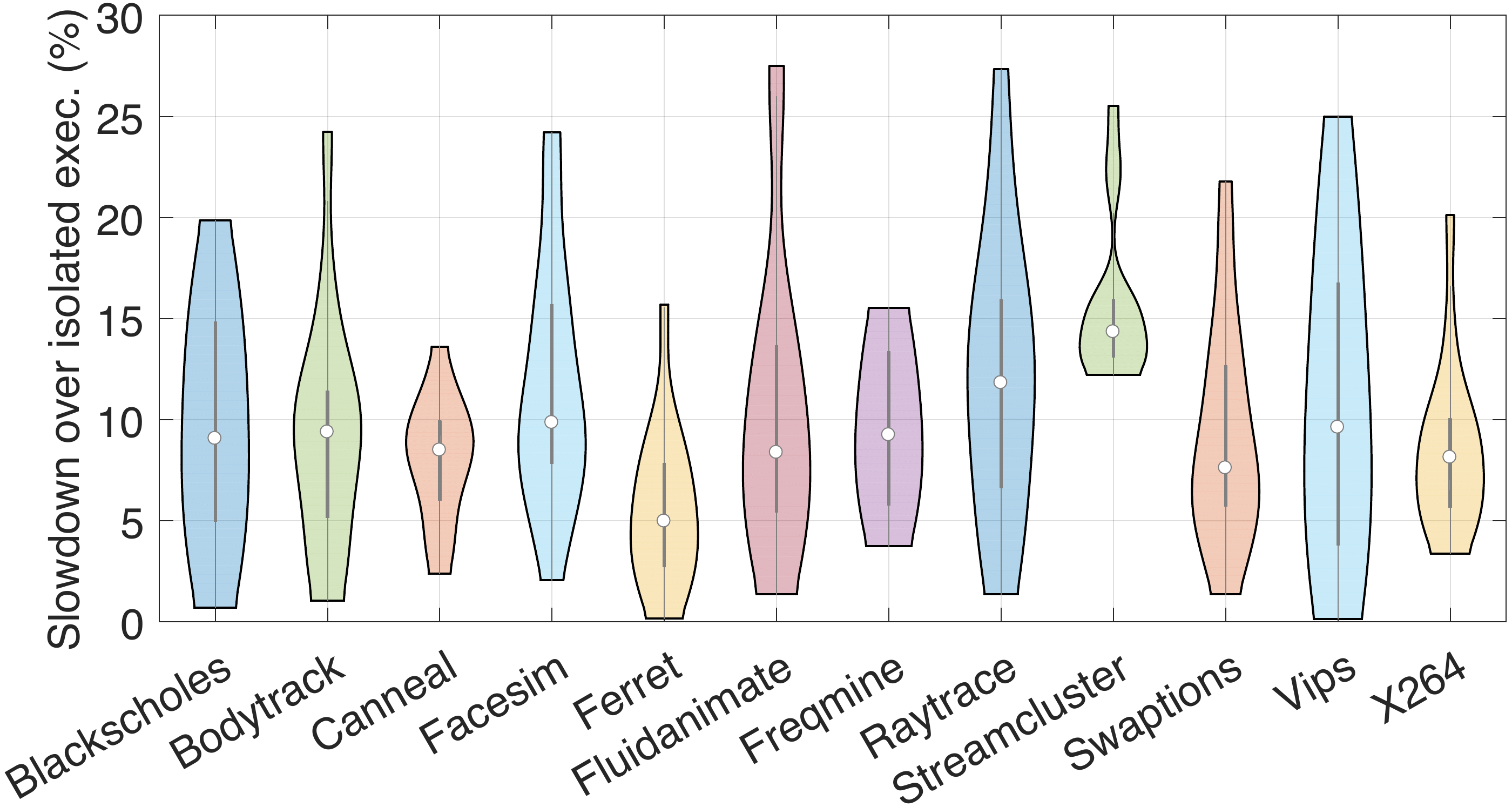}\\
  \caption{
  The slowdown distribution of computation-intensive PARSEC benchmarks when they run with a Spark task under our scheme. The baseline is running the target application in isolation.
}\label{fig:parsec}
\end{figure}

\paragraph{Interferences to PARSEC Applications.}
We further extend our experiments to investigate the impact for co-locating Spark tasks with other
computation-intensive applications. For this purpose, we run some computation-intensive C/C++ applications from the PARSEC benchmark suite (v3.0)~\cite{Bienia2011PARSEC}
using the large, native input provided by the suite. Figure~\ref{fig:parsec} shows the slowdown distribution
of each PARSEC benchmark when they run together with each of the 44 Spark benchmarks under our scheme. As all PARSEC
benchmarks are share-memory programs, this experiment was conducted on a single host.
As expected, we observe some
slowdown to the computation-intensive PARSEC benchmark, but the slowdown is modest -- less
than 30\%. For most of cases, the slowdown is less than 20\%. Given the significant benefit on system
throughput and server utilization given by our approach, we argue that such a small slowdown is acceptable when maximizing the server utilization is desired (which is typical
for many data center applications). There are other schemes such as
Bubble-Flux~\cite{Yang:2013:BPO:2485922.2485974} for reducing the interference via dynamically pausing non-critical
tasks, which are orthogonal to our scheme.

\subsection{Model Analysis \label{sec:model_analysis}}

\begin{figure}
  \centering
  \includegraphics[width=0.45\textwidth]{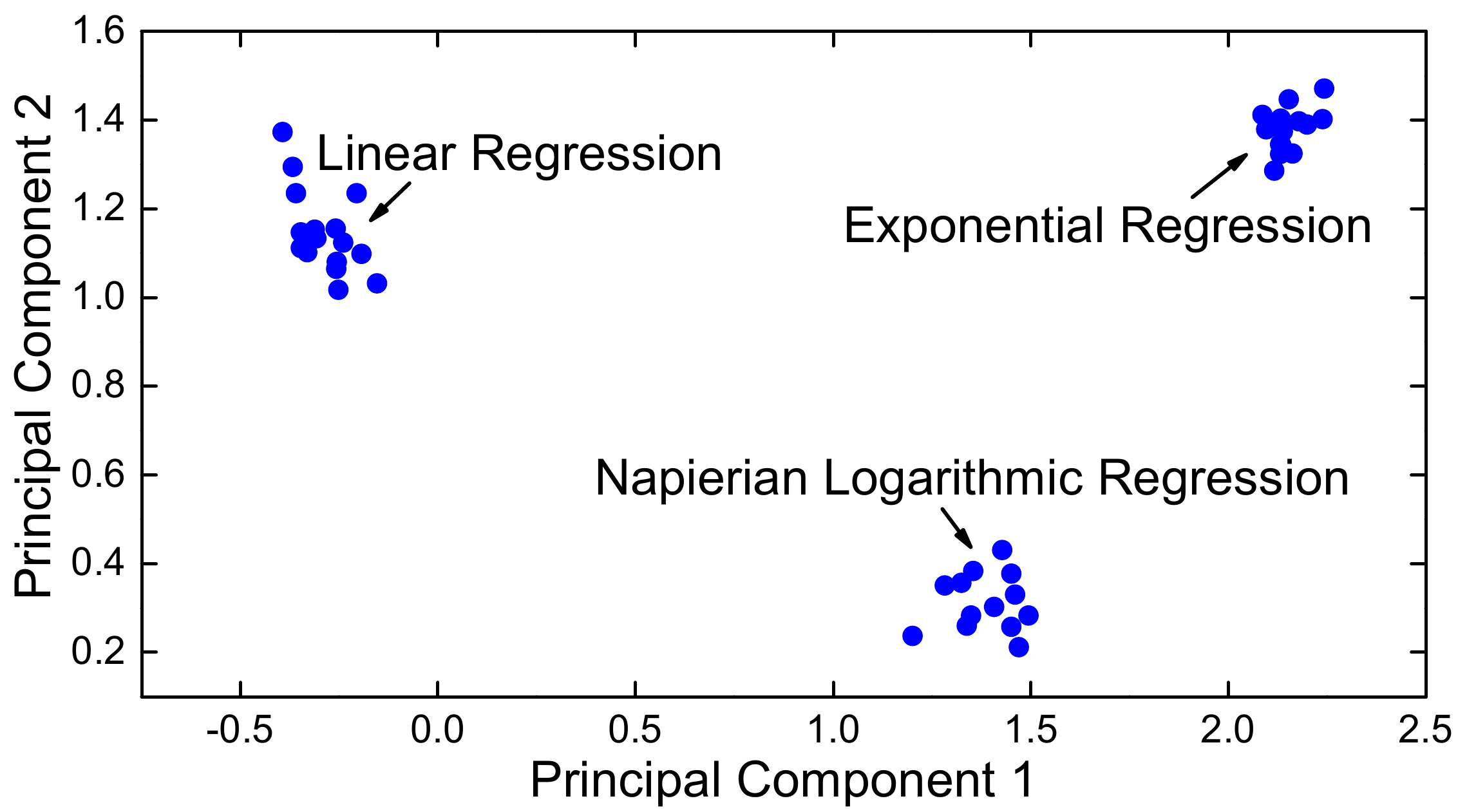}\\
  \caption{Program feature space. The original feature space is projected into 2 dimensions using
  \PCA. Programs can be grouped into three clusters and mapped to the three memory models described in Table~\ref{memory_function}.}
  \label{fig:pcaspace}
\end{figure}

\paragraph{Program Distribution.} Figure~\ref{fig:pcaspace} visually depicts the distribution of benchmarks on the feature space. To aid
clarity, we use \PCA to project the dimension of the original feature space down to two. Each point in the figure is one of the 44
benchmarks. This diagram clearly shows that the 44 benchmarks can be grouped into three clusters. After inspecting each cluster, we found
that we indeed use the same memory function (given on the figure) for all benchmarks in a cluster. This diagram justifies the chosen number
of memory functions. It also confirms our assumption that programs with similar features can be modeled using similar memory functions. To
measure the similarity of programs within each cluster, we calculate the Pearson correlation coefficient of each program to its cluster center on
the 2-d feature space shown in Figure~\ref{fig:pcaspace}. Our results show that the correlation coefficient is above 0.9999 for all program, with most
programs have a correlation coefficient of 1.0 (the strongest correlation). This confirm that our features are effective in capturing the
similarity of programs that use the same memory function.

We want to highlight that one of the advantages of our \KNN classifier is that the distance used to choose the \emph{nearest neighbor}
program gives a confidence estimation of how good the predicted memory function will be. If the target application is far from any of the
clusters in the feature space, it suggests that a new memory modeling technique will be required (and our approach allows new memory
functions to be easily inserted), or a conservative co-location policy should be used to
avoid saturating the memory system.

\begin{figure}
	\centering
	\includegraphics[width=0.48\textwidth]{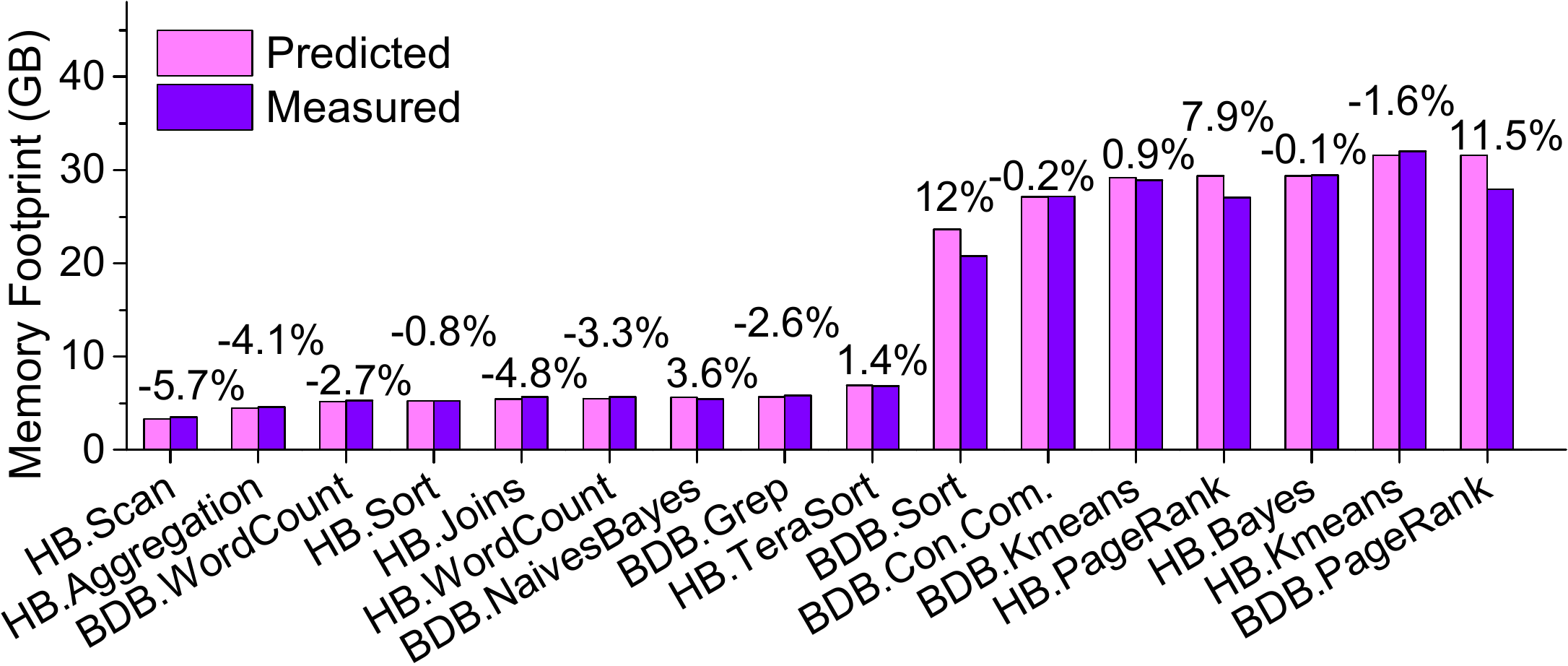}
	\caption{Predicted memory footprints vs measured values for HiBench (HB) and BigDataBench (BDB).}
	\label{fig_MemoryUsedVSMemoryPredicted}
\end{figure}

\paragraph{Prediction Accuracy.}
Figure~\ref{fig_MemoryUsedVSMemoryPredicted} compares the predicted optimal memory allocation against the measured value, using an input
size of around 280GB. All the models are trained and evaluated using ``leave-one-out-cross-validation" (see also
Section~\ref{sec:eval_method})). The prediction error of our approach is less than 5\% in most cases except for \texttt{HB.PageRank},
\texttt{BDB.PageRank} and \texttt{BDB.Sort}  for which our approach over-provisions around 8\% to 12\% of the memory. This translates to
1.5GB to 2GB of memory. Our approach also slightly under-estimates the memory requirement for some benchmarks, but the difference is
small so it does not significantly affect the performance. In general, the accuracy can be improved by using more training programs and
more sophisticated modeling techniques to better capture the application memory requirement, which is our future work. In practical terms,
one can also slightly over-provision (e.g. 10\%) the memory allocation to applications with higher priorities to tolerate potential
prediction errors. Overall, our approach can accurately predict the optimal memory allocation, with an average prediction
error of 5\%.

\begin{figure*}[t]
	\centering
    \scriptsize
	\begin{tabular}{cccc}
	\subfloat[HB.TeraSort, HB.Wordcount]{\includegraphics[scale=0.25]{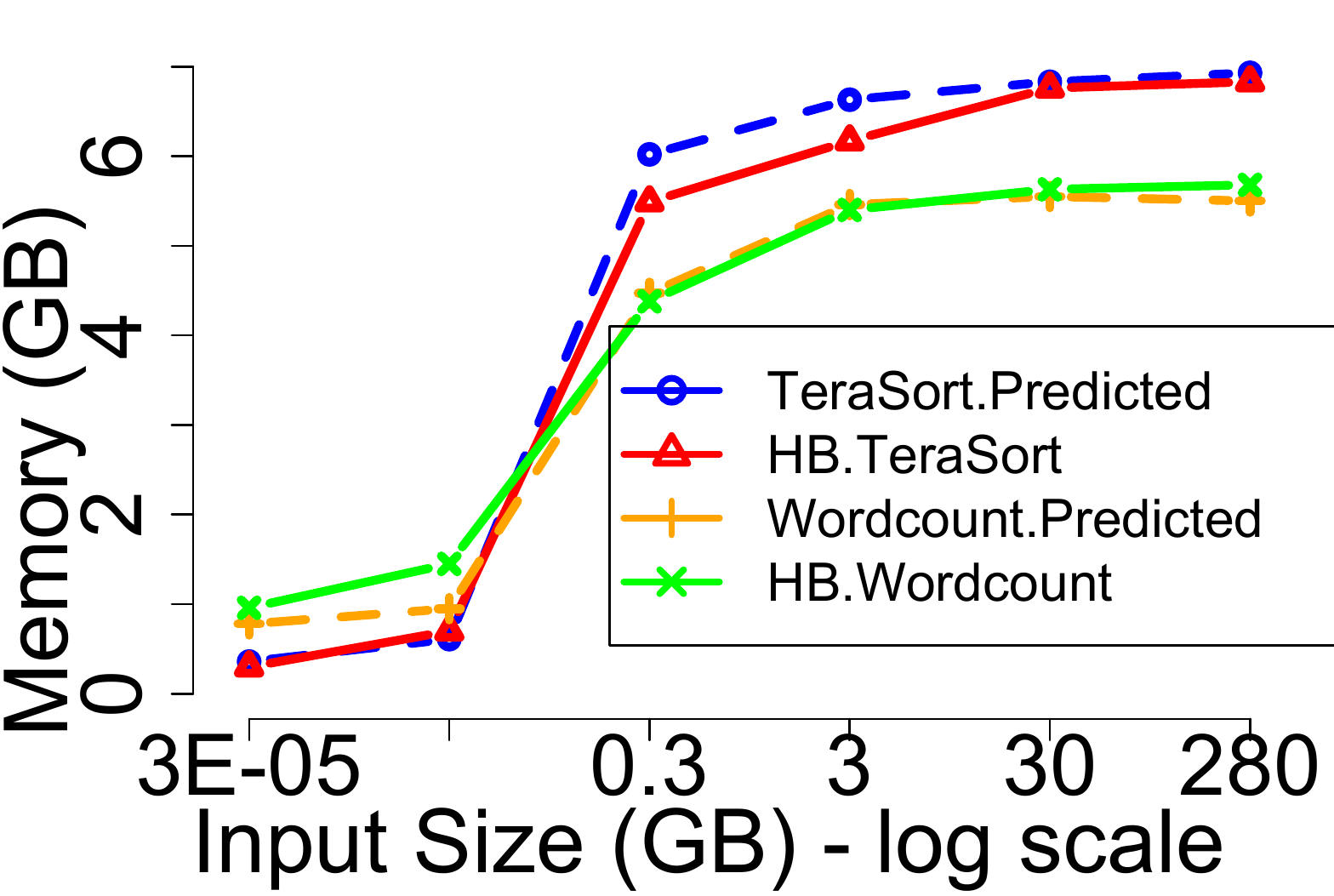}}
    &\subfloat[HB.Sort, BDB.Grep]{\includegraphics[scale=0.25]{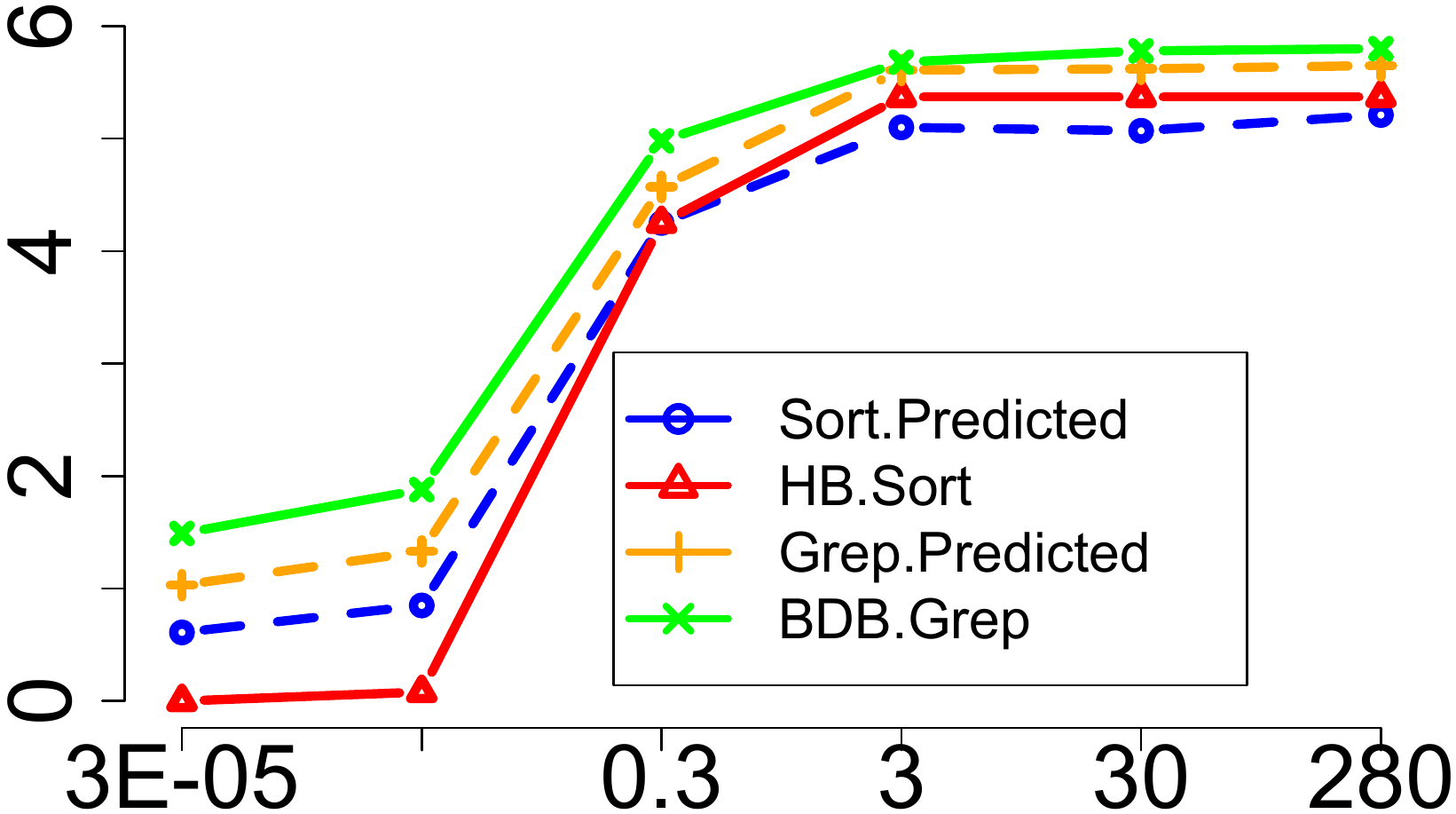}}
    &\subfloat[HB.Aggre, HB.Scan]{\includegraphics[scale=0.25]{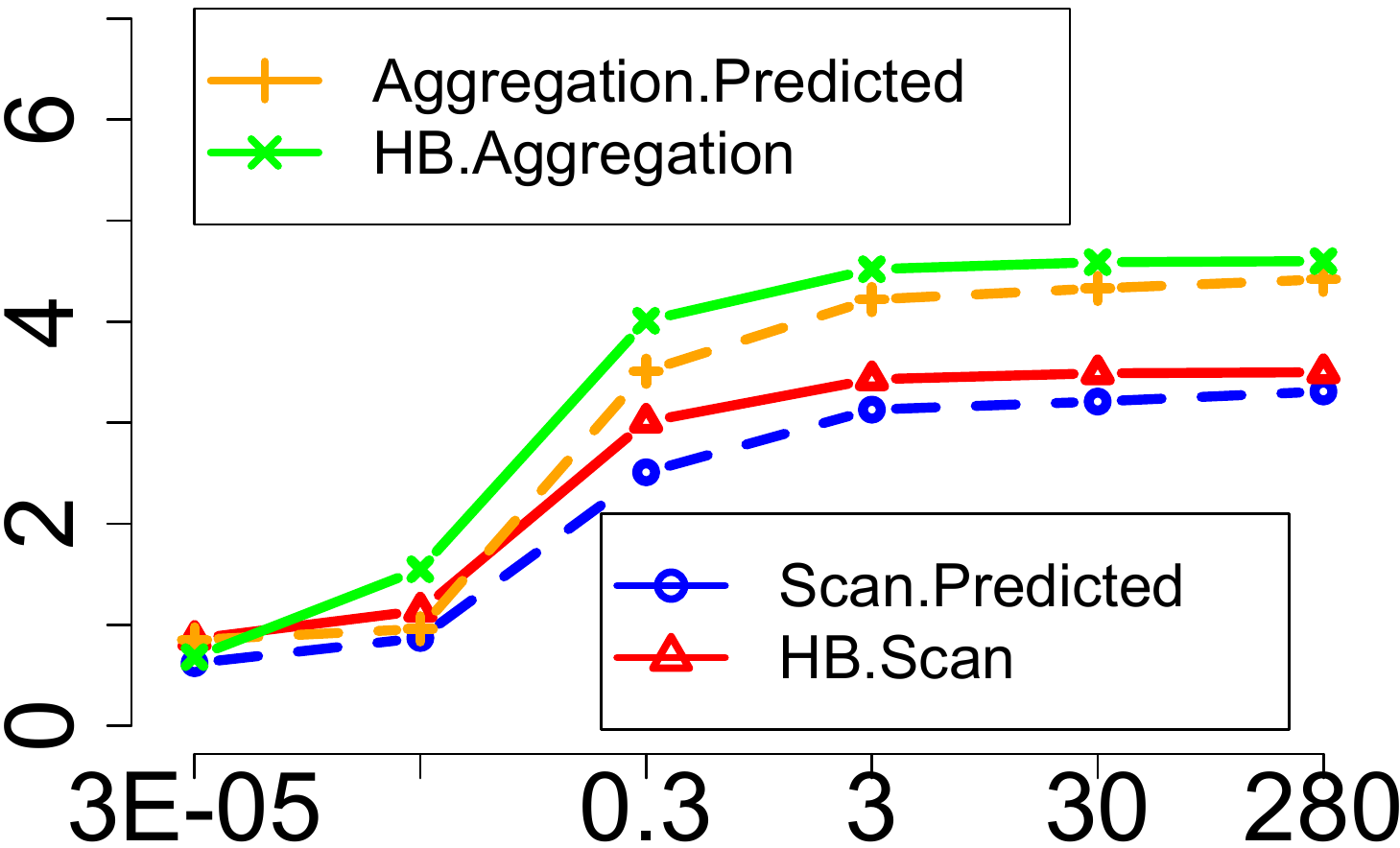}}
    & \subfloat[HB.Joins, BDB.Wordcount]{\includegraphics[scale=0.25]{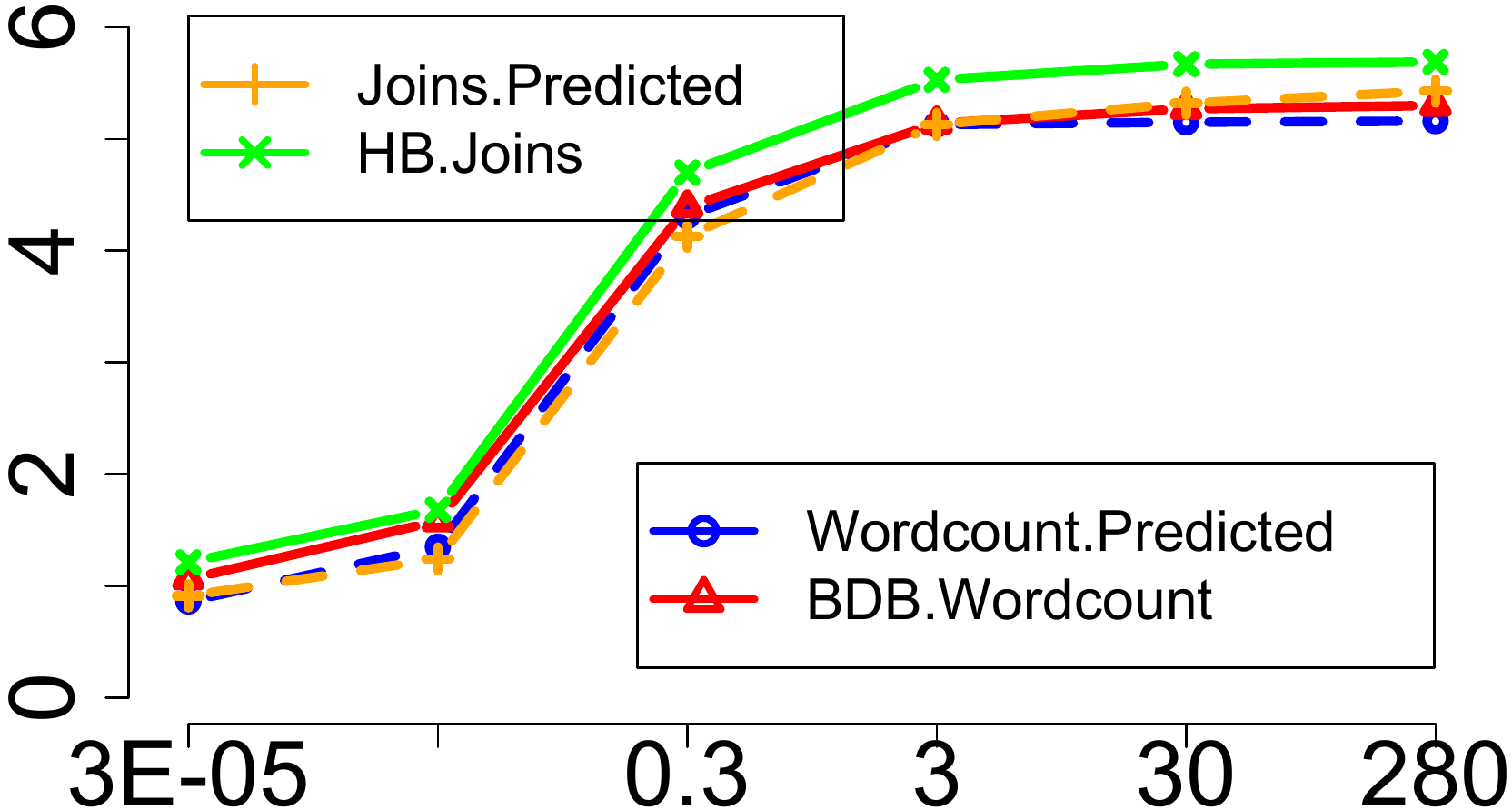}}
    \\
    \subfloat[BDB.Con.Comp, BDB.Kmeans]{\includegraphics[scale=0.28]{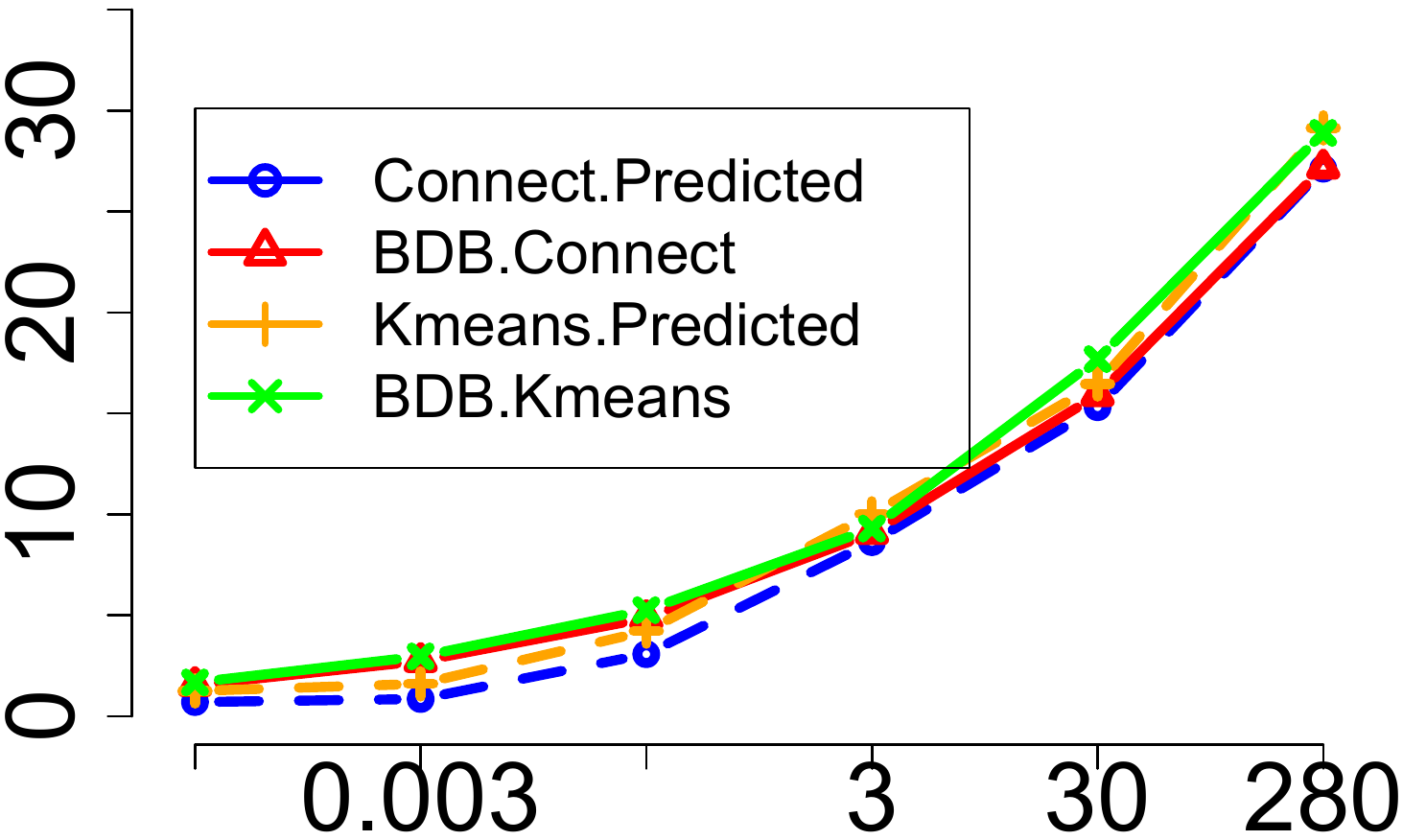}}
    &\subfloat[BDB.Sort, BDB.PageRank]{\includegraphics[scale=0.28]{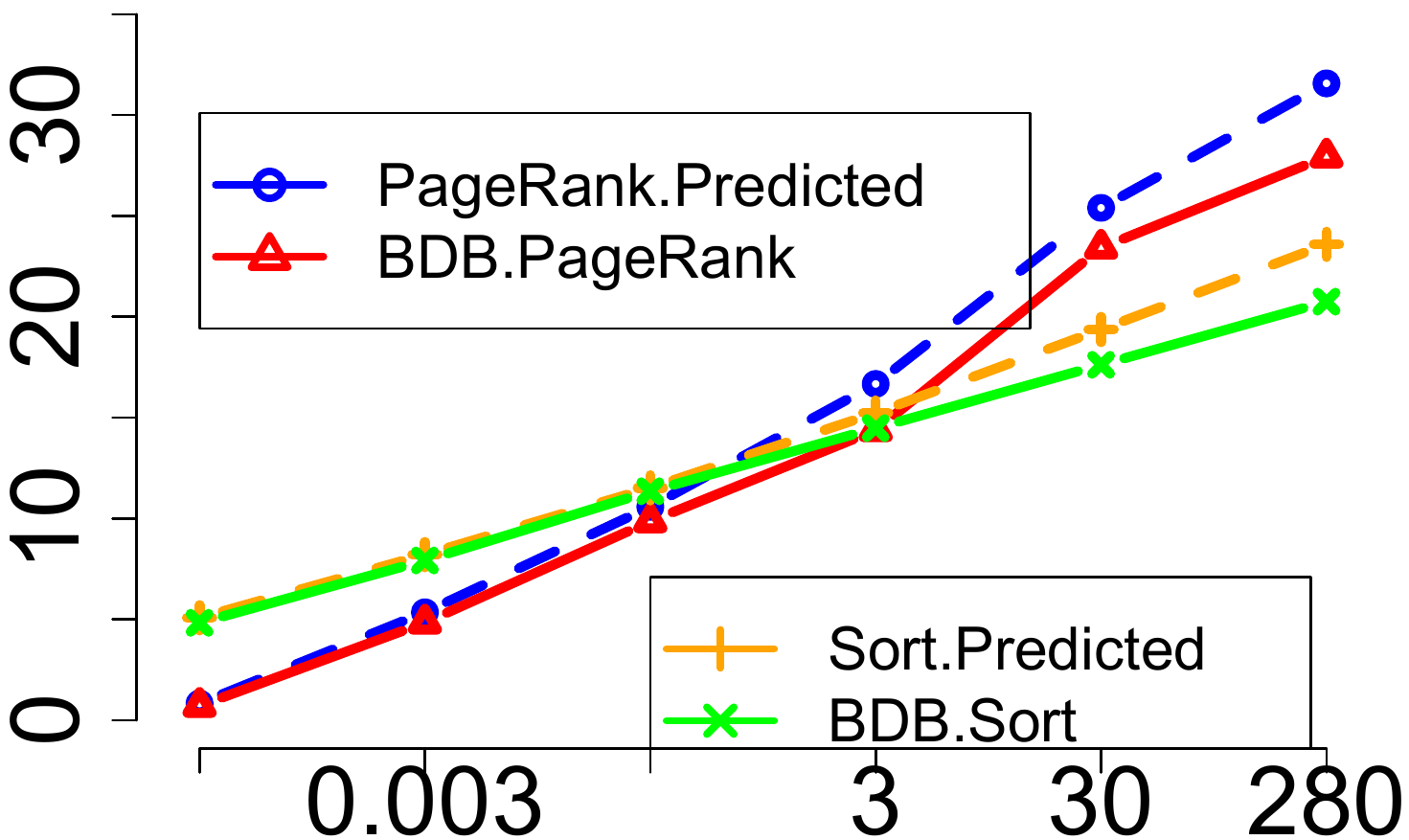}}
    & \subfloat[HB.Bayes, HB.Kmeans]{\includegraphics[scale=0.28]{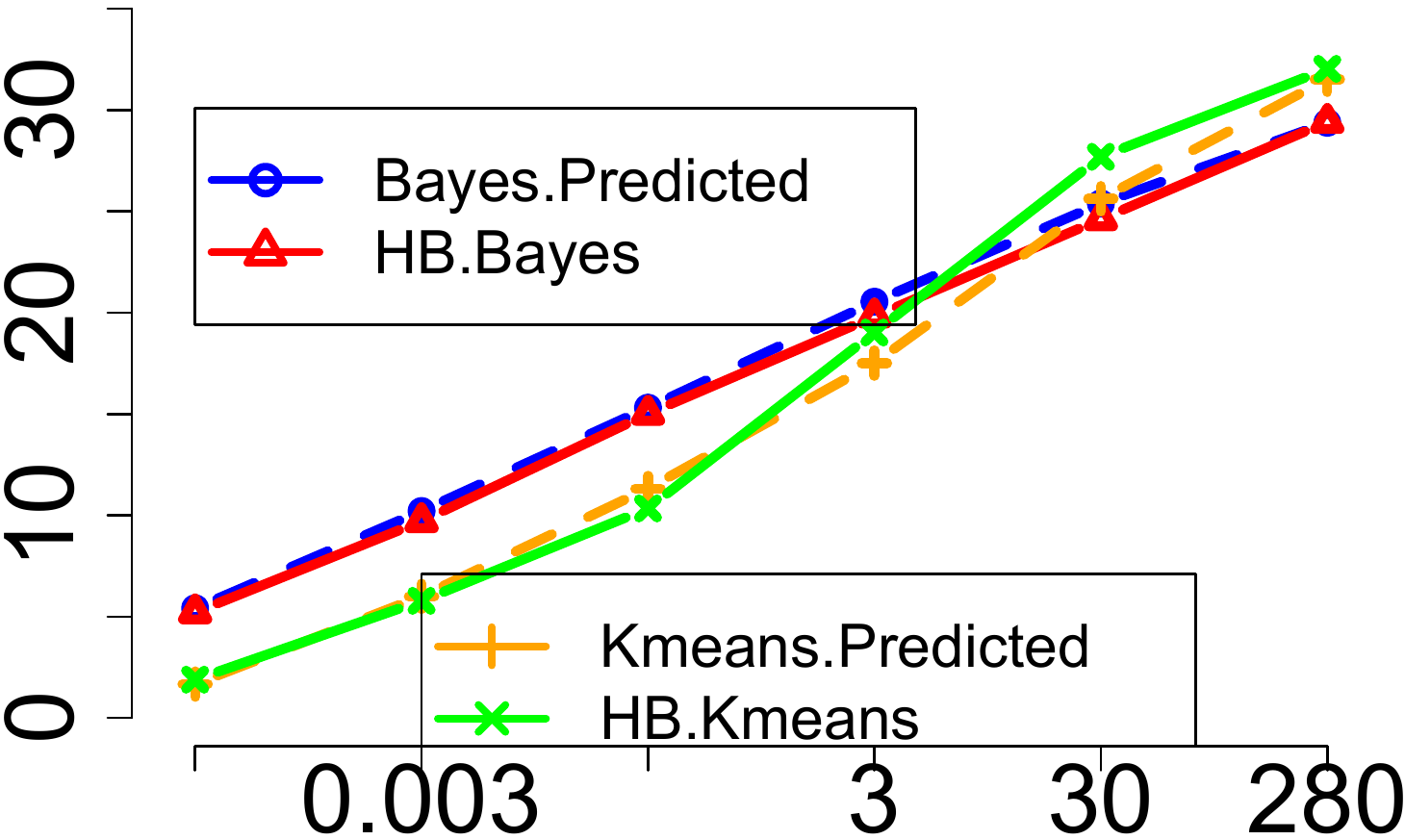}}
    & \subfloat[BDB.N.Bayes, HB.PageRank]{\includegraphics[scale=0.28]{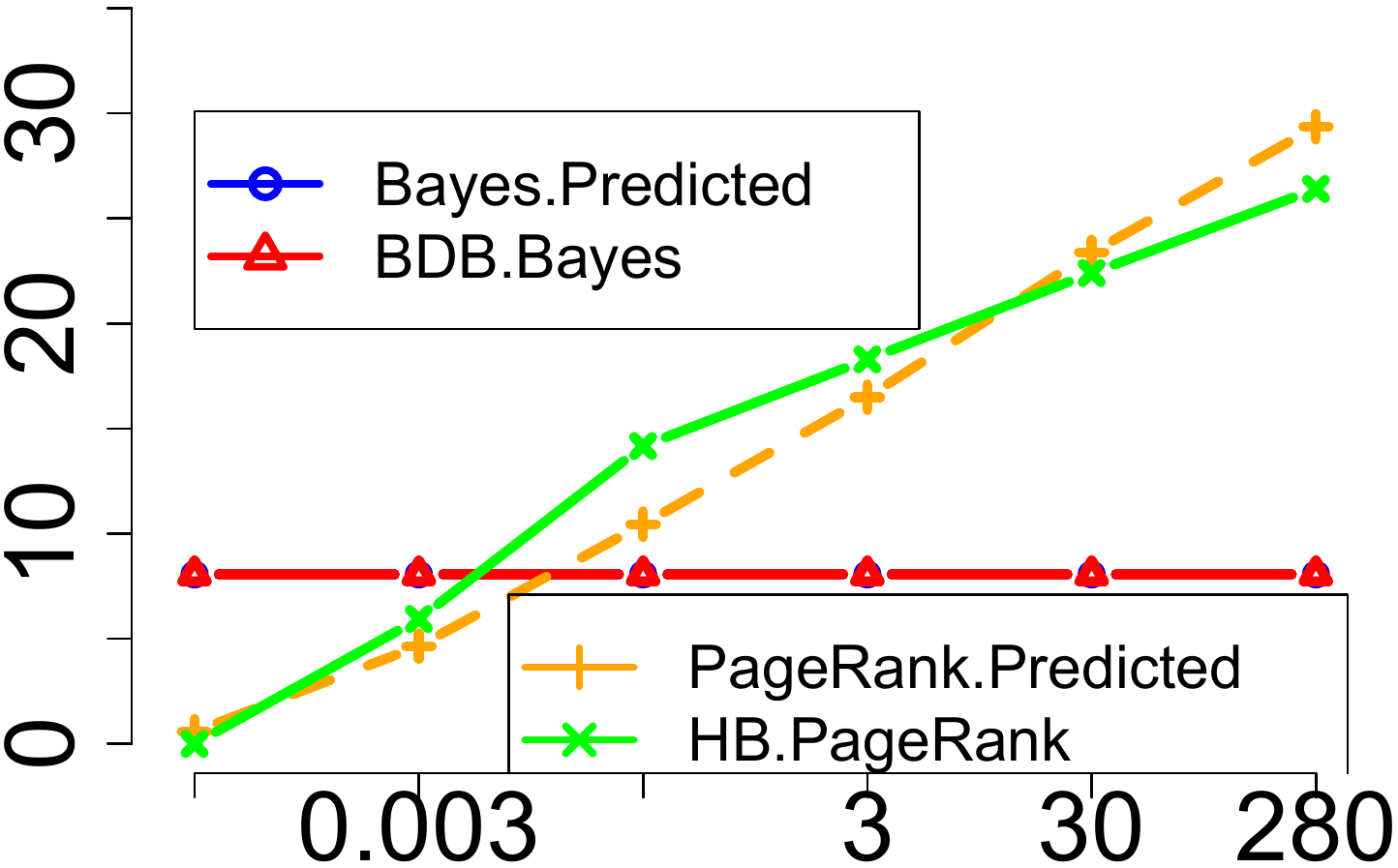}}

    \end{tabular}
    \caption{Comparisons of the predicted memory footprint to the measured value.
    This set of memory functions can precisely capture the memory requirement of our benchmarks.
    }
    \label{fig_benchmarksModels}
\end{figure*}

\paragraph{Compare to Alternative Classifiers.}
Table~\ref{table:pclass} gives the memory function prediction accuracy (averaged across benchmarks and inputs) of various alternative  classification techniques and our \KNN model.
The alternative models were built using the same features and training data. Thanks to the high-quality features, all classifiers are highly accurate in predicting the memory function.
We choose \KNN because its accuracy is comparable to alternative techniques but does not require re-training when a new memory function is added.

\paragraph{Memory Functions.}
Figure~\ref{fig_benchmarksModels} compares the predicted memory footprint to the measured values for HiBench and BigDataBench,
showing that our memory functions can precisely capture the application's memory footprint. This figure also shows that a single model is
unlikely to capture diverse application behaviors. We address this by developing an extensible framework into which we
can easily plug-in multiple models to capture different application behaviors.

\begin{table}[!t]
	\caption{Prediction accuracy for different classifiers}
	\label{table:pclass}
    \small
	\centering
	\begin{tabular}{llll}
		\toprule
		\textbf{Classifier} & \textbf{Accuracy (\%)} & \textbf{Classifier} & \textbf{Accuracy (\%)}\\
		\midrule
		 \rowcolor{Gray} Naive Bayes & 92.5 & SVM & 	95.4\\

        MLP &	94.1 &  Random Forests	& 95.5\\

         \rowcolor{Gray} Decision Tree	& 96.8 &  ANN	& 96.9 \\

        KNN &	97.4 & & \\
		\bottomrule
	\end{tabular}
\end{table}


\section{Related Work}
\label{sect_relatedwork}
Our work lies at the intersection between big data workload tuning and machine learning based system optimization. There is no existing
work which is similar to ours, in respect to co-locating big data applications with optimal memory allocation using predictive modeling.

\subsection{Optimizing Big Data Workloads}
\vspace{+2mm}
\paragraph{Domain-specific Optimization.} There exists a large
body of work  focusing on
optimizing a single application using domain-specific knowledge.
Prior work in domain-specific optimizations for single big data applications includes query
optimization~\cite{Binnig2013,Condie:2010:MO:1855711.1855732,Xu2015}, graph or data flow optimization~\cite{Fokoue2016, 180252, Borkar:2011:HFE:2004686.2005632,6729619}, task tuning~\cite{Cheng2014}
and personal assistant and deep learning services ~\cite{DBLP:conf/asplos/HauswaldLZLRKDM15}.
By contrast, we target resource modeling of Spark applications and
demonstrate that this technique is useful for scheduling multiple application tasks.

\paragraph{Memory Management.} Numerous techniques have been proposed to manage memory resources of big data applications~\cite{Tang2011}.
Many of the prior works  require using dedicated APIs to rewrite the
application~~\cite{Nguyen:2015:FCR:2694344.2694345,Fang:2015:ITT:2815400.2815407}. Fang \emph{et al.} introduce Interruptible Tasks, a
parallel data task that can be interrupted upon memory pressure. Their work aims to solve the out-of-memory problem when processing large
amounts of data on a single host~\cite{Fang:2015:ITT:2815400.2815407}. This work is thus orthogonal to our work and can be used to address
the problem of occasional over-subscription of memory resources. MemTune is a recent work on heap management for Spark
applications~\cite{ipdps16}. It detects memory contention and dynamically adjusts the memory partitions between Spark processes, but it
does not address the problem of precise memory allocation.



\paragraph{Application Scheduling.}
Verma \emph{et al.} use profiling information to schedule jobs within a MapReduce application~\cite{6298160}. Mashayekhy \emph{et al.}
develop energy-aware heuristics to map tasks of a big data application to servers to minimize energy usage~\cite{6899694}. Unlike our work,
all these works target scheduling jobs within a single application, and allocate all physical memory of a machine to one single
application. Other work looks at mapping parallelism by determining the number of cores and process time to be allocated to an
application~\cite{DBLP:conf/isca/HauswaldKLCLMDM15}. Our method promotes memory utilization on a local host, allowing the system to perform
more tasks than previously allowed with current methods. Consequently, higher multi-tasking levels may lead to an increase in non-local
data accesses within each task;
the scheduling framework in~\cite{DBLP:conf/isca/HauswaldKLCLMDM15} is therefore complementary to our work.

\paragraph{Task Co-location.}  Prior studies in task co-location include Bubble-Flux~\cite{Yang:2013:BPO:2485922.2485974},
Quasar~\cite{Delimitrou:2014:QRQ:2541940.2541941}, Tetris~\cite{Grandl:2014:MPC:2619239.2626334} and Cooper~\cite{Cooper},   which
co-locate tasks across machines. Other studies schedule workloads on multi-core processors~\cite{Zhuravlev:2010:ASR:1736020.1736036,
Liu:2014:OVM:2665671.2665720}. All the approaches mentioned above employ a single monolithic function to model the resource requirement of
application tasks. There is little ability to examine whether the function fits the application under the current runtime scenario. Other
fine-grained scheduling frameworks, like Mesos~\cite{Hindman:2011:MPF:1972457.1972488}, rely on the user to provide the resource
requirement of the application~\cite{Hindman:2011:MPF:1972457.1972488}. By contrast, we develop an extensive framework that uses multiple
modeling techniques to automatically estimate the resource requirement. Our approach allows new models to be added over time to target a
wider range of applications. Experimental results show that our approach yields better performance than a single model based approach.
On the other hand, the co-location policies developed in these prior works for determining which two applications should co-locate are complementary to our work.

\subsection{Predictive Modeling} Recent studies have shown that machine learning based predictive modeling is effective in code
optimization~\cite{Tournavitis:2009:THA:1542476.1542496,stock.12.taco,Simon:2013:ACI:2495258.2495914,Ganapathi:2009:CML:1855591.1855592,Wang:2014:IPP:2591460.2579561,ogilvie2017minimizing},
parallelism
mapping~\cite{Wang:2009:MPM:1504176.1504189,7851530,Wang:2013:UML:2509420.2512436,wang2015automatic,Taylor:2017:AOO:3078633.3081040}, task
scheduling~\cite{gdwzlcpc13,infocom17}, and processor resource allocation~\cite{c4ce9ce754bf431fba9c3ce457e6f28d}. In
\cite{Emani:2015:CDM:2737924.2737999}, a mixture-of-experts approach is proposed to schedule OpenMP programs on multi-cores. Their approach
uses multiple linear regression models to predict the optimal number of threads to use for a given program on a single machine. Our
approach differs from \cite{Emani:2015:CDM:2737924.2737999} in two aspects. First, we target a different problem (determining the memory
footprint vs the number of threads) and a different scale (multiple vs a single node). Secondly, we use different modeling techniques, both
linear and non-linear, to capture the memory behaviors of different applications. No work so far has used predictive modeling to model an
application's memory requirement to co-locate big data application tasks. This work is the first to do so.

\section{Conclusions}
\label{sect_conclusions} This paper has presented a novel scheme based on a mixture-of-experts approach to estimate the memory footprint of
a Spark applications for a given dataset.  Our approach determines at runtime, which of the off-line learned functions should be used to
model the application's memory resource demand. One of the advantages of our approach is that it provides a mechanism to gracefully add
additional expertise knowledge to target a wider range of applications. We combine our resource prediction framework with a runtime task
scheduler to co-locate latency-insensitive Spark applications, aiming to improve system throughput and application turnaround time. Using
the accurate prediction given by our framework, a runtime task scheduler can efficiently dispatch multiple applications to run concurrently
on a single host to improve the system's throughput and at the same time to ensure the total memory consumption does not exceed the
physical memory of the host. Our approach is applied to 44~representative big data applications built upon Apache Spark. On a 40-node
cluster, our approach achieves, on average, 83.9\% and 93.4\% of the \Oracle performance on system throughput and turnaround time,
respectively.

\section*{Acknowledgement}
We thank the anonymous reviewers for their insightful feedback. This work was partly supported by the UK EPSRC under grants EP/M01567X/1
(SANDeRs), EP/M029603/1 (Deep Online Cognition) and EP/M015793/1 (DIVIDEND). The corresponding author of this paper is Zheng Wang (Email:
z.wang@lancaster.ac.uk).

\newpage

\bibliographystyle{ACM-Reference-Format}
\bibliography{paper,relatedwork}


\begin{thebibliography}{00}


\ifx \showCODEN    \undefined \def \showCODEN     #1{\unskip}     \fi
\ifx \showDOI      \undefined \def \showDOI       #1{#1}\fi
\ifx \showISBNx    \undefined \def \showISBNx     #1{\unskip}     \fi
\ifx \showISBNxiii \undefined \def \showISBNxiii  #1{\unskip}     \fi
\ifx \showISSN     \undefined \def \showISSN      #1{\unskip}     \fi
\ifx \showLCCN     \undefined \def \showLCCN      #1{\unskip}     \fi
\ifx \shownote     \undefined \def \shownote      #1{#1}          \fi
\ifx \showarticletitle \undefined \def \showarticletitle #1{#1}   \fi
\ifx \showURL      \undefined \def \showURL       {\relax}        \fi
\providecommand\bibfield[2]{#2}
\providecommand\bibinfo[2]{#2}
\providecommand\natexlab[1]{#1}
\providecommand\showeprint[2][]{arXiv:#2}

\bibitem[\protect\citeauthoryear{Alpaydin}{Alpaydin}{2010}]%
        {Alpaydin:2010:IML:1734076}
\bibfield{author}{\bibinfo{person}{Ethem Alpaydin}.}
  \bibinfo{year}{2010}\natexlab{}.
\newblock \bibinfo{booktitle}{{\em Introduction to Machine Learning\/}
  (\bibinfo{edition}{2nd} ed.)}.
\newblock \bibinfo{publisher}{The MIT Press}.
\newblock


\bibitem[\protect\citeauthoryear{Awan, Brorsson, Vlassov, and Ayguade}{Awan
  et~al\mbox{.}}{2015}]%
        {7310708}
\bibfield{author}{\bibinfo{person}{Ahsan~Javed Awan}, \bibinfo{person}{Mats
  Brorsson}, \bibinfo{person}{Vladimir Vlassov}, {and} \bibinfo{person}{Eduard
  Ayguade}.} \bibinfo{year}{2015}\natexlab{}.
\newblock \showarticletitle{Performance characterization of in-memory data
  analytics on a modern cloud server}. In \bibinfo{booktitle}{{\em Big Data and
  Cloud Computing (BDCloud), 2015 IEEE Fifth International Conference on}}.
  \bibinfo{pages}{1--8}.
\newblock


\bibitem[\protect\citeauthoryear{Bienia}{Bienia}{2011}]%
        {Bienia2011PARSEC}
\bibfield{author}{\bibinfo{person}{Christian Bienia}.}
  \bibinfo{year}{2011}\natexlab{}.
\newblock {\em \bibinfo{title}{Benchmarking Modern Multiprocessors}}.
\newblock \bibinfo{thesistype}{Ph.D. Dissertation}. \bibinfo{school}{Princeton
  University}.
\newblock


\bibitem[\protect\citeauthoryear{Binnig, May, and Mindnich}{Binnig
  et~al\mbox{.}}{2013}]%
        {Binnig2013}
\bibfield{author}{\bibinfo{person}{Carsten Binnig}, \bibinfo{person}{Norman
  May}, {and} \bibinfo{person}{Tobias Mindnich}.}
  \bibinfo{year}{2013}\natexlab{}.
\newblock \showarticletitle{{SQLScript: Efficiently analyzing big enterprise
  data in SAP HANA}}. In \bibinfo{booktitle}{{\em Lecture Notes in Informatics
  (LNI), Proceedings - Series of the Gesellschaft fur Informatik (GI)}},
  Vol.~\bibinfo{volume}{P-214}. \bibinfo{pages}{363--382}.
\newblock


\bibitem[\protect\citeauthoryear{Borkar, Carey, Grover, Onose, and
  Vernica}{Borkar et~al\mbox{.}}{2011}]%
        {Borkar:2011:HFE:2004686.2005632}
\bibfield{author}{\bibinfo{person}{Vinayak Borkar}, \bibinfo{person}{Michael
  Carey}, \bibinfo{person}{Raman Grover}, \bibinfo{person}{Nicola Onose}, {and}
  \bibinfo{person}{Rares Vernica}.} \bibinfo{year}{2011}\natexlab{}.
\newblock \showarticletitle{Hyracks: A flexible and extensible foundation for
  data-intensive computing}. In \bibinfo{booktitle}{{\em Data Engineering
  (ICDE), 2011 IEEE 27th International Conference on}}.
  \bibinfo{pages}{1151--1162}.
\newblock


\bibitem[\protect\citeauthoryear{Cavazos, Fursin, Agakov, Bonilla, O'Boyle, and
  Temam}{Cavazos et~al\mbox{.}}{2007}]%
        {Cavazos:2007:RSG:1251974.1252540}
\bibfield{author}{\bibinfo{person}{John Cavazos}, \bibinfo{person}{Grigori
  Fursin}, \bibinfo{person}{Felix Agakov}, \bibinfo{person}{Edwin Bonilla},
  \bibinfo{person}{Michael~FP O'Boyle}, {and} \bibinfo{person}{Olivier Temam}.}
  \bibinfo{year}{2007}\natexlab{}.
\newblock \showarticletitle{Rapidly selecting good compiler optimizations using
  performance counters}. In \bibinfo{booktitle}{{\em Proceedings of the
  International Symposium on Code Generation and Optimization}}. IEEE Computer
  Society, \bibinfo{pages}{185--197}.
\newblock


\bibitem[\protect\citeauthoryear{Cheng, Rao, Guo, and Zhou}{Cheng
  et~al\mbox{.}}{2014}]%
        {Cheng2014}
\bibfield{author}{\bibinfo{person}{Dazhao Cheng}, \bibinfo{person}{Jia Rao},
  \bibinfo{person}{Yanfei Guo}, {and} \bibinfo{person}{Xiaobo Zhou}.}
  \bibinfo{year}{2014}\natexlab{}.
\newblock \showarticletitle{Improving MapReduce Performance in Heterogeneous
  Environments with Adaptive Task Tuning}. In \bibinfo{booktitle}{{\em
  Proceedings of the 15th International Middleware Conference}} {\em
  (\bibinfo{series}{Middleware '14})}. \bibinfo{publisher}{ACM},
  \bibinfo{address}{New York, NY, USA}, \bibinfo{pages}{97--108}.
\newblock
\showISBNx{978-1-4503-2785-5}


\bibitem[\protect\citeauthoryear{Condie, Conway, Alvaro, Hellerstein,
  Elmeleegy, and Sears}{Condie et~al\mbox{.}}{2010}]%
        {Condie:2010:MO:1855711.1855732}
\bibfield{author}{\bibinfo{person}{Tyson Condie}, \bibinfo{person}{Neil
  Conway}, \bibinfo{person}{Peter Alvaro}, \bibinfo{person}{Joseph~M.
  Hellerstein}, \bibinfo{person}{Khaled Elmeleegy}, {and}
  \bibinfo{person}{Russell Sears}.} \bibinfo{year}{2010}\natexlab{}.
\newblock \showarticletitle{MapReduce Online}. In \bibinfo{booktitle}{{\em
  Proceedings of the 7th USENIX Conference on Networked Systems Design and
  Implementation}} {\em (\bibinfo{series}{NSDI'10})}.
\newblock


\bibitem[\protect\citeauthoryear{Cummins, Petoumenos, Wang, and
  Leather}{Cummins et~al\mbox{.}}{2017a}]%
        {pact17}
\bibfield{author}{\bibinfo{person}{Chris Cummins}, \bibinfo{person}{Pavlos
  Petoumenos}, \bibinfo{person}{Zheng Wang}, {and} \bibinfo{person}{Hugh
  Leather}.} \bibinfo{year}{2017}\natexlab{a}.
\newblock \showarticletitle{End-to-end Deep Learning of Optimization
  Heuristics}. In \bibinfo{booktitle}{{\em The 26th International Conference on
  Parallel Architectures and Compilation Techniques (PACT)}}.
\newblock


\bibitem[\protect\citeauthoryear{Cummins, Petoumenos, Wang, and
  Leather}{Cummins et~al\mbox{.}}{2017b}]%
        {cummins2017synthesizing}
\bibfield{author}{\bibinfo{person}{Chris Cummins}, \bibinfo{person}{Pavlos
  Petoumenos}, \bibinfo{person}{Zheng Wang}, {and} \bibinfo{person}{Hugh
  Leather}.} \bibinfo{year}{2017}\natexlab{b}.
\newblock \showarticletitle{Synthesizing benchmarks for predictive modeling}.
  In \bibinfo{booktitle}{{\em Code Generation and Optimization (CGO), 2017
  IEEE/ACM International Symposium on}}. \bibinfo{pages}{86--99}.
\newblock


\bibitem[\protect\citeauthoryear{Databricks}{Databricks}{2016}]%
        {SparkPerf2016}
\bibfield{author}{\bibinfo{person}{Databricks}.}
  \bibinfo{year}{2016}\natexlab{}.
\newblock \bibinfo{title}{Spark-Perf}.
\newblock   (\bibinfo{year}{2016}).
\newblock
\showURL{%
\url{https://github.com/databricks/spark-perf}}


\bibitem[\protect\citeauthoryear{Delimitrou and Kozyrakis}{Delimitrou and
  Kozyrakis}{2014}]%
        {Delimitrou:2014:QRQ:2541940.2541941}
\bibfield{author}{\bibinfo{person}{Christina Delimitrou} {and}
  \bibinfo{person}{Christos Kozyrakis}.} \bibinfo{year}{2014}\natexlab{}.
\newblock \showarticletitle{Quasar: Resource-efficient and QoS-aware Cluster
  Management}. In \bibinfo{booktitle}{{\em Proceedings of the 19th
  International Conference on Architectural Support for Programming Languages
  and Operating Systems}} {\em (\bibinfo{series}{ASPLOS '14})}.
\newblock


\bibitem[\protect\citeauthoryear{Emani and Boyle}{Emani and Boyle}{2015}]%
        {Emani:2015:CDM:2737924.2737999}
\bibfield{author}{\bibinfo{person}{Murali~Krishna Emani} {and}
  \bibinfo{person}{Michael Boyle}.} \bibinfo{year}{2015}\natexlab{}.
\newblock \showarticletitle{Celebrating Diversity: A Mixture of Experts
  Approach for Runtime Mapping in Dynamic Environments}.
\newblock \bibinfo{journal}{{\em SIGPLAN Not.\/}} \bibinfo{volume}{50},
  \bibinfo{number}{6}, \bibinfo{pages}{499--508}.
\newblock
\showISSN{0362-1340}


\bibitem[\protect\citeauthoryear{Eyerman and Eeckhout}{Eyerman and
  Eeckhout}{2010}]%
        {Eyerman:2010:PJS:1736020.1736033}
\bibfield{author}{\bibinfo{person}{Stijn Eyerman} {and} \bibinfo{person}{Lieven
  Eeckhout}.} \bibinfo{year}{2010}\natexlab{}.
\newblock \showarticletitle{Probabilistic Job Symbiosis Modeling for SMT
  Processor Scheduling}.
\newblock \bibinfo{journal}{{\em SIGPLAN Not.\/}} \bibinfo{volume}{45},
  \bibinfo{number}{3}, \bibinfo{pages}{91--102}.
\newblock
\showISSN{0362-1340}


\bibitem[\protect\citeauthoryear{Fang, Nguyen, Xu, Demsky, and Lu}{Fang
  et~al\mbox{.}}{2015}]%
        {Fang:2015:ITT:2815400.2815407}
\bibfield{author}{\bibinfo{person}{Lu Fang}, \bibinfo{person}{Khanh Nguyen},
  \bibinfo{person}{Guoqing Xu}, \bibinfo{person}{Brian Demsky}, {and}
  \bibinfo{person}{Shan Lu}.} \bibinfo{year}{2015}\natexlab{}.
\newblock \showarticletitle{Interruptible Tasks: Treating Memory Pressure As
  Interrupts for Highly Scalable Data-parallel Programs}. In
  \bibinfo{booktitle}{{\em Proceedings of the 25th Symposium on Operating
  Systems Principles}} {\em (\bibinfo{series}{SOSP '15})}.
\newblock


\bibitem[\protect\citeauthoryear{Fokoue, Hassanzadeh, Sadoghi, and
  Zhang}{Fokoue et~al\mbox{.}}{2016}]%
        {Fokoue2016}
\bibfield{author}{\bibinfo{person}{Achille Fokoue}, \bibinfo{person}{Oktie
  Hassanzadeh}, \bibinfo{person}{Mohammad Sadoghi}, {and} \bibinfo{person}{Ping
  Zhang}.} \bibinfo{year}{2016}\natexlab{}.
\newblock \showarticletitle{Predicting Drug-Drug Interactions Through
  Similarity-Based Link Prediction Over Web Data}. In \bibinfo{booktitle}{{\em
  Proceedings of the 25th International Conference Companion on World Wide
  Web}} {\em (\bibinfo{series}{WWW '16 Companion})}.
  \bibinfo{publisher}{International World Wide Web Conferences Steering
  Committee}, \bibinfo{address}{Republic and Canton of Geneva, Switzerland},
  \bibinfo{pages}{175--178}.
\newblock
\showISBNx{978-1-4503-4144-8}


\bibitem[\protect\citeauthoryear{Ganapathi, Datta, Fox, and
  Patterson}{Ganapathi et~al\mbox{.}}{2009}]%
        {Ganapathi:2009:CML:1855591.1855592}
\bibfield{author}{\bibinfo{person}{Archana Ganapathi}, \bibinfo{person}{Kaushik
  Datta}, \bibinfo{person}{Armando Fox}, {and} \bibinfo{person}{David
  Patterson}.} \bibinfo{year}{2009}\natexlab{}.
\newblock \showarticletitle{A Case for Machine Learning to Optimize Multicore
  Performance}. In \bibinfo{booktitle}{{\em Proceedings of the First USENIX
  Conference on Hot Topics in Parallelism}} {\em
  (\bibinfo{series}{HotPar'09})}.
\newblock


\bibitem[\protect\citeauthoryear{Gao, Zhu, Jia, Luo, and Wang}{Gao
  et~al\mbox{.}}{2013}]%
        {Gao2013}
\bibfield{author}{\bibinfo{person}{Wanling Gao}, \bibinfo{person}{Yuqing Zhu},
  \bibinfo{person}{Zhen Jia}, \bibinfo{person}{Chunjie Luo}, {and}
  \bibinfo{person}{Lei Wang}.} \bibinfo{year}{2013}\natexlab{}.
\newblock \showarticletitle{{Bigdatabench: a big data benchmark suite from web
  search engines}}. \bibinfo{pages}{1--7}.
\newblock
\showISBNx{9781479930975}


\bibitem[\protect\citeauthoryear{Grandl, Ananthanarayanan, Kandula, Rao, and
  Akella}{Grandl et~al\mbox{.}}{2014}]%
        {Grandl:2014:MPC:2619239.2626334}
\bibfield{author}{\bibinfo{person}{Robert Grandl}, \bibinfo{person}{Ganesh
  Ananthanarayanan}, \bibinfo{person}{Srikanth Kandula},
  \bibinfo{person}{Sriram Rao}, {and} \bibinfo{person}{Aditya Akella}.}
  \bibinfo{year}{2014}\natexlab{}.
\newblock \showarticletitle{Multi-resource Packing for Cluster Schedulers}, In
  \bibinfo{booktitle}{Proceedings of the 2014 ACM Conference on SIGCOMM}.
\newblock \bibinfo{journal}{{\em SIGCOMM Comput. Commun. Rev.\/}}
  \bibinfo{volume}{44}, \bibinfo{number}{4}, \bibinfo{pages}{455--466}.
\newblock
\showISSN{0146-4833}


\bibitem[\protect\citeauthoryear{Grewe, Wang, and O’Boyle}{Grewe
  et~al\mbox{.}}{2013}]%
        {gdwzlcpc13}
\bibfield{author}{\bibinfo{person}{Dominik Grewe}, \bibinfo{person}{Zheng
  Wang}, {and} \bibinfo{person}{Michael~FP O’Boyle}.}
  \bibinfo{year}{2013}\natexlab{}.
\newblock \showarticletitle{OpenCL task partitioning in the presence of GPU
  contention}. In \bibinfo{booktitle}{{\em International Workshop on Languages
  and Compilers for Parallel Computing}}. \bibinfo{pages}{87--101}.
\newblock


\bibitem[\protect\citeauthoryear{Hauswald, Kang, Laurenzano, Chen, Li, Mudge,
  Dreslinski, Mars, and Tang}{Hauswald et~al\mbox{.}}{2015a}]%
        {DBLP:conf/isca/HauswaldKLCLMDM15}
\bibfield{author}{\bibinfo{person}{Johann Hauswald}, \bibinfo{person}{Yiping
  Kang}, \bibinfo{person}{Michael~A. Laurenzano}, \bibinfo{person}{Quan Chen},
  \bibinfo{person}{Cheng Li}, \bibinfo{person}{Trevor Mudge},
  \bibinfo{person}{Ronald~G. Dreslinski}, \bibinfo{person}{Jason Mars}, {and}
  \bibinfo{person}{Lingjia Tang}.} \bibinfo{year}{2015}\natexlab{a}.
\newblock \showarticletitle{DjiNN and Tonic: DNN As a Service and Its
  Implications for Future Warehouse Scale Computers}, In
  \bibinfo{booktitle}{ISCA '15}.
\newblock \bibinfo{journal}{{\em SIGARCH Comput. Archit. News\/}}
  \bibinfo{volume}{43}, \bibinfo{number}{3}, \bibinfo{pages}{27--40}.
\newblock
\showISSN{0163-5964}


\bibitem[\protect\citeauthoryear{Hauswald, Laurenzano, Zhang, Li, Rovinski,
  Khurana, Dreslinski, Mudge, Petrucci, Tang, and Mars}{Hauswald
  et~al\mbox{.}}{2015b}]%
        {DBLP:conf/asplos/HauswaldLZLRKDM15}
\bibfield{author}{\bibinfo{person}{Johann Hauswald},
  \bibinfo{person}{Michael~A. Laurenzano}, \bibinfo{person}{Yunqi Zhang},
  \bibinfo{person}{Cheng Li}, \bibinfo{person}{Austin Rovinski},
  \bibinfo{person}{Arjun Khurana}, \bibinfo{person}{Ronald~G. Dreslinski},
  \bibinfo{person}{Trevor Mudge}, \bibinfo{person}{Vinicius Petrucci},
  \bibinfo{person}{Lingjia Tang}, {and} \bibinfo{person}{Jason Mars}.}
  \bibinfo{year}{2015}\natexlab{b}.
\newblock \showarticletitle{Sirius: An Open End-to-End Voice and Vision
  Personal Assistant and Its Implications for Future Warehouse Scale
  Computers}, In \bibinfo{booktitle}{ASPLOS '15}.
\newblock \bibinfo{journal}{{\em SIGPLAN Not.\/}} \bibinfo{volume}{50},
  \bibinfo{number}{4}, \bibinfo{pages}{223--238}.
\newblock
\showISSN{0362-1340}


\bibitem[\protect\citeauthoryear{Hindman, Konwinski, Zaharia, Ghodsi, Joseph,
  Katz, Shenker, and Stoica}{Hindman et~al\mbox{.}}{2011}]%
        {Hindman:2011:MPF:1972457.1972488}
\bibfield{author}{\bibinfo{person}{Benjamin Hindman}, \bibinfo{person}{Andy
  Konwinski}, \bibinfo{person}{Matei Zaharia}, \bibinfo{person}{Ali Ghodsi},
  \bibinfo{person}{Anthony~D. Joseph}, \bibinfo{person}{Randy Katz},
  \bibinfo{person}{Scott Shenker}, {and} \bibinfo{person}{Ion Stoica}.}
  \bibinfo{year}{2011}\natexlab{}.
\newblock \showarticletitle{Mesos: A Platform for Fine-grained Resource Sharing
  in the Data Center}. In \bibinfo{booktitle}{{\em Proceedings of the 8th
  USENIX Conference on Networked Systems Design and Implementation}} {\em
  (\bibinfo{series}{NSDI'11})}. \bibinfo{pages}{295--308}.
\newblock


\bibitem[\protect\citeauthoryear{Huang, Huang, Dai, Xie, and Huang}{Huang
  et~al\mbox{.}}{2011}]%
        {Huang2011}
\bibfield{author}{\bibinfo{person}{Shengsheng Huang}, \bibinfo{person}{Jie
  Huang}, \bibinfo{person}{Jinquan Dai}, \bibinfo{person}{Tao Xie}, {and}
  \bibinfo{person}{Bo Huang}.} \bibinfo{year}{2011}\natexlab{}.
\newblock \showarticletitle{The HiBench Benchmark Suite: Characterization of
  the MapReduce-Based Data Analysis}. In \bibinfo{booktitle}{{\em New Frontiers
  in Information and Software as Services: Service and Application Design
  Challenges in the Cloud}}, \bibfield{editor}{\bibinfo{person}{Divyakant
  Agrawal}, \bibinfo{person}{K.~Sel{\c{c}}uk Candan}, {and}
  \bibinfo{person}{Wen-Syan Li}} (Eds.). \bibinfo{publisher}{Springer Berlin
  Heidelberg}, \bibinfo{address}{Berlin, Heidelberg},
  \bibinfo{pages}{209--228}.
\newblock


\bibitem[\protect\citeauthoryear{Jacobs, Jordan, Nowlan, and Hinton}{Jacobs
  et~al\mbox{.}}{1991}]%
        {Jacobs:1991:AML:1351011.1351018}
\bibfield{author}{\bibinfo{person}{Robert~A. Jacobs},
  \bibinfo{person}{Michael~I. Jordan}, \bibinfo{person}{Steven~J. Nowlan},
  {and} \bibinfo{person}{Geoffrey~E. Hinton}.} \bibinfo{year}{1991}\natexlab{}.
\newblock \showarticletitle{Adaptive Mixtures of Local Experts}.
\newblock \bibinfo{journal}{{\em Neural Comput.\/}} \bibinfo{volume}{3},
  \bibinfo{number}{1} (\bibinfo{date}{March} \bibinfo{year}{1991}),
  \bibinfo{pages}{79--87}.
\newblock
\showISSN{0899-7667}


\bibitem[\protect\citeauthoryear{Jiang, Zhang, Hou, Chai, Mckee, Jia, and
  Sun}{Jiang et~al\mbox{.}}{2014}]%
        {Jiang2014}
\bibfield{author}{\bibinfo{person}{Tao Jiang}, \bibinfo{person}{Qianlong
  Zhang}, \bibinfo{person}{Rui Hou}, \bibinfo{person}{Lin Chai},
  \bibinfo{person}{Sally~A Mckee}, \bibinfo{person}{Zhen Jia}, {and}
  \bibinfo{person}{Ninghui Sun}.} \bibinfo{year}{2014}\natexlab{}.
\newblock \showarticletitle{Understanding the behavior of in-memory computing
  workloads}. In \bibinfo{booktitle}{{\em Workload Characterization (IISWC),
  2014 IEEE International Symposium on Workload Characterization}}. IEEE,
  \bibinfo{pages}{22--30}.
\newblock


\bibitem[\protect\citeauthoryear{Keller and Gray}{Keller and Gray}{1985}]%
        {Keller1985}
\bibfield{author}{\bibinfo{person}{James~M. Keller} {and}
  \bibinfo{person}{Michael~R. Gray}.} \bibinfo{year}{1985}\natexlab{}.
\newblock \showarticletitle{{A Fuzzy K-Nearest Neighbor Algorithm}}.
\newblock \bibinfo{journal}{{\em IEEE Transactions on Systems, Man and
  Cybernetics\/}} (\bibinfo{year}{1985}).
\newblock


\bibitem[\protect\citeauthoryear{Kyrola, Blelloch, and Guestrin}{Kyrola
  et~al\mbox{.}}{2012}]%
        {180252}
\bibfield{author}{\bibinfo{person}{Aapo Kyrola}, \bibinfo{person}{Guy
  Blelloch}, {and} \bibinfo{person}{Carlos Guestrin}.}
  \bibinfo{year}{2012}\natexlab{}.
\newblock \showarticletitle{GraphChi: Large-scale Graph Computation on Just a
  PC}. In \bibinfo{booktitle}{{\em Proceedings of the 10th USENIX Conference on
  Operating Systems Design and Implementation}} {\em
  (\bibinfo{series}{OSDI'12})}.
\newblock


\bibitem[\protect\citeauthoryear{Li, Tan, Wang, Zhang, and Salapura}{Li
  et~al\mbox{.}}{2015}]%
        {Li2015}
\bibfield{author}{\bibinfo{person}{Min Li}, \bibinfo{person}{Jian Tan},
  \bibinfo{person}{Yandong Wang}, \bibinfo{person}{Li Zhang}, {and}
  \bibinfo{person}{Valentina Salapura}.} \bibinfo{year}{2015}\natexlab{}.
\newblock \showarticletitle{SparkBench: A Comprehensive Benchmarking Suite for
  in Memory Data Analytic Platform Spark}. In \bibinfo{booktitle}{{\em
  Proceedings of the 12th ACM International Conference on Computing Frontiers}}
  {\em (\bibinfo{series}{CF '15})}.
\newblock


\bibitem[\protect\citeauthoryear{Liu and Li}{Liu and Li}{2014}]%
        {Liu:2014:OVM:2665671.2665720}
\bibfield{author}{\bibinfo{person}{Ming Liu} {and} \bibinfo{person}{Tao Li}.}
  \bibinfo{year}{2014}\natexlab{}.
\newblock \showarticletitle{Optimizing Virtual Machine Consolidation
  Performance on NUMA Server Architecture for Cloud Workloads}.
\newblock \bibinfo{journal}{{\em SIGARCH Comput. Archit. News\/}}
  \bibinfo{volume}{42}, \bibinfo{number}{3}, \bibinfo{pages}{325--336}.
\newblock
\showISSN{0163-5964}


\bibitem[\protect\citeauthoryear{Llull, Fan, Zahedi, and Lee}{Llull
  et~al\mbox{.}}{2017}]%
        {Cooper}
\bibfield{author}{\bibinfo{person}{Qiuyun Llull}, \bibinfo{person}{Songchun
  Fan}, \bibinfo{person}{Seyed~Majid Zahedi}, {and} \bibinfo{person}{Benjamin~C
  Lee}.} \bibinfo{year}{2017}\natexlab{}.
\newblock \showarticletitle{Cooper: Task Colocation with Cooperative Games}. In
  \bibinfo{booktitle}{{\em High Performance Computer Architecture (HPCA), 2017
  IEEE International Symposium on}}. \bibinfo{pages}{421--432}.
\newblock


\bibitem[\protect\citeauthoryear{Lo, Cheng, Govindaraju, Ranganathan, and
  Kozyrakis}{Lo et~al\mbox{.}}{2015}]%
        {Lo:2015:HIR:2749469.2749475}
\bibfield{author}{\bibinfo{person}{David Lo}, \bibinfo{person}{Liqun Cheng},
  \bibinfo{person}{Rama Govindaraju}, \bibinfo{person}{Parthasarathy
  Ranganathan}, {and} \bibinfo{person}{Christos Kozyrakis}.}
  \bibinfo{year}{2015}\natexlab{}.
\newblock \showarticletitle{Heracles: Improving Resource Efficiency at Scale}.
  In \bibinfo{booktitle}{{\em Proceedings of the 42Nd Annual International
  Symposium on Computer Architecture}} {\em (\bibinfo{series}{ISCA '15})}.
\newblock


\bibitem[\protect\citeauthoryear{Luk, Hong, and Kim}{Luk et~al\mbox{.}}{2009}]%
        {Luk:2009:QEP:1669112.1669121}
\bibfield{author}{\bibinfo{person}{Chi-Keung Luk}, \bibinfo{person}{Sunpyo
  Hong}, {and} \bibinfo{person}{Hyesoon Kim}.} \bibinfo{year}{2009}\natexlab{}.
\newblock \showarticletitle{Qilin: Exploiting Parallelism on Heterogeneous
  Multiprocessors with Adaptive Mapping}. In \bibinfo{booktitle}{{\em
  Proceedings of the 42Nd Annual IEEE/ACM International Symposium on
  Microarchitecture}} {\em (\bibinfo{series}{MICRO 42})}.
  \bibinfo{publisher}{ACM}, \bibinfo{pages}{45--55}.
\newblock


\bibitem[\protect\citeauthoryear{Manly}{Manly}{2004}]%
        {manly2004multivariate}
\bibfield{author}{\bibinfo{person}{Bryan~FJ Manly}.}
  \bibinfo{year}{2004}\natexlab{}.
\newblock \bibinfo{booktitle}{{\em Multivariate statistical methods: a
  primer}}.
\newblock \bibinfo{publisher}{CRC Press}.
\newblock


\bibitem[\protect\citeauthoryear{Mashayekhy, Nejad, Grosu, Zhang, and
  Shi}{Mashayekhy et~al\mbox{.}}{2015}]%
        {6899694}
\bibfield{author}{\bibinfo{person}{L. Mashayekhy}, \bibinfo{person}{M.~M.
  Nejad}, \bibinfo{person}{D. Grosu}, \bibinfo{person}{Q. Zhang}, {and}
  \bibinfo{person}{W. Shi}.} \bibinfo{year}{2015}\natexlab{}.
\newblock \showarticletitle{Energy-Aware Scheduling of MapReduce Jobs for Big
  Data Applications}.
\newblock \bibinfo{journal}{{\em IEEE Transactions on Parallel and Distributed
  Systems\/}} \bibinfo{volume}{26}, \bibinfo{number}{10}
  (\bibinfo{year}{2015}), \bibinfo{pages}{2720--2733}.
\newblock


\bibitem[\protect\citeauthoryear{Nguyen, Wang, Bu, Fang, Hu, and Xu}{Nguyen
  et~al\mbox{.}}{2015}]%
        {Nguyen:2015:FCR:2694344.2694345}
\bibfield{author}{\bibinfo{person}{Khanh Nguyen}, \bibinfo{person}{Kai Wang},
  \bibinfo{person}{Yingyi Bu}, \bibinfo{person}{Lu Fang},
  \bibinfo{person}{Jianfei Hu}, {and} \bibinfo{person}{Guoqing Xu}.}
  \bibinfo{year}{2015}\natexlab{}.
\newblock \showarticletitle{FACADE: A Compiler and Runtime for (Almost)
  Object-Bounded Big Data Applications}. In \bibinfo{booktitle}{{\em
  Proceedings of the Twentieth International Conference on Architectural
  Support for Programming Languages and Operating Systems}} {\em
  (\bibinfo{series}{ASPLOS '15})}.
\newblock


\bibitem[\protect\citeauthoryear{Ogilvie, Petoumenos, Wang, and
  Leather}{Ogilvie et~al\mbox{.}}{2017}]%
        {ogilvie2017minimizing}
\bibfield{author}{\bibinfo{person}{William~F Ogilvie}, \bibinfo{person}{Pavlos
  Petoumenos}, \bibinfo{person}{Zheng Wang}, {and} \bibinfo{person}{Hugh
  Leather}.} \bibinfo{year}{2017}\natexlab{}.
\newblock \showarticletitle{Minimizing the cost of iterative compilation with
  active learning}. In \bibinfo{booktitle}{{\em Code Generation and
  Optimization (CGO), 2017 IEEE/ACM International Symposium on}}.
  \bibinfo{pages}{245--256}.
\newblock


\bibitem[\protect\citeauthoryear{Ousterhout, Rasti, Ratnasamy, Shenker, and
  Chun}{Ousterhout et~al\mbox{.}}{2015}]%
        {Ousterhout:2015:MSP:2789770.2789791}
\bibfield{author}{\bibinfo{person}{Kay Ousterhout}, \bibinfo{person}{Ryan
  Rasti}, \bibinfo{person}{Sylvia Ratnasamy}, \bibinfo{person}{Scott Shenker},
  {and} \bibinfo{person}{Byung-Gon Chun}.} \bibinfo{year}{2015}\natexlab{}.
\newblock \showarticletitle{Making Sense of Performance in Data Analytics
  Frameworks}. In \bibinfo{booktitle}{{\em Proceedings of the 12th USENIX
  Conference on Networked Systems Design and Implementation}} {\em
  (\bibinfo{series}{NSDI'15})}. \bibinfo{publisher}{USENIX Association},
  \bibinfo{address}{Berkeley, CA, USA}, \bibinfo{pages}{293--307}.
\newblock
\showISBNx{978-1-931971-218}


\bibitem[\protect\citeauthoryear{Ren, Gao, Wang, and Wang}{Ren
  et~al\mbox{.}}{2017}]%
        {infocom17}
\bibfield{author}{\bibinfo{person}{Jie Ren}, \bibinfo{person}{Ling Gao},
  \bibinfo{person}{Hai Wang}, {and} \bibinfo{person}{Zheng Wang}.}
  \bibinfo{year}{2017}\natexlab{}.
\newblock \showarticletitle{Optimise web browsing on heterogeneous mobile
  platforms: a machine learning based approach}. In \bibinfo{booktitle}{{\em
  IEEE International Conference on Computer Communications (INFOCOM), 2017}}
  {\em (\bibinfo{series}{INFOCOM 2017})}.
\newblock


\bibitem[\protect\citeauthoryear{Rupprecht, Culhane, and Pietzuch}{Rupprecht
  et~al\mbox{.}}{2017}]%
        {DBLP:journals/pvldb/RupprechtCP17}
\bibfield{author}{\bibinfo{person}{Lukas Rupprecht}, \bibinfo{person}{William
  Culhane}, {and} \bibinfo{person}{Peter~R. Pietzuch}.}
  \bibinfo{year}{2017}\natexlab{}.
\newblock \showarticletitle{SquirrelJoin: Network-Aware Distributed Join
  Processing with Lazy Partitioning}.
\newblock \bibinfo{journal}{{\em {PVLDB}\/}} \bibinfo{volume}{10},
  \bibinfo{number}{11} (\bibinfo{year}{2017}), \bibinfo{pages}{1250--1261}.
\newblock


\bibitem[\protect\citeauthoryear{Shvachko, Kuang, Radia, and Chansler}{Shvachko
  et~al\mbox{.}}{2010}]%
        {Shvachko2010}
\bibfield{author}{\bibinfo{person}{Konstantin Shvachko},
  \bibinfo{person}{Hairong Kuang}, \bibinfo{person}{Sanjay Radia}, {and}
  \bibinfo{person}{Robert Chansler}.} \bibinfo{year}{2010}\natexlab{}.
\newblock \showarticletitle{The Hadoop Distributed File System}. In
  \bibinfo{booktitle}{{\em Proceedings of the 2010 IEEE 26th Symposium on Mass
  Storage Systems and Technologies (MSST)}} {\em (\bibinfo{series}{MSST '10})}.
  \bibinfo{pages}{1--10}.
\newblock


\bibitem[\protect\citeauthoryear{Simon, Cavazos, Wimmer, and Kulkarni}{Simon
  et~al\mbox{.}}{2013}]%
        {Simon:2013:ACI:2495258.2495914}
\bibfield{author}{\bibinfo{person}{Douglas Simon}, \bibinfo{person}{John
  Cavazos}, \bibinfo{person}{Christian Wimmer}, {and} \bibinfo{person}{Sameer
  Kulkarni}.} \bibinfo{year}{2013}\natexlab{}.
\newblock \showarticletitle{Automatic Construction of Inlining Heuristics Using
  Machine Learning}. In \bibinfo{booktitle}{{\em Proceedings of the 2013
  IEEE/ACM International Symposium on Code Generation and Optimization (CGO)}}
  {\em (\bibinfo{series}{CGO '13})}. \bibinfo{pages}{1--12}.
\newblock


\bibitem[\protect\citeauthoryear{Singh, Bhadauria, and McKee}{Singh
  et~al\mbox{.}}{2009}]%
        {Singh:2009:RTP:1577129.1577137}
\bibfield{author}{\bibinfo{person}{Karan Singh}, \bibinfo{person}{Major
  Bhadauria}, {and} \bibinfo{person}{Sally~A. McKee}.}
  \bibinfo{year}{2009}\natexlab{}.
\newblock \showarticletitle{Real Time Power Estimation and Thread Scheduling
  via Performance Counters}.
\newblock \bibinfo{journal}{{\em SIGARCH Comput. Archit. News\/}}
  \bibinfo{volume}{37}, \bibinfo{number}{2} (\bibinfo{date}{jul}
  \bibinfo{year}{2009}), \bibinfo{pages}{46--55}.
\newblock
\showISSN{0163-5964}


\bibitem[\protect\citeauthoryear{Sparks, Talwalkar, Smith, Kottalam, Pan,
  Gonzalez, Franklin, Jordan, and Kraska}{Sparks et~al\mbox{.}}{2013}]%
        {6729619}
\bibfield{author}{\bibinfo{person}{Evan~R Sparks}, \bibinfo{person}{Ameet
  Talwalkar}, \bibinfo{person}{Virginia Smith}, \bibinfo{person}{Jey Kottalam},
  \bibinfo{person}{Xinghao Pan}, \bibinfo{person}{Joseph Gonzalez},
  \bibinfo{person}{Michael~J Franklin}, \bibinfo{person}{Michael~I Jordan},
  {and} \bibinfo{person}{Tim Kraska}.} \bibinfo{year}{2013}\natexlab{}.
\newblock \showarticletitle{MLI: An API for distributed machine learning}. In
  \bibinfo{booktitle}{{\em Data Mining (ICDM), 2013 IEEE 13th International
  Conference on}}. \bibinfo{pages}{1187--1192}.
\newblock


\bibitem[\protect\citeauthoryear{Stock, Pouchet, and Sadayappan}{Stock
  et~al\mbox{.}}{2012}]%
        {stock.12.taco}
\bibfield{author}{\bibinfo{person}{Kevin Stock},
  \bibinfo{person}{Louis-No\"{e}l Pouchet}, {and} \bibinfo{person}{P.
  Sadayappan}.} \bibinfo{year}{2012}\natexlab{}.
\newblock \showarticletitle{Using Machine Learning to Improve Automatic
  Vectorization}.
\newblock \bibinfo{journal}{{\em ACM Trans. Archit. Code Optim.\/}}
  \bibinfo{volume}{8}, \bibinfo{number}{4}, Article \bibinfo{articleno}{50}
  (\bibinfo{date}{jan} \bibinfo{year}{2012}), \bibinfo{numpages}{23}~pages.
\newblock
\showISSN{1544-3566}


\bibitem[\protect\citeauthoryear{Tang, Mars, Vachharajani, Hundt, and
  Soffa}{Tang et~al\mbox{.}}{2011}]%
        {Tang2011}
\bibfield{author}{\bibinfo{person}{Lingjia Tang}, \bibinfo{person}{Jason Mars},
  \bibinfo{person}{Neil Vachharajani}, \bibinfo{person}{Robert Hundt}, {and}
  \bibinfo{person}{Mary~Lou Soffa}.} \bibinfo{year}{2011}\natexlab{}.
\newblock \showarticletitle{The Impact of Memory Subsystem Resource Sharing on
  Datacenter Applications}.
\newblock \bibinfo{journal}{{\em SIGARCH Comput. Archit. News\/}}
  \bibinfo{volume}{39}, \bibinfo{number}{3}, \bibinfo{pages}{283--294}.
\newblock
\showISSN{0163-5964}


\bibitem[\protect\citeauthoryear{Taylor, Marco, and Wang}{Taylor
  et~al\mbox{.}}{2017}]%
        {Taylor:2017:AOO:3078633.3081040}
\bibfield{author}{\bibinfo{person}{Ben Taylor}, \bibinfo{person}{Vicent~Sanz
  Marco}, {and} \bibinfo{person}{Zheng Wang}.} \bibinfo{year}{2017}\natexlab{}.
\newblock \showarticletitle{Adaptive Optimization for OpenCL Programs on
  Embedded Heterogeneous Systems}. In \bibinfo{booktitle}{{\em Proceedings of
  the 18th ACM SIGPLAN/SIGBED Conference on Languages, Compilers, and Tools for
  Embedded Systems}} {\em (\bibinfo{series}{LCTES 2017})}.
  \bibinfo{pages}{11--20}.
\newblock


\bibitem[\protect\citeauthoryear{Thusoo, Sarma, Jain, Shao, Chakka, Anthony,
  Liu, Wyckoff, and Murthy}{Thusoo et~al\mbox{.}}{2009}]%
        {Thusoo2009}
\bibfield{author}{\bibinfo{person}{Ashish Thusoo}, \bibinfo{person}{Joydeep~Sen
  Sarma}, \bibinfo{person}{Namit Jain}, \bibinfo{person}{Zheng Shao},
  \bibinfo{person}{Prasad Chakka}, \bibinfo{person}{Suresh Anthony},
  \bibinfo{person}{Hao Liu}, \bibinfo{person}{Pete Wyckoff}, {and}
  \bibinfo{person}{Raghotham Murthy}.} \bibinfo{year}{2009}\natexlab{}.
\newblock \showarticletitle{Hive: A Warehousing Solution over a Map-reduce
  Framework}.
\newblock \bibinfo{journal}{{\em Proc. VLDB Endow.\/}} \bibinfo{volume}{2},
  \bibinfo{number}{2}, \bibinfo{pages}{1626--1629}.
\newblock


\bibitem[\protect\citeauthoryear{Tournavitis, Wang, Franke, and
  O'Boyle}{Tournavitis et~al\mbox{.}}{2009}]%
        {Tournavitis:2009:THA:1542476.1542496}
\bibfield{author}{\bibinfo{person}{Georgios Tournavitis},
  \bibinfo{person}{Zheng Wang}, \bibinfo{person}{Bj\"{o}rn Franke}, {and}
  \bibinfo{person}{Michael~F.P. O'Boyle}.} \bibinfo{year}{2009}\natexlab{}.
\newblock \showarticletitle{Towards a Holistic Approach to
  Auto-parallelization: Integrating Profile-driven Parallelism Detection and
  Machine-learning Based Mapping}. In \bibinfo{booktitle}{{\em Proceedings of
  the 30th ACM SIGPLAN Conference on Programming Language Design and
  Implementation}} {\em (\bibinfo{series}{PLDI '09})}.
  \bibinfo{pages}{177--187}.
\newblock


\bibitem[\protect\citeauthoryear{Vavilapalli, Murthy, Douglas, Agarwal, Konar,
  Evans, Graves, Lowe, Shah, Seth, Saha, Curino, O'Malley, Radia, Reed, and
  Baldeschwieler}{Vavilapalli et~al\mbox{.}}{2013}]%
        {Vavilapalli:2013:AHY:2523616.2523633}
\bibfield{author}{\bibinfo{person}{Vinod~Kumar Vavilapalli},
  \bibinfo{person}{Arun~C. Murthy}, \bibinfo{person}{Chris Douglas},
  \bibinfo{person}{Sharad Agarwal}, \bibinfo{person}{Mahadev Konar},
  \bibinfo{person}{Robert Evans}, \bibinfo{person}{Thomas Graves},
  \bibinfo{person}{Jason Lowe}, \bibinfo{person}{Hitesh Shah},
  \bibinfo{person}{Siddharth Seth}, \bibinfo{person}{Bikas Saha},
  \bibinfo{person}{Carlo Curino}, \bibinfo{person}{Owen O'Malley},
  \bibinfo{person}{Sanjay Radia}, \bibinfo{person}{Benjamin Reed}, {and}
  \bibinfo{person}{Eric Baldeschwieler}.} \bibinfo{year}{2013}\natexlab{}.
\newblock \showarticletitle{Apache Hadoop YARN: Yet Another Resource
  Negotiator}. In \bibinfo{booktitle}{{\em Proceedings of the 4th Annual
  Symposium on Cloud Computing}} {\em (\bibinfo{series}{SOCC '13})}.
\newblock


\bibitem[\protect\citeauthoryear{Verma, Cherkasova, and Campbell}{Verma
  et~al\mbox{.}}{2012}]%
        {6298160}
\bibfield{author}{\bibinfo{person}{A. Verma}, \bibinfo{person}{L. Cherkasova},
  {and} \bibinfo{person}{R.~H. Campbell}.} \bibinfo{year}{2012}\natexlab{}.
\newblock \showarticletitle{Two Sides of a Coin: Optimizing the Schedule of
  MapReduce Jobs to Minimize Their Makespan and Improve Cluster Performance}.
  In \bibinfo{booktitle}{{\em 2012 IEEE 20th International Symposium on
  Modeling, Analysis and Simulation of Computer and Telecommunication
  Systems}}. \bibinfo{pages}{11--18}.
\newblock


\bibitem[\protect\citeauthoryear{Wang, Grewe, and O’boyle}{Wang
  et~al\mbox{.}}{2015}]%
        {wang2015automatic}
\bibfield{author}{\bibinfo{person}{Zheng Wang}, \bibinfo{person}{Dominik
  Grewe}, {and} \bibinfo{person}{Michael~FP O’boyle}.}
  \bibinfo{year}{2015}\natexlab{}.
\newblock \showarticletitle{Automatic and portable mapping of data parallel
  programs to opencl for gpu-based heterogeneous systems}.
\newblock \bibinfo{journal}{{\em ACM Transactions on Architecture and Code
  Optimization (TACO)\/}} \bibinfo{volume}{11}, \bibinfo{number}{4}
  (\bibinfo{year}{2015}), \bibinfo{pages}{42}.
\newblock


\bibitem[\protect\citeauthoryear{Wang and O'Boyle}{Wang and O'Boyle}{2009}]%
        {Wang:2009:MPM:1504176.1504189}
\bibfield{author}{\bibinfo{person}{Zheng Wang} {and} \bibinfo{person}{Michael
  O'Boyle}.} \bibinfo{year}{2009}\natexlab{}.
\newblock \showarticletitle{Mapping Parallelism to Multi-cores: A Machine
  Learning Based Approach}. In \bibinfo{booktitle}{{\em Proceedings of the 14th
  ACM SIGPLAN Symposium on Principles and Practice of Parallel Programming}}
  {\em (\bibinfo{series}{PPoPP '09})}. \bibinfo{pages}{75--84}.
\newblock


\bibitem[\protect\citeauthoryear{Wang and O'Boyle}{Wang and O'Boyle}{2010}]%
        {7851530}
\bibfield{author}{\bibinfo{person}{Z. Wang} {and} \bibinfo{person}{M.~F.~P.
  O'Boyle}.} \bibinfo{year}{2010}\natexlab{}.
\newblock \showarticletitle{Partitioning streaming parallelism for multi-cores:
  A machine learning based approach}. In \bibinfo{booktitle}{{\em 19th
  International Conference on Parallel Architectures and Compilation Techniques
  (PACT)}} {\em (\bibinfo{series}{PACT '10})}. \bibinfo{pages}{307--318}.
\newblock


\bibitem[\protect\citeauthoryear{Wang and O'boyle}{Wang and O'boyle}{2013}]%
        {Wang:2013:UML:2509420.2512436}
\bibfield{author}{\bibinfo{person}{Zheng Wang} {and} \bibinfo{person}{Michael
  F.~P. O'boyle}.} \bibinfo{year}{2013}\natexlab{}.
\newblock \showarticletitle{Using Machine Learning to Partition Streaming
  Programs}.
\newblock \bibinfo{journal}{{\em ACM Trans. Archit. Code Optim.\/}}
  \bibinfo{volume}{10}, \bibinfo{number}{3} (\bibinfo{year}{2013}),
  \bibinfo{pages}{20:1--20:25}.
\newblock


\bibitem[\protect\citeauthoryear{Wang, Tournavitis, Franke, and O'boyle}{Wang
  et~al\mbox{.}}{2014}]%
        {Wang:2014:IPP:2591460.2579561}
\bibfield{author}{\bibinfo{person}{Zheng Wang}, \bibinfo{person}{Georgios
  Tournavitis}, \bibinfo{person}{Bj\"{o}rn Franke}, {and}
  \bibinfo{person}{Michael F.~P. O'boyle}.} \bibinfo{year}{2014}\natexlab{}.
\newblock \showarticletitle{Integrating Profile-driven Parallelism Detection
  and Machine-learning-based Mapping}.
\newblock \bibinfo{journal}{{\em ACM Trans. Archit. Code Optim.\/}}
  \bibinfo{volume}{11}, \bibinfo{number}{1} (\bibinfo{year}{2014}),
  \bibinfo{pages}{2:1--2:26}.
\newblock


\bibitem[\protect\citeauthoryear{Wen, Wang, and O'Boyle}{Wen
  et~al\mbox{.}}{2014}]%
        {c4ce9ce754bf431fba9c3ce457e6f28d}
\bibfield{author}{\bibinfo{person}{Yuan Wen}, \bibinfo{person}{Zheng Wang},
  {and} \bibinfo{person}{Michael~FP O'Boyle}.} \bibinfo{year}{2014}\natexlab{}.
\newblock \showarticletitle{Smart multi-task scheduling for OpenCL programs on
  CPU/GPU heterogeneous platforms}. In \bibinfo{booktitle}{{\em High
  Performance Computing (HiPC), 2014 21st International Conference on}}.
  \bibinfo{publisher}{IEEE}, \bibinfo{pages}{1--10}.
\newblock


\bibitem[\protect\citeauthoryear{Xu, Saltaformaggio, Gamage, Kompella, and
  Xu}{Xu et~al\mbox{.}}{2015}]%
        {Xu2015}
\bibfield{author}{\bibinfo{person}{Cong Xu}, \bibinfo{person}{Brendan
  Saltaformaggio}, \bibinfo{person}{Sahan Gamage}, \bibinfo{person}{Ramana~Rao
  Kompella}, {and} \bibinfo{person}{Dongyan Xu}.}
  \bibinfo{year}{2015}\natexlab{}.
\newblock \showarticletitle{vRead: Efficient Data Access for Hadoop in
  Virtualized Clouds}. In \bibinfo{booktitle}{{\em Proceedings of the 16th
  Annual Middleware Conference}} {\em (\bibinfo{series}{Middleware '15})}.
  \bibinfo{publisher}{ACM}, \bibinfo{address}{New York, NY, USA},
  \bibinfo{pages}{125--136}.
\newblock
\showISBNx{978-1-4503-3618-5}


\bibitem[\protect\citeauthoryear{Xu, Li, Zhang, Butt, Wang, and Hu}{Xu
  et~al\mbox{.}}{2016}]%
        {ipdps16}
\bibfield{author}{\bibinfo{person}{L. Xu}, \bibinfo{person}{M. Li},
  \bibinfo{person}{L. Zhang}, \bibinfo{person}{A.~R. Butt}, \bibinfo{person}{Y.
  Wang}, {and} \bibinfo{person}{Z.~Z. Hu}.} \bibinfo{year}{2016}\natexlab{}.
\newblock \showarticletitle{MEMTUNE: Dynamic Memory Management for In-Memory
  Data Analytic Platforms}. In \bibinfo{booktitle}{{\em 2016 IEEE International
  Parallel and Distributed Processing Symposium (IPDPS)}}.
  \bibinfo{publisher}{IEEE}, \bibinfo{pages}{383--392}.
\newblock
\showISSN{1530-2075}


\bibitem[\protect\citeauthoryear{Yang, Breslow, Mars, and Tang}{Yang
  et~al\mbox{.}}{2013}]%
        {Yang:2013:BPO:2485922.2485974}
\bibfield{author}{\bibinfo{person}{Hailong Yang}, \bibinfo{person}{Alex
  Breslow}, \bibinfo{person}{Jason Mars}, {and} \bibinfo{person}{Lingjia
  Tang}.} \bibinfo{year}{2013}\natexlab{}.
\newblock \showarticletitle{Bubble-flux: Precise Online QoS Management for
  Increased Utilization in Warehouse Scale Computers}. In
  \bibinfo{booktitle}{{\em Proceedings of the 40th Annual International
  Symposium on Computer Architecture}} {\em (\bibinfo{series}{ISCA '13})}.
\newblock


\bibitem[\protect\citeauthoryear{Yang, Blackburn, and McKinley}{Yang
  et~al\mbox{.}}{2015}]%
        {Yang:2015:CPM:2749469.2750401}
\bibfield{author}{\bibinfo{person}{Xi Yang}, \bibinfo{person}{Stephen~M.
  Blackburn}, {and} \bibinfo{person}{Kathryn~S. McKinley}.}
  \bibinfo{year}{2015}\natexlab{}.
\newblock \showarticletitle{Computer Performance Microscopy with Shim}.
\newblock \bibinfo{journal}{{\em SIGARCH Comput. Archit. News\/}}
  \bibinfo{volume}{43}, \bibinfo{number}{3}, \bibinfo{pages}{170--184}.
\newblock
\showISSN{0163-5964}


\bibitem[\protect\citeauthoryear{Zaharia, Chowdhury, Franklin, Shenker, and
  Stoica}{Zaharia et~al\mbox{.}}{2010}]%
        {Zaharia2010}
\bibfield{author}{\bibinfo{person}{Matei Zaharia}, \bibinfo{person}{Mosharaf
  Chowdhury}, \bibinfo{person}{Michael~J. Franklin}, \bibinfo{person}{Scott
  Shenker}, {and} \bibinfo{person}{Ion Stoica}.}
  \bibinfo{year}{2010}\natexlab{}.
\newblock \showarticletitle{Spark: Cluster Computing with Working Sets}. In
  \bibinfo{booktitle}{{\em Proceedings of the 2Nd USENIX Conference on Hot
  Topics in Cloud Computing}} {\em (\bibinfo{series}{HotCloud'10})}.
  \bibinfo{pages}{10--10}.
\newblock


\bibitem[\protect\citeauthoryear{Zhuravlev, Blagodurov, and Fedorova}{Zhuravlev
  et~al\mbox{.}}{2010}]%
        {Zhuravlev:2010:ASR:1736020.1736036}
\bibfield{author}{\bibinfo{person}{Sergey Zhuravlev}, \bibinfo{person}{Sergey
  Blagodurov}, {and} \bibinfo{person}{Alexandra Fedorova}.}
  \bibinfo{year}{2010}\natexlab{}.
\newblock \showarticletitle{Addressing Shared Resource Contention in Multicore
  Processors via Scheduling}.
\newblock \bibinfo{journal}{{\em SIGPLAN Not.\/}} \bibinfo{volume}{45},
  \bibinfo{number}{3}, \bibinfo{pages}{129--142}.
\newblock
\showISSN{0362-1340}


\end{thebibliography}
\balance
\end{document}